\documentclass[useAMS,usenatbib]{style/mnras}

\usepackage{times}

\usepackage{setspace}
\usepackage{graphics}
\usepackage{graphicx}
\usepackage{amssymb}
\usepackage{longtable}
\usepackage{amsmath}
\usepackage{amsbsy,fixmath}
\usepackage{epstopdf}
\usepackage{color}
\usepackage[T1]{fontenc}
\usepackage{aecompl}
\usepackage[none]{hyphenat}
\usepackage[normalem]{ulem}
\usepackage{physics}
\usepackage{mathrsfs}

\bibpunct[,]{(}{)}{;}{a}{}{,}

\usepackage{colortbl}
\usepackage{xcolor}
\usepackage{bm}
\usepackage{float}
\usepackage{multicol}

\usepackage{tabularx}

\usepackage{csquotes}
\usepackage{multirow}
\usepackage{rotating}

\definecolor{gray97}{gray}{.97}
\definecolor{gray75}{gray}{.75}
\definecolor{gray45}{gray}{.45}

\newcommand{\msun}{\ensuremath{M_{\odot}}}

\newcommand{\F}[1]{\ensuremath{F=#1}}
\newcommand{\runa}{\texttt{RunA}}
\newcommand{\runb}{\texttt{RunB}}

\defcitealias{Goicovic2016}{Paper~I}
\newcommand{\paperI}{\citetalias{Goicovic2016}}

\defcitealias{Goicovic2017}{Paper~II}
\newcommand{\paperII}{\citetalias{Goicovic2017}}
\title[Clumpy accretion onto MBHBs]{Accretion of clumpy cold gas onto massive black hole binaries: a possible fast route to binary coalescence}

\author[Goicovic et al.]{Felipe G. Goicovic$^{1,2}$\thanks{E-mail:
felipe.goicovic@gmail.com}, Cristi\'an Maureira-Fredes$^{3}$, Alberto Sesana$^{4}$,\newauthor Pau Amaro-Seoane$^{5,6,7,8}$, and Jorge Cuadra$^{2,9}$\\
$^{1}$Heidelberg Institute for Theoretical Studies (HITS), Schloss-Wolfsbrunnenweg 35, D-69118 Heidelberg, Germany\\
$^{2}$Instituto de Astrof\'isica, Pontificia Universidad Cat\'olica de Chile, Av. Vicu\~na Mackenna 4860, 7820436 Macul, Santiago, Chile\\
$^{3}$Max Planck Institute for Gravitational Physics (Albert Einstein Institute), Am M\"ulenberg 1, 14476, Potsdam-Golm, Germany\\
$^{4}$School of Physics and Astronomy and Institute of Gravitational Wave Astronomy, University of Birmingham, Edgbaston B15 2TT, United Kingdom\\
$^{5}$Institute of Space Sciences (ICE, CSIC) \& Institut d'Estudis Espacials de Catalunya (IEEC)
          at Campus UAB, Carrer de Can Magrans s/n 08193 Barcelona, Spain\\
$^{6}$Institute of Applied Mathematics,\
           Academy of Mathematics and Systems Science,\
           CAS, Beijing 100190, China\\
$^{7}$Kavli Institute for Astronomy and Astrophysics,\
           Beijing 100871, China\\
$^{8}$Zentrum f\"ur Astronomie und Astrophysik, TU Berlin,\
           Hardenbergstra{\ss}e 36, 10623 Berlin, Germany\\
$^{9}$Max Planck Institute f\"ur extraterrestriche Physik (MPE), D-85748 Garching, Germany\\
}

\begin{document}

\date{\today}

\pagerange{\pageref{firstpage}--\pageref{lastpage}} \pubyear{2017}

\maketitle

\label{firstpage}

\begin{abstract}
In currently favoured hierarchical cosmologies, the formation of massive black hole binaries (MBHBs) following galaxy mergers is unavoidable. Still, due the complex physics governing the (hydro)dynamics of the post-merger dense environment of stars and gas in galactic nuclei, the final fate of those MBHBs is still unclear. In gas-rich environments, it is plausible that turbulence and gravitational instabilities feed gas to the nucleus in the form of a series of cold incoherent clumps, thus providing a way to exchange energy and angular momentum between the MBHB and its surroundings. Within this context, we present a suite of smoothed-particle-hydrodynamical models to study the evolution of a sequence of near-radial turbulent gas clouds as they infall towards equal-mass, circular MBHBs. We focus on the dynamical response of the binary orbit to different levels of anisotropy of the incoherent accretion events. Compared to a model extrapolated from a set of individual cloud-MBHB interactions, we find that accretion increases considerably and the binary evolution is faster. This occurs because the continuous infall of clouds drags inwards circumbinary gas left behind by previous accretion events, thus promoting a more  effective exchange of angular momentum between the MBHB and the gas. These results suggest that sub-parsec MBHBs efficiently evolve towards coalescence during the interaction with a sequence of individual gas pockets.

\end{abstract}


\newcommand{\papergas}{\citet{MaureiraFredes2018}}
\newcommand{\tojorge}[1]{\textcolor{cyan}{\textbf{To Jorge:} #1}}
\newcommand{\tosesa}[1]{\textcolor{cyan}{\textbf{To Alberto:} #1}}

\section{Introduction}

Massive black holes (MBHs) with masses between $\sim$$10^6-10^{10}M_\odot$
\citep{Kelly2013} have been observed to inhabit the nuclei of massive
galaxies. And even though these objects are tiny with respect to their hosts,
they are believed to play a fundamental, yet not fully understood, role in
galaxy evolution. In fact, the observed tight correlations between the MBH
masses and the properties of their galaxies
\citep[e.g.][]{1998AJ....115.2285M,2000ApJ...539L...9F,2000ApJ...539L..13G}
strongly suggest a coevolution scenario through feedback processes
\citep[][ and references therein]{Kormendy2013}.

Furthermore, according to the currently favoured $\Lambda$CDM cosmology,
galaxies assemble hierarchically through mergers following their parent dark
matter haloes, building up from smaller structures \citep{1978MNRAS.183..341W}.
A natural outcome of this process is the formation of massive black hole
binaries (hereafter MBHBs). Following a merger of two galaxies, we expect
both MBHs harboured in their nuclei to sink toward the centre of the merger
remnant and eventually form a binary \citep{Beg80}.
Therefore, understanding MBHB formation and dynamical evolution is
a necessary piece for reconstructing the puzzle of the hierarchical
growth of structures in the Universe. This is particularly interesting if we
consider that MBHBs are powerful sources of gravitational waves (GWs),
detectable by the future space-based laser interferometer space antenna
\citep[LISA,][]{Consortium2013,2017arXiv170200786A}. For GWs to carry away
enough energy, however, the two MBHs have to reach a separation of
$a_{\rm GW}\sim 10^{-3}$ pc. This is extremely small compared to the
$\approx 1$pc separation at which dynamical friction, initially
driving the evolution of the two MBHs in the merger remnant, ceases to be
efficient \citep{2008gady.book.....B}.

The exact evolution and the final fate of the MBHs strongly depends on the
content and distribution of both gas and stars in the galactic nucleus, and
different processes can change dramatically the merging timescale, even
resulting in values higher than the Hubble time. Although there is some
indirect evidence that these binaries coalesce into a single object
\citep[e.g., the low scatter in the $M-\sigma$ relation,][]{Jahnke2011},
the processes involved are not completely understood.

For instance, due to its dissipative nature, gas can be very efficient in
absorbing and transporting outwards the angular momentum of the pair, likely
leading to a rapid evolution and eventual coalescence. From observations, as
well as numerical simulations, it has been established that in gas-rich galaxy
mergers there is a large inflow of gaseous material to the central kiloparsec
of the galactic remnant, often resulting in a massive circumnuclear disc
\citep{Barnes1992, SandersMirabel1996,2007Sci...316.1874M}.
Driven by dynamical friction and global torques from this disc,
the pair of MBHs decays very
efficiently down to separations of the order of $\sim$ 1--0.1 pc, where it
forms a gravitationally bound binary \citep{2004ApJ...607..765E,
2005ApJ...630..152E, 2007Sci...316.1874M, Fiacconi2013, Roskar2015,
DelValle2015}.
At these sub-parsec scales, most theoretical and numerical studies have
focused on the evolution of binaries surrounded by a gaseous circumbinary
disc, often either co-rotating \citep[see e.g.][]{Ivanov1999,ArmNat05,C09, Haiman2009,
Lod09,Nix11a,Roedig2011,Roedig2012,Kocsis2012,Pau2013,DOrazio2013,
Munoz2016,Miranda2017,Tang2017}
or conuter-rotating \citep[see e.g.][]{Nixon2015,Roedig2014} with respect to the
binary's orbital motion. These discs are generally assumed to be well-defined,
smooth, and relaxed, with no attempt to link their
presence to the gaseous environment around the binary, nor to the fuelling
mechanisms that bring gas to the nucleus. Furthermore, all these idealised
scenarios are subject to the disc consumption problem, namely, if the disc
dissolves through some process \citep[e.g. star formation, AGN/supernovae
feedback,][]{Lupi2015}, the evolution of the binary orbit stops. In fact,
the evolution of MBHBs in gas-rich environments is intimately
related to the unsolved problem of gas supply to the centre of galactic
nuclei.

The fuelling of gaseous material onto galactic nuclei is currently a critical,
yet uncertain piece in the galaxy formation puzzle. The wide range of physical
scales involved makes the fuelling a very complex process -- the gas needs
to lose its angular momentum efficiently to be transported from galactic
scales down to the nuclear region \citep{Hopkins2011}.
A plausible mechanism that could break this angular momentum barrier is
provided by chaotic feeding or ballistic accretion of cold streams when
the gas is cold enough to form dense clouds or filaments \citep{Hobbs11}.

Cold accretion onto black holes has long been predicted by both theory
and simulations. The possible relevance of discrete and randomly
oriented accretion events was first highlighted by \citet{KP06}
\citep[see also][]{King2007, Nayakshin2007, Nayakshin2012}, which they referred
to as \enquote{chaotic accretion} scenario. They argue that a series of gas
pockets falling from uncorrelated directions could explain the rapid growth
of MBHs across cosmic time, and can reproduce some of the observed properties
(e.g. the $M_{\rm BH}-\sigma$ relation).
One of the most extensive efforts to study the evolution of multiphase gaseous
haloes and its impact on MBH accretion is the work of Gaspari and collaborators
\citep{Gaspari2012, Gaspari2013, Gaspari2015, Gaspari2017a, Gaspari2017b,
Gaspari2018}.
By means of numerical simulations, they have shown that realistic
turbulence, cooling, and heating affecting the hot halo, can dramatically change
the accretion flow onto black holes, departing from the idealised picture
of the Bondi prescription \citep{Bondi1952}. Under certain physical conditions,
cold clouds and filaments condense out of the hot phase due to thermal
instabilities. Chaotic collisions then promote the funnelling of
the cold phase towards the MBH, leading to episodic spikes in the
accretion rate.

A likely example of this kind of discrete accretion events is the putative
molecular cloud that resulted in the unusual distribution of stars orbiting
our Galaxy's MBH. Several numerical studies have shown that portions of a
near-radial gas infall can be captured by a MBH to form one or more eccentric
discs that eventually fragment to form stars
\citep{BR08, HobNay09, Alig2011, Gaburov2012, Map12, Luc13,Mapelli2016},
roughly reproducing the observed stellar distribution
\citep{Paumard2006, Lu2009}.
An interesting feature of these observed stellar discs is that they appear to
be misaligned with respect to the Galactic plane, indicating that the
infalling gas had an angular momentum direction unrelated to that of the
large scale structure of the Galaxy.

From an observational perspective, the development of multi-wavelength
observations has started to unveil the multiphase structure of massive galaxies.
These galaxies have been observed to harbour neutral and molecular gas down
to $\sim$10 K \citep[e.g.][]{Combes2007}. The Atacama Large Millimetre Array
(ALMA) has opened the gate to high-resolution detection of molecular gas
in early-type galaxies. For instance, there have been observations of giant
molecular associations in the NGC5044 group within $r\lesssim$ 4 kpc,
inferred to have chaotic dynamics \citep{David2014}.
In massive galaxy clusters, ALMA has detected molecular hydrogen in the
core of the Abell~1835 galaxy cluster \citep{McNamara2014},
which may  be supported in a rotating, turbulent disc oriented nearly face-on.
In the Abell~1664 cluster, the molecular gas shows asymmetric velocity
structure, with two gas clumps flowing into the nucleus \citep{Russell2014}.
More importantly, in a recent study, \citet{Tremblay2016} reported observations
that reveal a cold, clumpy accretion flow towards a massive black hole fuel
reservoir in the nucleus of the Abell~2597 galaxy cluster, a nearby giant
elliptical galaxy surrounded by a dense halo of hot plasma.
They infer that these cold molecular clouds are within the innermost hundred
parsecs, moving inwards at about 300 kilometres per second, likely fuelling
the black hole with gaseous material. The authors claim that this is the first
time a distribution of molecular clouds has been unambiguously linked with
MBH growth.
{Finally, \citet{Temi2017} have presented the detection of
molecular clouds in three group centred elliptical galaxies
(NGC 5846, NGC 4636, and NGC 5044) which are likely the result
of condensation of the hot gas, and are consistent with a turbulent
dynamics, as predicted by the cold chaotic accretion scenario
\citep{Gaspari2013}.}

Despite these several clues suggesting that stochastic condensation of cold
gas and its accretion onto the central MBH is essential for active galactic
nuclei, it has been barely explored as a possible source of gas for sub-parsec
MBHBs. Notable exceptions are the studies of \citet{Dunhill2014},
\citet[][ hereafter Paper~I]{Goicovic2016} and \citet[][ hereafter Paper~II]{Goicovic2017},
where the Authors have numerically modelled the interaction of a single
gaseous cloud with MBHBs for a variety of orbital configurations\footnote{
  {\citet{ArcaSedda2017} and \citet{Bortolas2018} took
    a similar approach in modelling the  infall of stellar clusters onto
    MBHBs. Although the phenomenology is  different, they find similar
    results to our \paperII, whereby retrograde encounters shrink more
    efficiently the binary orbit compared to prograde ones.}}.
In particular, \paperII~extrapolated the results for these single cloud models
to estimate the long-term evolution of a binary interacting with a sequence
of near-radial infalling clouds, finding a very efficient shrinking of its
orbit. However, the approach presented in \paperII~is subject to a series of
limitations, mainly, considering only the prompt evolution during the first
impact of the clouds without the secular effects of the remaining gas.
These effects can become very important after a series of accretion events
due to the formation of circumbinary structures that can continue extracting
angular momentum and increasing accretion onto the binary. With the goal of
overcoming these limitations, we present a suite of hydrodynamical simulations
of MBHBs interacting with a sequence of clouds, infalling incoherently
onto the binary with different levels of anisotropy.
These models now combine the `discrete' evolution due to the prompt
interaction of each individual cloud, with the secular torques exerted by
the left-over gas.

This paper is organised as follows.
We first describe our numerical model in Section~\ref{sec:simulation},
highlighting the main differences with respect
to the studies presented in \paperI~and \paperII. In
Section~\ref{sec:evolution} we show the dynamical evolution of the binary
in response to the different levels of anisotropy of the accretion events,
and by comparing with the single cloud extrapolations, we identify the effects
of the non-accreted material. In Section~\ref{sec:comparison} we study the
same effects on a different suite of runs featuring an higher cloud
supply rate.
We finally summarise and discuss the
implications of our results in Section~\ref{sec:summary}.
Unless stated otherwise, throughout this paper we use units
such that $G=a_0=M_0=1$, where $G$ is the gravitational constant,
and $a_0$, $M_0$ are the initial binary separation and mass, respectively.


\section{The numerical model}
\label{sec:simulation}

We base our model of multiple clouds on the simulations presented
in \paperI~and \paperII. The new suite of simulations is
fully described in the companion paper \papergas.
We model the interaction between the gas clouds and the MBHBs using the
smoothed-particle-hydrodynamics (SPH) code \textsc{gadget}-3, an updated
version of the public code \textsc{gadget}-2 \citep{Springel2005}.

As described in \paperI, we modified the standard version of the code to
include a deterministic accretion recipe, where the sink particles
representing the MBHs accrete all bound SPH particles within a fixed radius
($r_{\rm sink}$) as implemented by \cite{Cuadra2006}; we set this radius to
$r_{\rm sink}=0.1a_0$. Additionally, we treat the thermodynamics of the gas
using a barotropic equation of state (i.e. pressure is a function of density
only) with two regimes; for densities lower than some critical value
($\rho_c$), the gas evolves isothermally, while for higher densities the gas
is adiabatic. This prescription captures the expected temperature dependence
with density of the gas without an explicit implementation of cooling, and has
the effect of halting the collapse of the densest gas, avoiding excessively
small time-steps that would stall the simulations.
For the simulations presented in this paper we have chosen
$\rho_c=10^{-2} M_0 a_0^{-3}$, which is two orders of magnitude smaller than
the value presented in \paperI~and \paperII. This change is necessary to
avoid stalling the simulations due to the formation of a large amount of
clumps in some configurations. Since we are not following in detail the
evolution of these clumps, stopping their collapse in a lower density should
not affect greatly our results. This fact only becomes important when scaling
our simulations to physical units -- for the fiducial model presented in
\paperI, where the binary has $M_0=10^6M_\odot$ and $a_0=0.2$ pc, the value of
the critical density is now $\rho_c =10^{-16}$ g cm$^{-3}$, which is not
necessarily an accurate representation of the collapse of molecular hydrogen
\citep{Masunaga2000}.

To improve the orbit integration accuracy we have made changes with respect
to the single cloud models presented in \paperII. First, we have modified
the code to remove the sink particles from the gravity tree approximation,
and their forces are instead summed directly. In simple words, each sink
particle sees all other particles individually. Additionally, the orbit of
the black holes is integrated using a fixed time-step, set equal to
$\Delta t_{\rm sink}=0.02P_0$.

A more critical modification with respect to the previous simulations is the
frequency at which the code makes the domain decomposition and full tree
reconstruction. The single cloud models of \paperII~had sufficient resolution
to study the transient evolution of the binary during the first impact of the
gas. Conversely, the typically much smaller secular torques exerted by the
remaining gas were not resolved.
The main source of this numerical noise is the aforementioned frequency,
more specifically, when the tree reconstruction is not done at the same time
with the black hole accretion, the sudden disappearance of particles during
an update creates spurious jumps in the total angular momentum,
which are generally larger than the small torques that we are trying to
resolve with these models. In order to improve the conservation of angular
momentum, we modified \textsc{gadget} to enforce the domain decomposition and
full tree reconstruction every time the sink particles are active. This
implementation is complementary to the standard criterion based on the amount
of active particles \citep{Springel2005}.
While this greatly increases the computational cost
(typically by a factor of $\sim$10), the numerical
fluctuations in the total angular momentum are vastly reduced, allowing a
robust study of its dynamical evolution during \textit{all} phases of the
interaction (see below).

Considering the aforementioned modifications, and to make these simulations
feasible, we have decreased the resolution elements of the individual clouds
with respect to the single cloud models. We model the gas clouds using
$5\times 10^4$, $2\times 10^5$, $5\times 10^5$ and $10^6$ particles (in
contrast with the $4\times 10^6$ used before) to evaluate the convergence of
our results with resolution. To identify these different resolutions
throughout this paper we use \texttt{50k}, \texttt{200k}, \texttt{500k} and
\texttt{1m}, respectively.

\subsection{External potential}

In the single cloud simulations presented in \paperI~and \paperII, the system
was treated in isolation, i.e. with no external forces acting upon it. In
reality, such a binary would be placed deep in the potential well of a galaxy,
where stars and dark matter exert their gravitational influence. To mimic the
presence of the galactic spheroid we include an external potential with a
Hernquist profile \citep{Hernquist1990} which can be written as
\begin{equation}
\rho(r) = \frac{M_*}{2\pi}\frac{a_*}{r}\frac{1}{(r+a_*)^3},
\end{equation}
and an enclosed mass given by
\begin{equation}
m(r) = M_*\frac{r^2}{(r+a_*)^2},
\end{equation}
where $M_*$ and $a_*$ are scaling constants that represent the total mass of
the spheroid and the size of core, respectively.
Based on a typical $M_{\rm BH}-M_{\rm bulge}$ relation \citep{1998AJ....115.2285M}
and a radius-to-stellar-mass relation \citep{Dabringhausen2008},
we choose $M_{\rm H}=4.78\times 10^2M_0$ and $r_{\rm H}=3.24\times 10^2a_0$,
thus implying
\begin{equation}
m(a_0) \approx 5\times10^{-3} M_0.
\end{equation}
As a consequence, this external force represents only a small perturbation to
the dynamics of the interaction that is still dominated by the binary's
gravitational potential. Moreover, because this external potential is
spherically symmetric, orbits experience precession only, where energy
and angular momentum are conserved.
  In principle, this precession could enhance stream collisions,
  prompting more gas onto the binary, although the timescales are too long
  with respect to the frequency of cloud infall, which is the dominant effect
  in our models. Based on this, we do not expect the oscillations induced
  by this potential to have a secular effect in the overall evolution of
  the system during the stages modelled.
In practice, this potential maintains the binary close to the centre of the
reference frame by applying a restoring force when it drifts away due to
the interaction with the incoming material. It is important to mention that
the gas also feels this external force.

For the single cloud simulations this potential was not needed, as drifts
produced by the interaction with the cloud or by numerical inaccuracies
did not affect the prompt phase studied.
For the simulations presented here, however, this restoring force guarantees
that the orbit of each cloud remains close
to the originally calculated values, as the pericentre distance for each cloud
was computed assuming the binary's centre of mass static at the origin of
the Cartesian coordinates.

\subsection{Initial conditions}
\label{sec:ics}

The binary consists of two sink particles, initially having equal masses and
a circular orbit in the $x-y$ plane.
On the other hand, all the clouds are initially spherical with uniform
density, a turbulent velocity field, and a total mass 100 times smaller
than the binary (i.e. $M_c=0.01M_0$). And similar to the single cloud models,
the turbulent velocity field is drawn from a random distribution with power
spectrum $P_v(k) \propto k^{-4}$. It is worth mentioning that the random
seed is exactly the same for all clouds. With these choices all clouds
are identical except for their initial orbit, which simplifies the
comparison between the different distributions.

\begin{figure}
\centering
\begin{picture}(0.5,0.8)
\put(0,0){\includegraphics[width=0.46\textwidth]{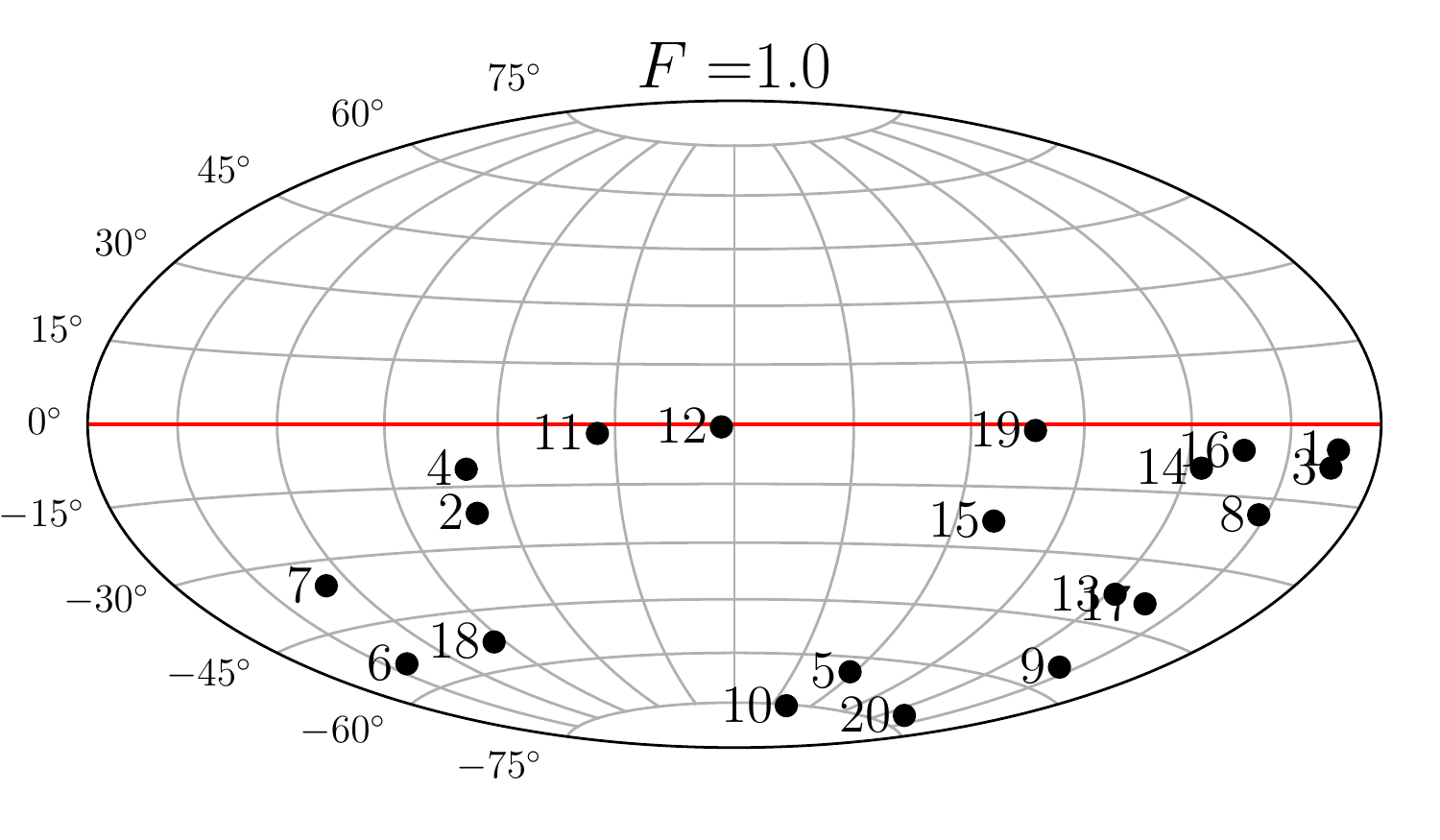}}
\put(0,0.267){\includegraphics[width=0.46\textwidth]{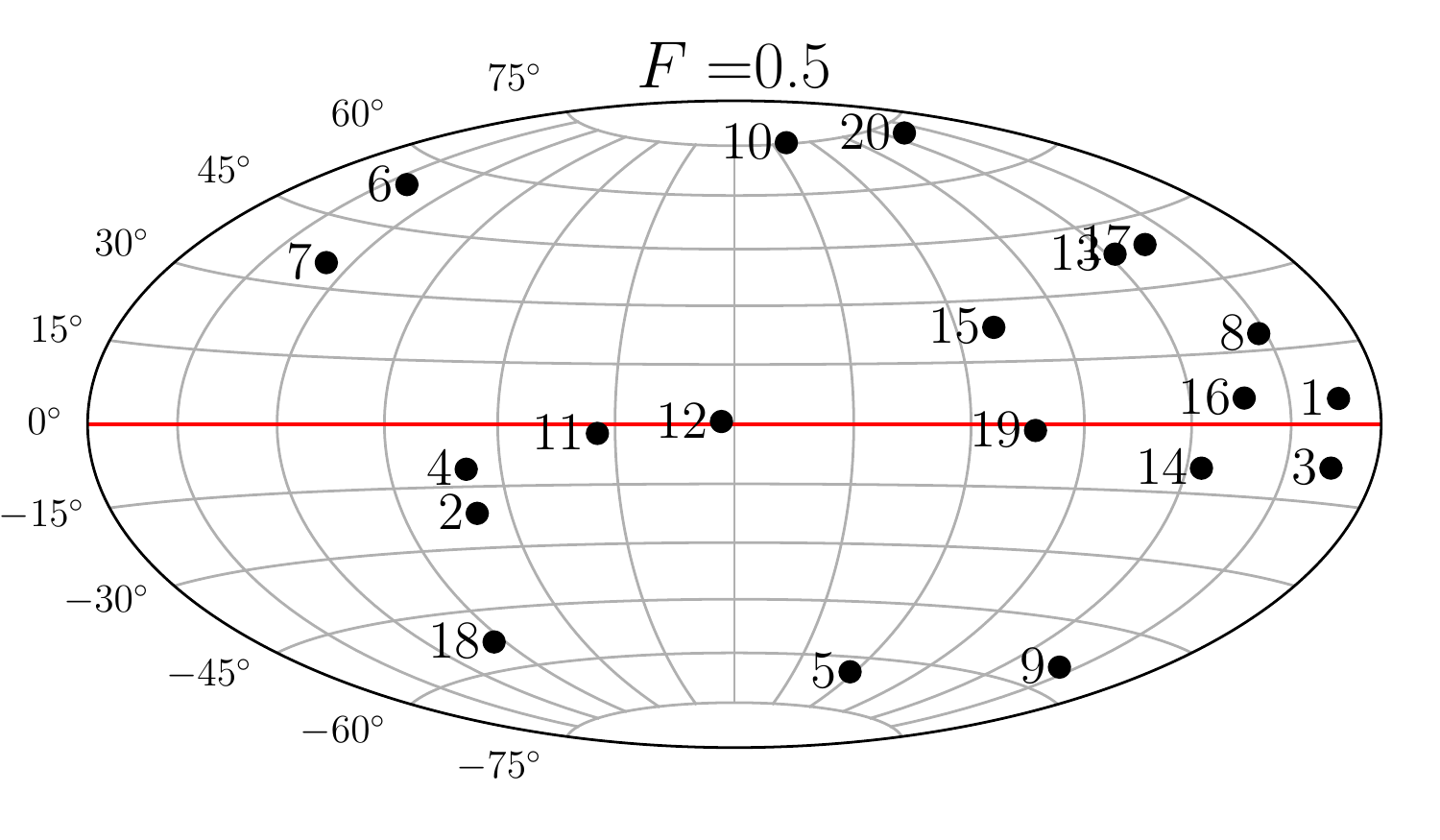}}
\put(0,0.533){\includegraphics[width=0.46\textwidth]{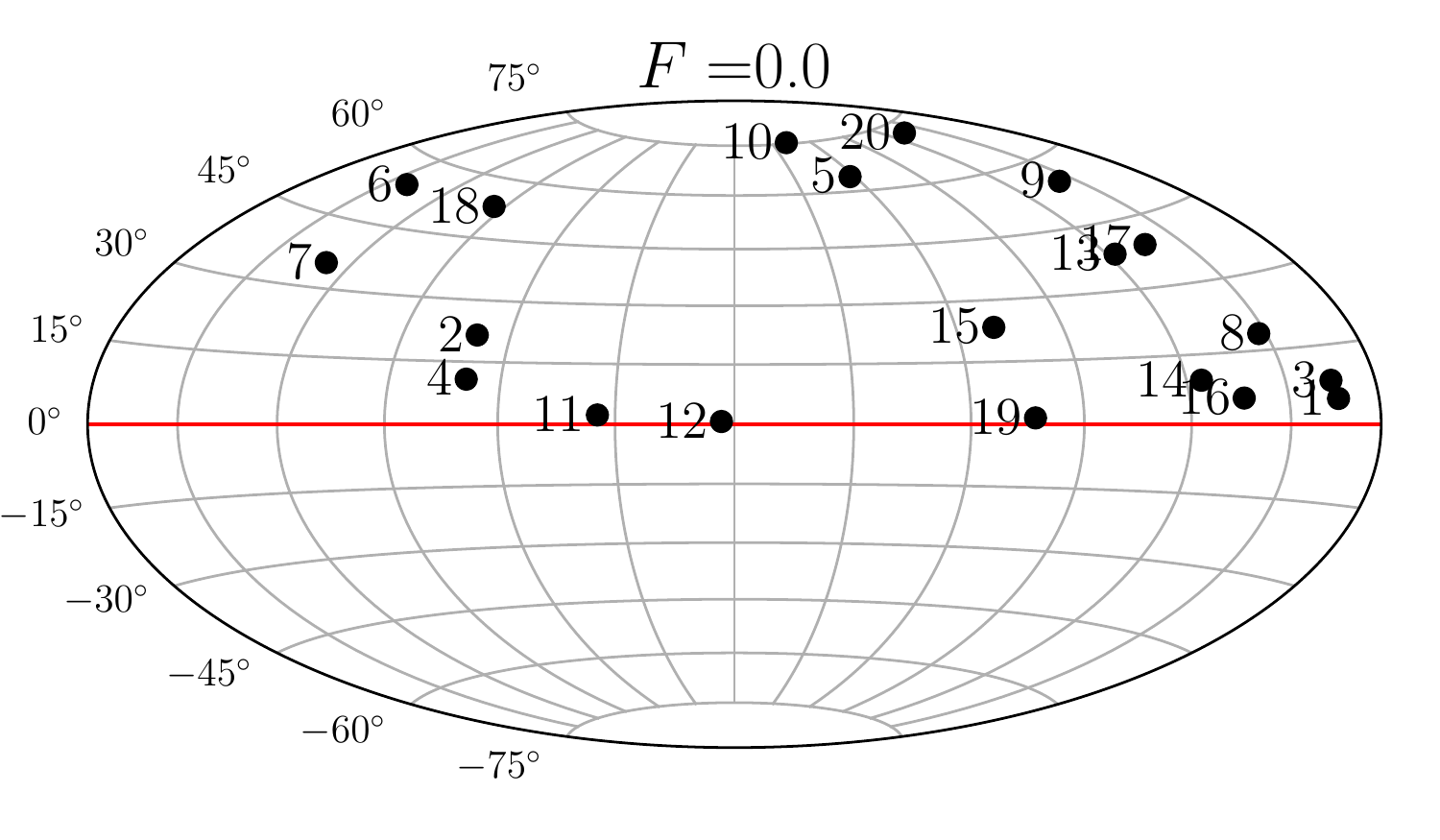}}
\end{picture}
\caption{Angular momentum orientations of the initial orbit for 20 randomly
  selected clouds from different $F$ distributions, as indicated at the top
  of each projection. The binary's angular momentum initially points
  to the north pole $(\theta=90^\circ)$ on these projections. Note that
  \F{0.0} and \F{1.0} cloud orbits are generated by mirroring the
  inclination angle $\theta$ from \F{0.5} with respect to the binary's
  orbital plane (red solid line) to swap between one hemisphere to the
  other. Labelled numbers on each circle indicate the order in which we
  introduce the clouds.}
\label{fig:ics}
\end{figure}

\begin{figure}
\centering
\includegraphics[width=0.46\textwidth]{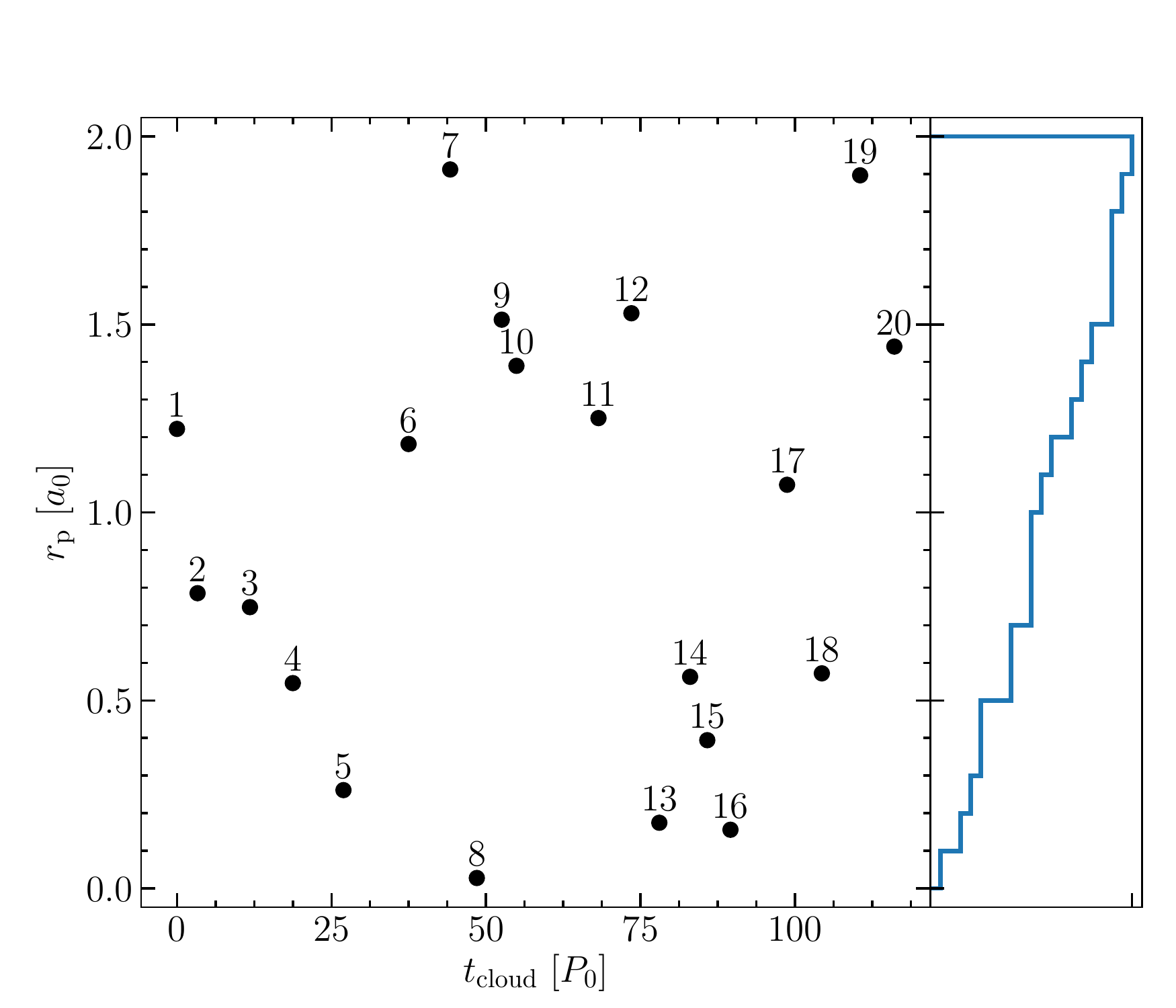}
\caption{Distribution of pericentre distances as a function of the time at
  which each cloud is included to the simulation. This distribution is used
  for the 3 levels of anisotropy $F$ shown in Fig.~\ref{fig:ics}.
  The labelled numbers indicate the order of the clouds.
  The right plot shows the cumulative distribution of pericentre distances.}
\label{fig:peris}
\end{figure}

The initial setup of these simulations is motivated by the Monte Carlo models
presented in \paperII, and is described in full detail in \papergas.
We use the $F$-distributions to model the
different levels of anisotropy of the cloud distribution, where the number $F$
represents the fraction of counter-rotating events with respect to the binary
(southern hemisphere of the projections shown in Fig.~\ref{fig:ics}).
On the other hand, we use a flat distribution of pericentre distances.
{This choice stems from the dynamics seen in simulations
  of cold clump dynamics \citet{Gaspari2013}.  Gaseous clumps
  condensing out of the interstellar medium suffer many chaotic collisions
  that redistribute their orbital angular momenta. At small orbital
  angular momenta values -- i.e. for clouds falling onto the galactic center --
  a uniform reshuffling results in a $p(b)\propto b\dd b$
  impact parameter distribution which, in turn, produces a flat distribution
  of pericentre distances due to gravitational focusing of the MBHB.}

We generate a total of 20 clouds with orientations shown in
Fig.~\ref{fig:ics} and periapsis shown in Fig.~\ref{fig:peris}.
Note that the same pericentre distances are used for the 3 different levels
of anisotropy~$F$. We set the maximum value of the pericentre distance
to $r_{\rm p, max}=2a_0$, to ensure a transient evolution with the first
impact independent of the cloud's relative inclination (see \paperII).
{Although the exact choice of maximum value is somewhat arbitrary,
\citet{Dunhill2014} demonstrated that the effect on the binary
semimajor axis produced by clouds with larger impact parameters is negligible.}

Finally, the time difference between
events exhibited on Fig.~\ref{fig:peris} was determined from
a Gamma distribution with shape 2 and a scale parameter of $\theta = 2.5 P_{0}$.
The average time between successive injections is $\Delta t\approx6P_0$.
For a binary with $M_0=10^6 M_\odot$ and $a_0=0.2$ pc, this rate of clouds
corresponds to a mass inflow of $\approx 0.2 M_\odot$~yr$^{-1}$ entering
the inner parsec.
{Note that the arrival rate of clouds is a free parameter
  in our model, and can have important consequences on the evolution
  of the binary (see comparison in \S\ref{sec:comparison}, and also
  \citealt{MaureiraFredes2018}). It is therefore important to fix it to
  a physically motivated value. Our choice ensures that the resulting
  mass inflow rate is in broad agreement with inflow rates onto single
  MBHs found by numerical studies about the formation and infall of
  cold clumps \citep[e.g.][]{Hobbs11,Gaspari2013, Gaspari2015, Fiacconi2018}.
  These models show, however, that the gaseous inflow critically depends
  on the different physical mechanisms acting on the interstellar medium,
  hence future observational campaigns are needed to better constrain
  the typical rate of these discrete accretion events on galactic nuclei.}

In order to make a robust comparison of the dynamical impact of each
distribution it is necessary to reduce some of the stochasticity produced
by the low number of events. With that goal in mind, the different angular
momentum distributions were not sampled independently from each other.
Only the angles from the isotropic distribution (\F{0.5}) are randomly drawn.
We then generate the co-rotating distribution (\F{0.0}) by reflecting
symmetrically the negative inclination angles ($\theta<0^\circ$) with
respect to the binary's orbital plane onto the northern hemisphere.
The same is done to get the counter-rotating distribution (\F{1.0}), but
mirroring the angles $\theta>0^\circ$ onto the southern hemisphere.
The azimuthal angles are kept unchanged. This means that the only difference
between distributions is the inclination angle $\theta$ of the cloud's orbits.

Because we are modelling 3 levels of anisotropy for the incoming clouds
  and we are performing runs at four different resolutions,
  we are presenting a total of 12 simulations, corresponding to \texttt{RunB}
  in \papergas. We concentrate our analysis on this set of runs because we
  were able to evolve all the three \texttt{50k} simulations up to the 20th
  cloud, thus providing a consistent dataset for the analysis. \texttt{RunA}
  of \papergas~yields similar results, which we summarise in
  \S\ref{sec:comparison}.

All the simulations described in this paper are run in different `segments'.
Initially one cloud is placed with the corresponding initial position and
velocity vectors. This system is evolved until the time set for the following
cloud to enter has been reached;
at this point we stop the simulation.
The new cloud is added to the final snapshot to generate the initial
conditions for the following section.


\section{Dynamical evolution of the binary}
\label{sec:evolution}

\subsection{Angular momentum}

\begin{figure*}
\begin{picture}(1,0.66666)
\put(0,0){\includegraphics[width=0.3333\textwidth]{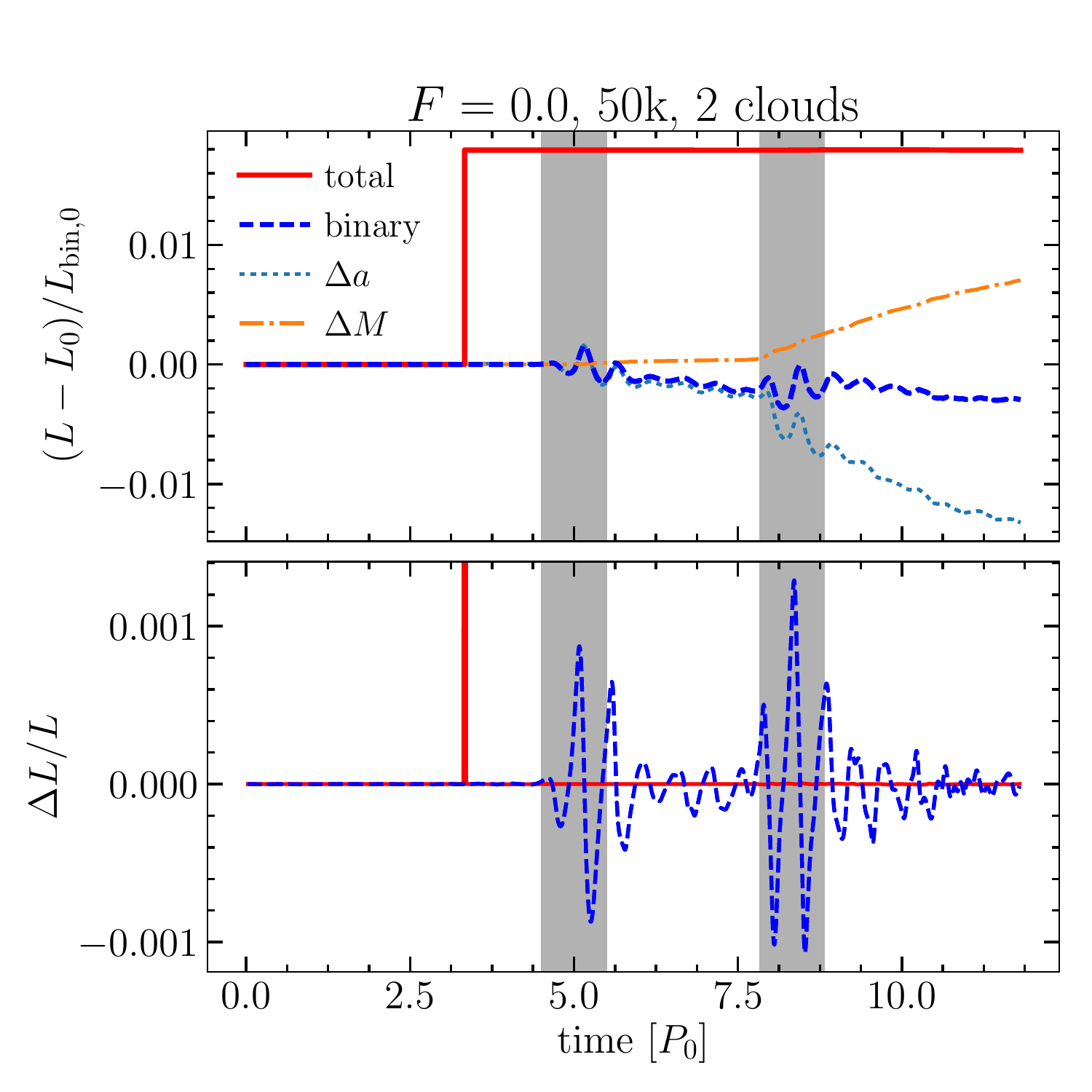}}
\put(0.3333,0){\includegraphics[width=0.3333\textwidth]{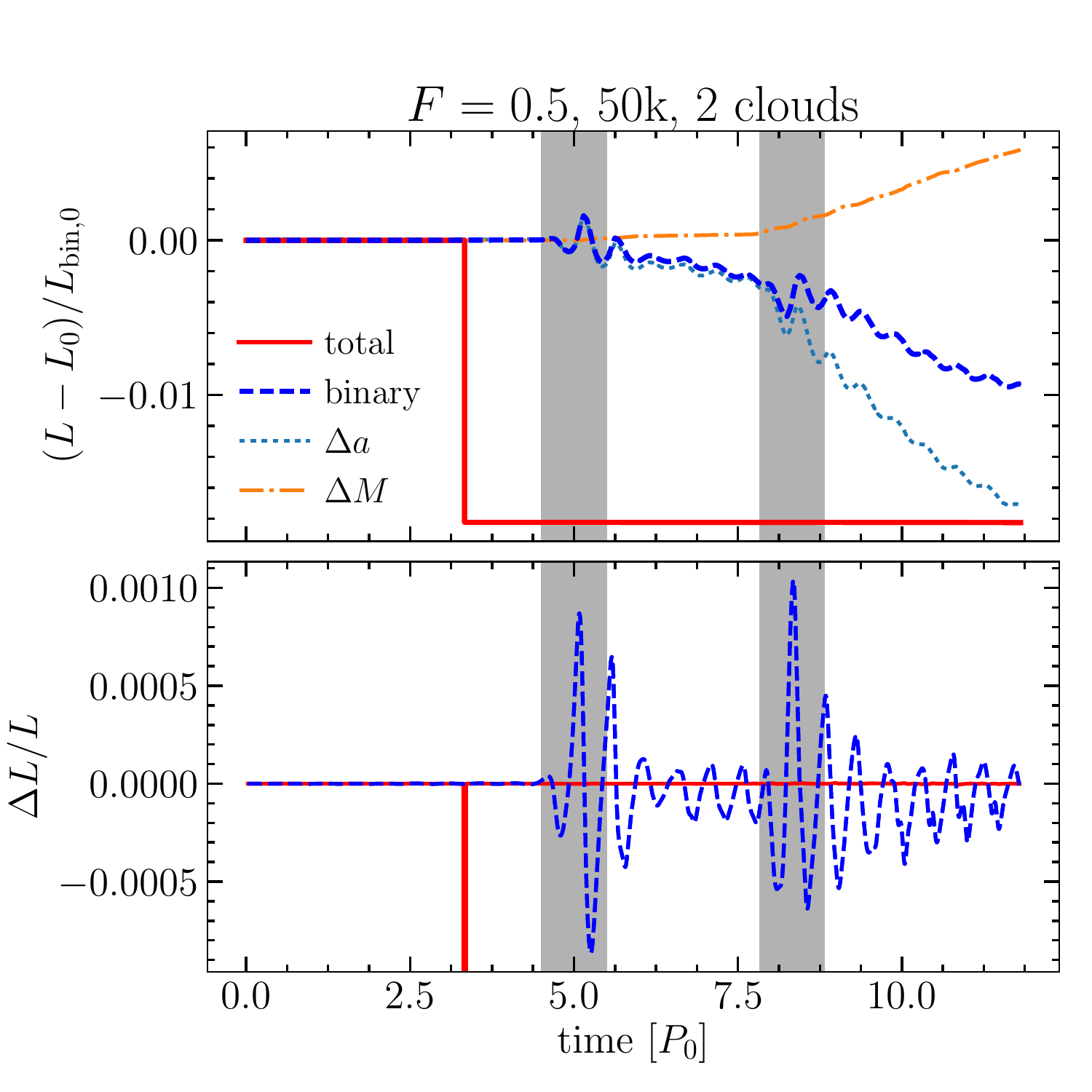}}
\put(0.6666,0){\includegraphics[width=0.3333\textwidth]{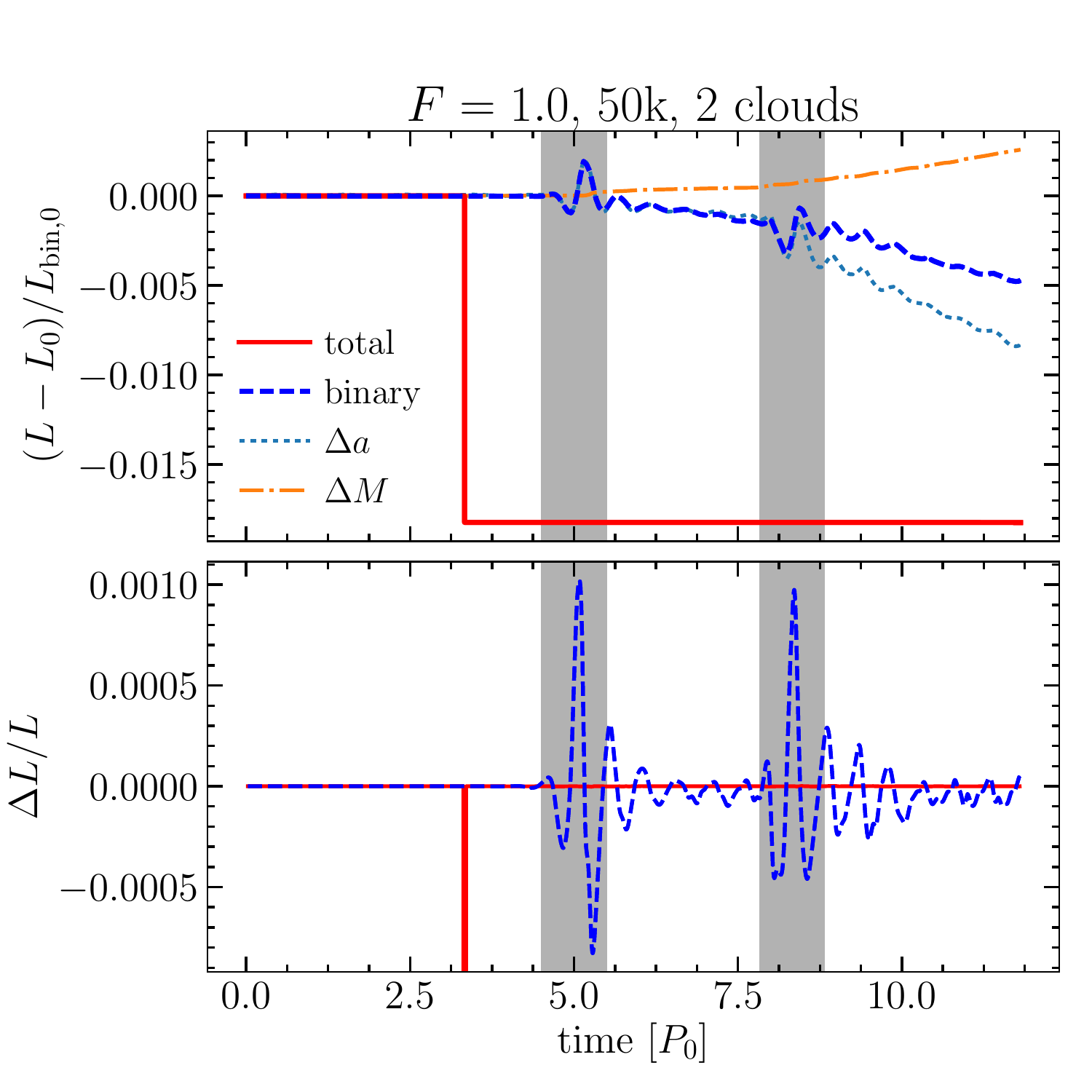}}
\put(0,0.3333){\includegraphics[width=0.3333\textwidth]{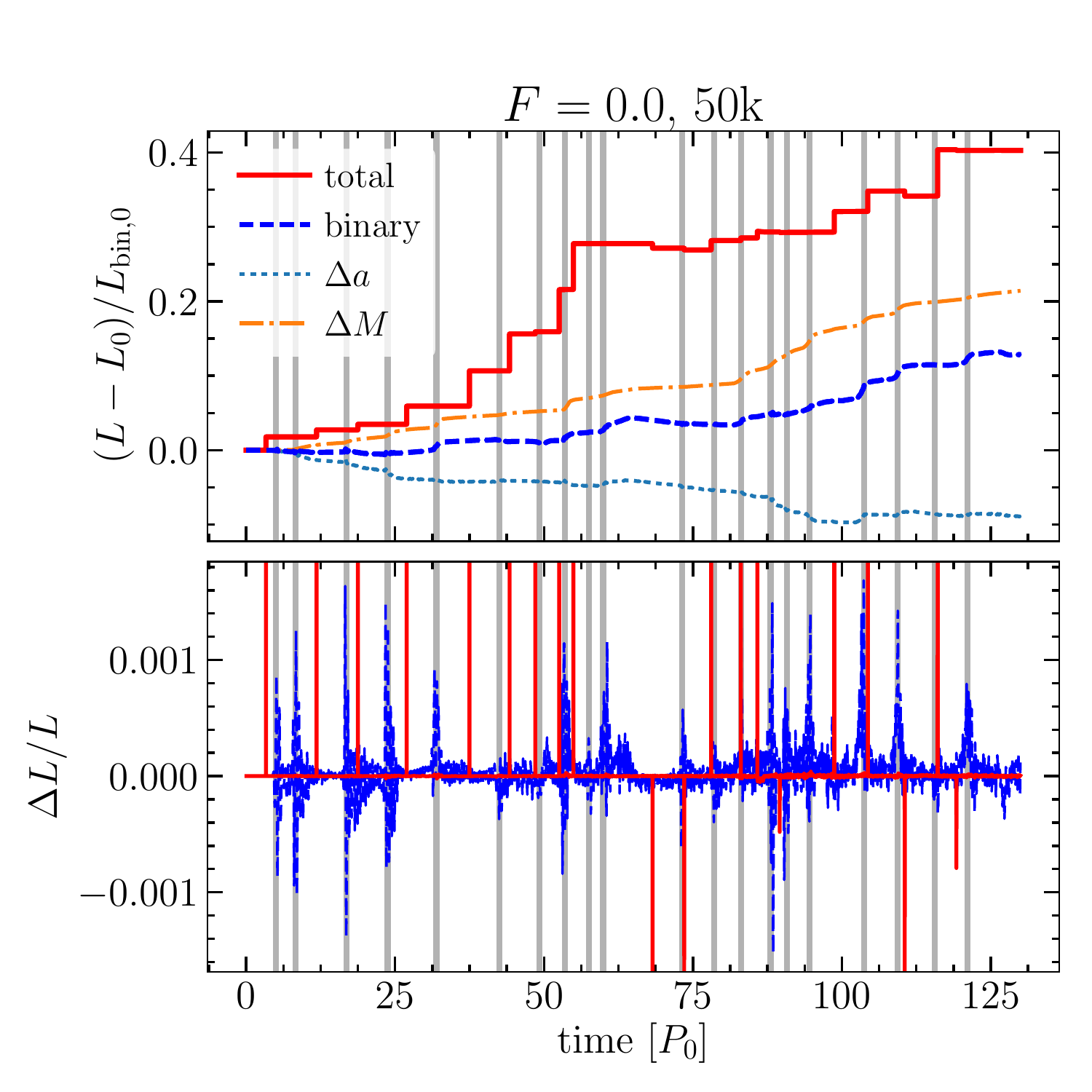}}
\put(0.3333,0.3333){\includegraphics[width=0.3333\textwidth]{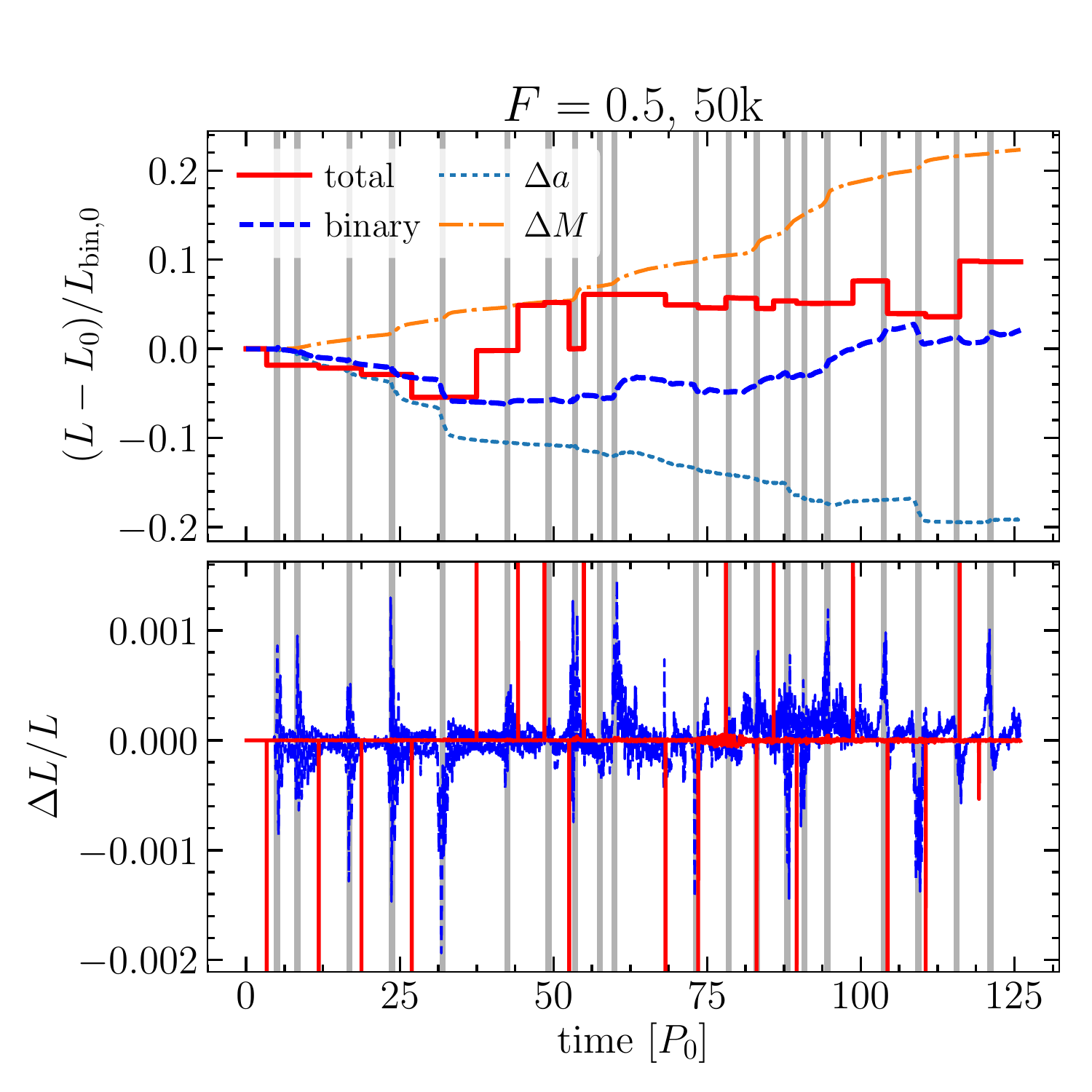}}
\put(0.6666,0.3333){\includegraphics[width=0.3333\textwidth]{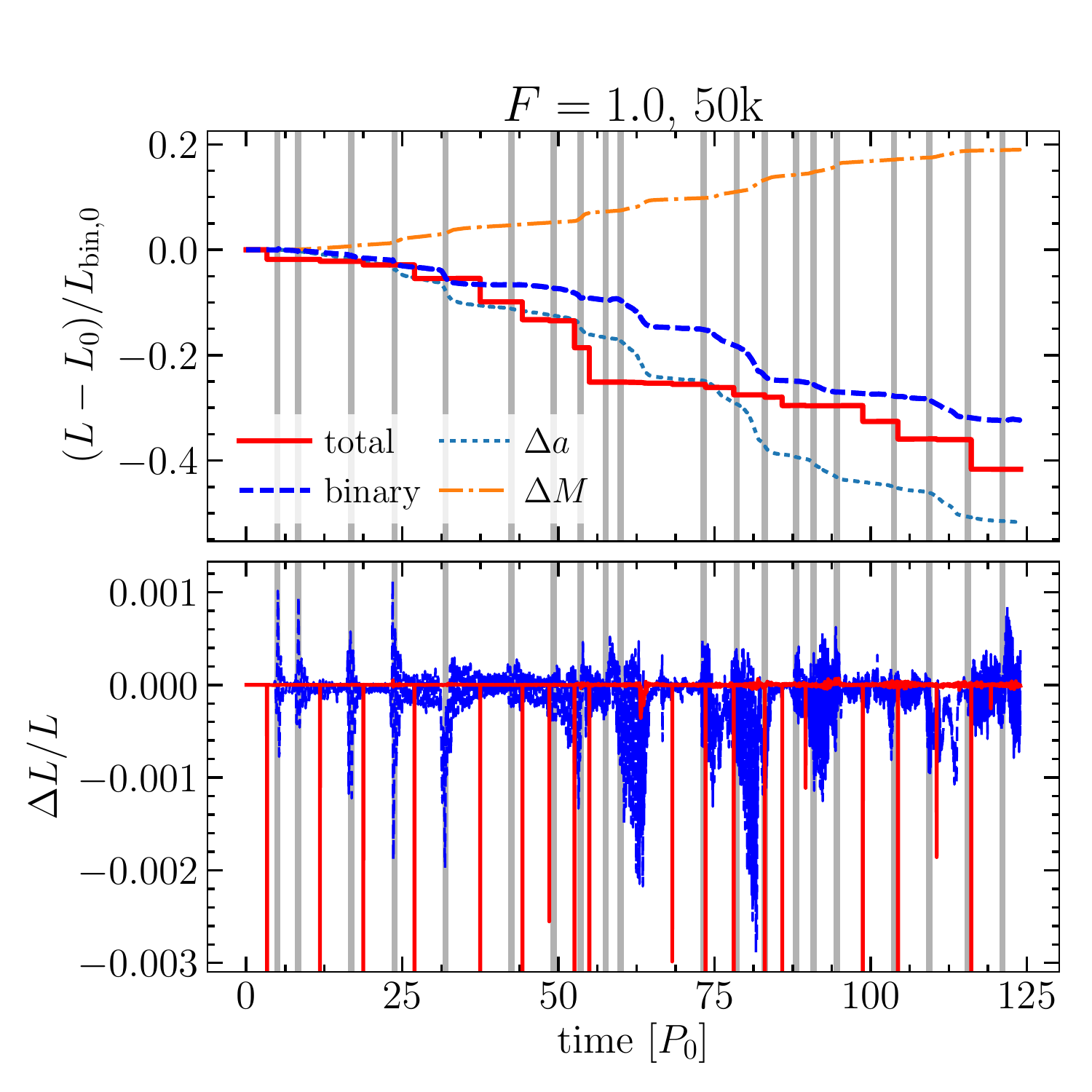}}
\end{picture}
\caption{Angular momentum evolution for the lower resolution runs
  (\texttt{50k}). The upper panels of each plot show the evolution of the
  angular momentum magnitude of the entire system (red solid lines) and the
  binary (dashed blue lines). The dotted light blue lines represent the
  contribution of the semimajor axis evolution to the binary's angular
  momentum, while the dotted-dashed orange lines are the contribution of the
  total mass change, as shown in equation~\eqref{deltaL}. The lower panels
  display the relative change between consecutive snapshots of the same
  lines. The vertical shaded areas indicate the arrival time of the clouds.
  The extreme ``jumps'' seen in the red lines are caused by new clouds
  entering the system, thus providing additional angular momentum.
  The top row of plots show the evolution during the entire simulations, while
  the bottom row is a zoom-in on the first $\approx12$ orbits, which shows
  in detail the interaction with 2 clouds.}
\label{fig:AngMom_Fs}
\end{figure*}

The first step is to establish whether the simulations have the accuracy to
measure the effects of the interaction between the binary and the sequence of
clouds. With this goal in mind, we measure the conservation of angular momentum
in the different models, as shown in Fig.~\ref{fig:AngMom_Fs}. This corresponds
to the evolution in the lower resolution simulation, but because the binary
evolution is almost independent from the number of particles (see below),
these examples illustrate the evolution of the higher resolution models as well.

Consistent with the single cloud models, the largest effect in the binary
angular momentum occurs during the interaction with the first incoming
material, depicted as vertical shaded areas. The fluctuations following the
first encounter with each cloud rapidly decrease. It is important to clarify
that the rather extreme jumps observed in the total angular momentum are not
numerically driven, but produced by the addition of new clouds into the system.
From the lower panels of Fig.~\ref{fig:AngMom_Fs} it is clear that this
does not translate into spurious jumps in the binary evolution, as there are
no such noticeable changes in its angular momentum.
Moreover, simulation's segments delimited by the addition of clouds are
extremely flat, i.e. the numerical fluctuations in the total momentum are
way below the observed binary evolution and consequently they can hardly be
seen at this scale. This implies that the new models, thanks to the frequent
tree updates, have the numerical accuracy to resolve the binary evolution
beyond the prompt phase, as opposed to the old single cloud simulations.

  {From the top panels of Fig.~\ref{fig:AngMom_Fs} it is possible to
  observe that the evolution of the binary is somewhat following the injection
  of momentum in the form of clouds. In other words, the binary increases its
  angular momentum in the \F{0.0} model, where as in the \F{1.0} model the
  opposite occurs and it decreases. For the isotropic case (\F{0.5}) the
  angular momentum increases and decreases depending on whether the new cloud
  is co-rotating or counter-rotating, respectively. Although we expect a
  monotonically increasing angular momentum in the \F{0.0} simulation,
  the small drops caused by the inclusion of clouds 11, 12, 16 and 19 occur
  because the binary inclination slightly changes during the simulations.
  Said clouds, being so close to the initial binary's orbital plane
  (see Fig.~\ref{fig:ics}, upper panel), are now located in the southern
  hemisphere of the binary at the time of inclusion.}

The behaviour of the binary's angular momentum is in agreement with our
previous results where the main effect of the infalling material is to
exchange its angular momentum through capture and accretion (\paperII).
It is important to bear in mind that an increase in the binary angular
momentum does not necessarily imply an expansion of the semimajor axis;
this only means that the contribution of the total accreted mass, which
is always positive, is much larger to that of the likely negative semimajor
axis. This can be understood similarly to what is shown in \paperII, where
we write the binary angular momentum in terms of its parameters as follows:
\begin{equation}
  L_{\rm bin}=\frac{q}{(1+q)^2}M^{3/2}\sqrt{Ga(1-e^2)},
\end{equation}
where $G$ is the gravitational constant. By differentiating we find that at
first order this expression implies
\begin{equation}
  \frac{\Delta{L_{\rm bin}}}{L_{\rm bin}}=\frac{1-q}{q(1+q)}\Delta{q}+\frac{3}{2}\frac{\Delta{M}}{M}+\frac{1}{2}\frac{\Delta{a}}{a}-\frac{e}{1-e^2}\Delta{e}.
  \label{deltaL}
\end{equation}
Similar to the single cloud models, during all of our simulations the binary
remains roughly equal-mass and circular. We can therefore calculate
approximately the total change in angular momentum based on the change in $a$
and $M$ only,
\begin{equation}
  \frac{\Delta{L_{\rm bin}}}{L_{\rm bin}}\approx\frac{3}{2}\frac{\Delta{M}}{M}+\frac{1}{2}\frac{\Delta{a}}{a}.
  \label{deltaL2}
\end{equation}

The contributions of both semimajor axis and mass terms are displayed in the
upper panels of Fig.~\ref{fig:AngMom_Fs}. From there it is clear that the
addition of these two terms is sufficient to reproduce the binary's angular
momentum evolution, justifying our approximation of ignoring the mass ratio
and eccentricity contributions. It is also evident that the $\Delta a$
contribution to the MBHB angular momentum change is always negative,
indicating that the binary shrinks also when $\Delta L_{\rm bin}>0$.

\subsection{Orbital elements}

\begin{figure*}
\centering
\begin{picture}(1,0.5)
\put(0,0){\includegraphics[width=0.3333\textwidth]{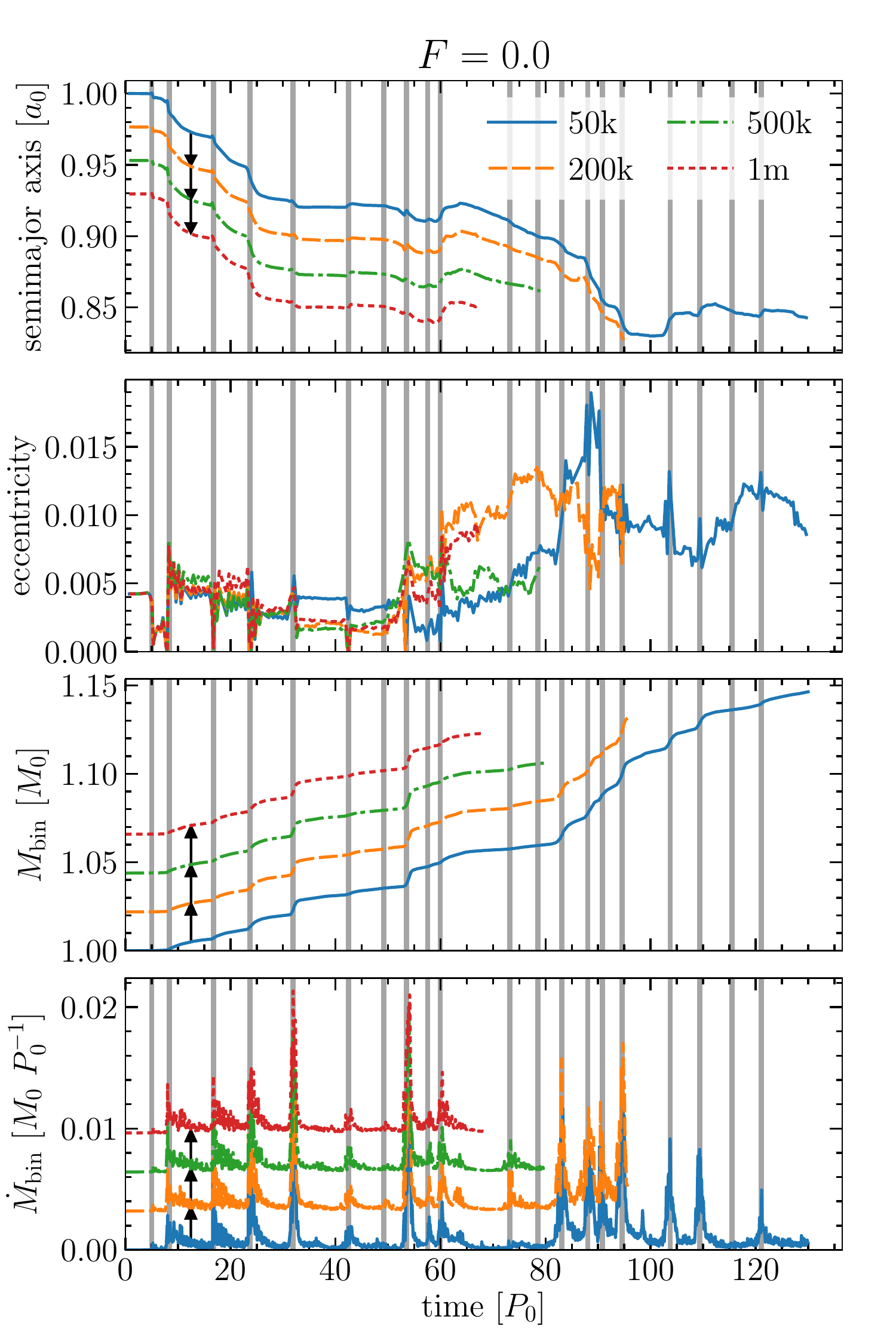}}
\put(0.333,0){\includegraphics[width=0.3333\textwidth]{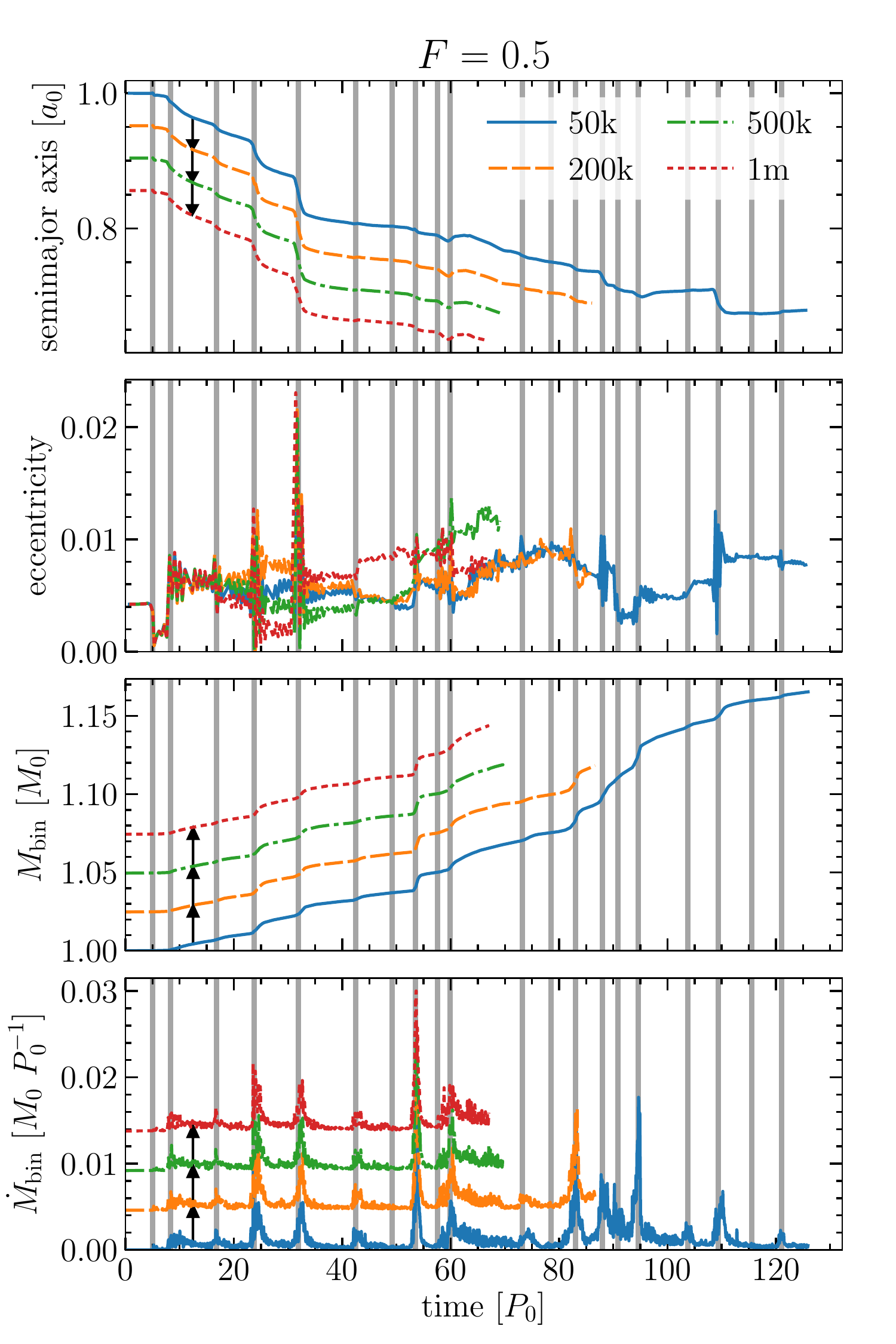}}
\put(0.666,0){\includegraphics[width=0.3333\textwidth]{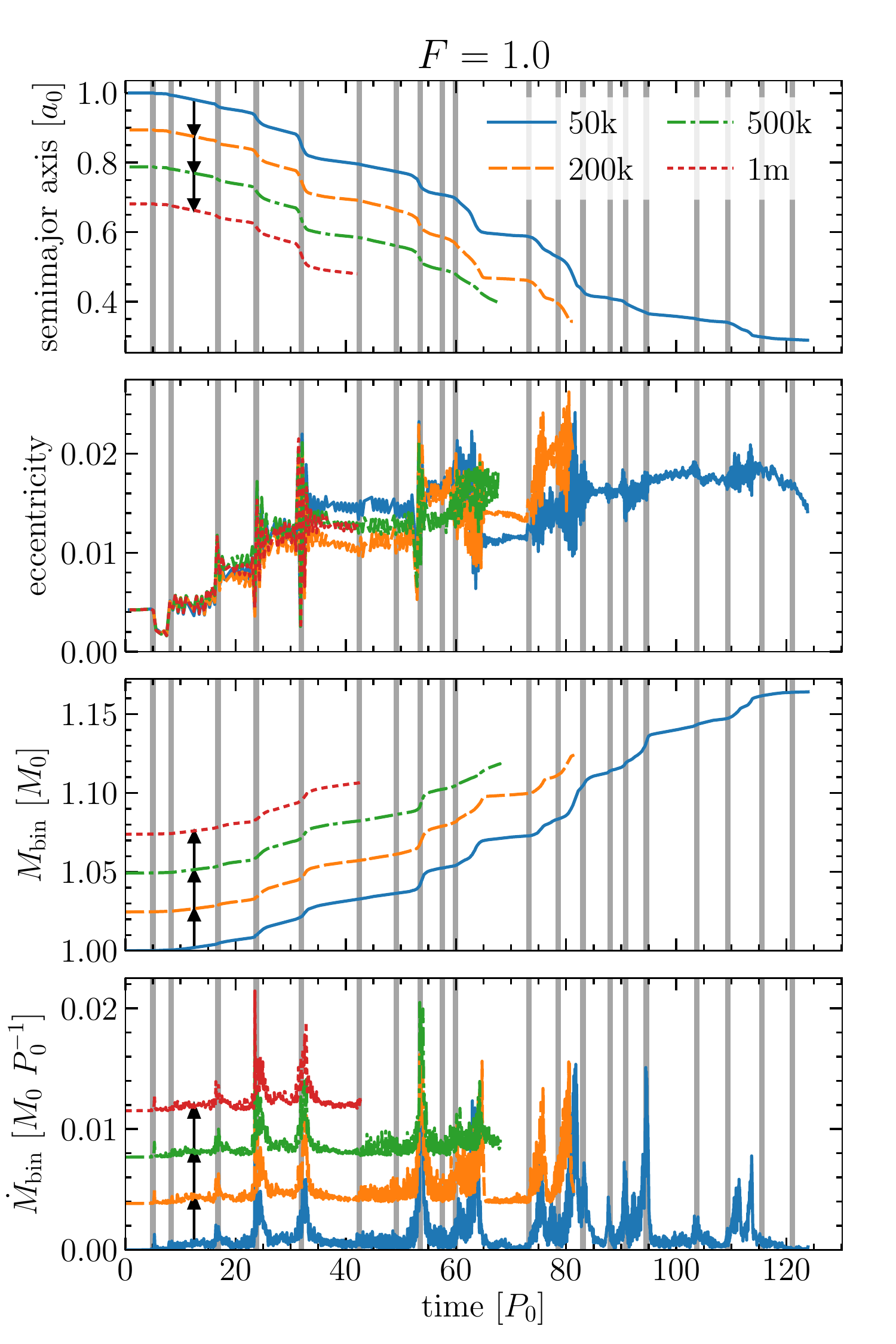}}
\end{picture}
\caption{Evolution of the binary semimajor axis (top panels), eccentricity
  (upper-middle panels), total accreted mass (lower-middle panels), and
  accretion rate (bottom panels) for all the simulations. The corresponding
  distribution is indicated at the top of each plot. The different lines in
  each panel represent the cloud resolutions, as indicated in the legend.
  The vertical shaded areas indicate the arrival times of the clouds.
  Note that the non-solid lines are arbitrarily shifted either upwards or
  downwards with respect to the solid line (\texttt{50k} resolution) in the
  semimajor axis, accreted mass and accretion rate panels for clarity.
  The virtually identical evolution during these stages indicates convergence with the
  number of particles per cloud.}
\label{fig:binary_Fs}
\end{figure*}

On the basis of the adequate angular momentum conservation, it is possible
to study the evolution of the binary orbital elements during the entirety
of the simulations. To properly include the influence of the external
potential on the binary, we compute its orbital elements ($a$ and $e$) using
the periapsis and apoapsis, instead of the relative energy and angular
momentum of the two MBHs. Figure~\ref{fig:binary_Fs} displays the time
evolution of the semimajor axis, eccentricity, total mass and accretion rate
comparing the different resolutions modelled.
  Because the different resolutions tend to lie on top of each other, we
  arbitrarily shifted the non-solid lines with respect to the solid line
  (\texttt{50k} resolution) for the semimajor axis, accreted mass and
  accretion rate, which is indicated by black arrows. The virtually identical
  orbital evolution during the modelled stages indicates convergence with the
  number of particles per cloud. This is because the dynamical evolution of
  the system is dominated by capture and gravitational slingshot, which are
  properly resolved even with a low number of particles.
The largest differences are present in the eccentricity evolution.
This is due to the typically low values ($e\lesssim 0.02$) of this quantity
that makes it much more sensitive to noise.
{Although the sources of this noise are difficult to identify, the
dependence with particle number suggests that the details of
the accretion are likely playing an important role. Due to the symmetries
of our configuration, the binary mass ratio remains robustly around
unity throughout the modelled stages, but with some stochastic small
deviations on top. Considering equation~\eqref{deltaL}, for a given
change of $L_{\rm bin}$, $M$ and $a$, changes in $\mu$ and $e$
are highly entangled.}
For this reason we restrict the
analysis to the semimajor axis and mass evolution. In any case, all our
binaries remain essentially circular over the modelled timescales (although
they consistently show a mild tendency to increase eccentricity),
and the small discrepancies in eccentricity do not affect the overall
evolution of the system.

{Similar to the results shown in \paperI, the accretion rate onto the
binary presents a very clear periodicity at half the
binary's orbital period (or twice the binary frequency).
This is mainly a dynamical effect, associated with
streams of gas that feed each black hole alternately when they intersect.
We do not identify any other persistent trends in the accretion rates.
Variability related to the binary orbital period is the type of behaviour that
we expect will help to identify and characterise these systems, as it could
appear in AGN light curves and be detected by future time-domain
surveys, and it has been extensively studied with numerical
simulations of quasi-steady circumbinary
discs \citep[e.g.][]{MacFadyen2008, C09, Roedig2011, Shi2012, DOrazio2013,
Dunhill2015, Farris2015, DOrazio2015, Shi2015, Munoz2016, Tang2017, Tang2018}.
Note that for our fiducial scaling the binary period is several thousands of
years, certainly too long to be observed (see Table~2 in \paperI, for different
scalings of these systems). However, we expect that the behaviour
we find is qualitatively representative of those more rapidly varying systems,
especially if there is a steady supply of clouds reaching the galactic nucleus.
}

\begin{table}
\centering
\caption[Total change of the binary semimajor axis and mass]{Total change of
  the binary semimajor axis ($\Delta a$) and mass ($\Delta M$) for the
  different models at the final time ($T_{\rm fin}$) of the simulations. The
  model name is composed of the $F$ distribution followed by the resolution of
  the individual clouds. $T_{\rm fin}$ is the final time reached by each
  simulation. The values with subscript `10th' correspond to the total change
  after the interaction with the 10th cloud, which is measured at $t=65 P_0$.
  This point was not reached by the higher resolution runs of \F{1.0}
  distribution.}
\label{tab:deltasF}
\begin{tabular}{lccccr}
\hline
\hline
Model & $T_{\rm fin}$ & $\Delta a$ & $\Delta M$ & $\Delta a_{\rm 10th}$ & $\Delta M_{\rm 10th}$\\
      & ($P_0$)     &   ($a_0$)    & ($M_0$)    & ($a_0$)               & ($M_0$) \\ \hline
\texttt{F0.0\_50k} & 145.4 & -0.160 & 0.158 & -0.079 & 0.056 \\
\texttt{F0.0\_200k} & 95.7 & -0.157 & 0.110 & -0.075 & 0.057 \\
\texttt{F0.0\_500k} & 79.7 & -0.094 & 0.062 & -0.079 & 0.057 \\ \medskip
\texttt{F0.0\_1m} & 68.2 & -0.083 & 0.057 & -0.078 & 0.056 \\
\texttt{F0.5\_50k} & 126.0 & -0.325 & 0.165 & -0.218 & 0.064 \\
\texttt{F0.5\_200k} & 86.7 & -0.266 & 0.094 & -0.222 & 0.065 \\
\texttt{F0.5\_500k} & 69.7 & -0.233 & 0.069 & -0.221 & 0.064 \\ \medskip
\texttt{F0.5\_1m} & 67.1 & -0.227 & 0.070 & -0.221 & 0.067 \\
\texttt{F1.0\_50k} & 124.0 & -0.714 & 0.164 & -0.405 & 0.070 \\
\texttt{F1.0\_200k} & 81.5 & -0.555 & 0.099 & -0.428 & 0.073 \\
\texttt{F1.0\_500k} & 68.2 & -0.398 & 0.070 & -0.371 & 0.065 \\
\texttt{F1.0\_1m} & 42.7 & -0.207 & 0.033 & - & - \\
\hline
\end{tabular}
\end{table}

In Table~\ref{tab:deltasF} we compile the total change in both binary
semimajor axis and total mass for all the simulations. The first 3 columns
show the values measured at the end of each run ($T_{\rm fin}$),
As expected, increasing the number of particles per cloud makes the simulations
more computationally expensive, forcing us to stop them earlier in the
evolution of the system. Additionally, the \F{1.0} models are more expensive
than the others because counter-rotating clouds tend to form more clumps.
In order to facilitate the comparison between different resolutions, the
values with subscript `10th' correspond to the total change after the
interaction with the 10th cloud, measured at $t=65 P_0$, point that was
reached by most of our simulations. Notice that, for a given $F$
  distribution, the discrepancies on the binary parameters at this point are
  very small, being usually less than 5\% with respect to the actual changes
  measured along the simulation.

Based on the results showed in Fig.~\ref{fig:binary_Fs} and the convergence of
the simulation with resolution, most of the analysis in this paper is based on
the \texttt{50k} runs because more clouds have interacted with the binary. We
expect the conclusions we will reach will apply for the higher resolution
simulations.

\begin{figure}
\centering
\includegraphics[width=0.5\textwidth]{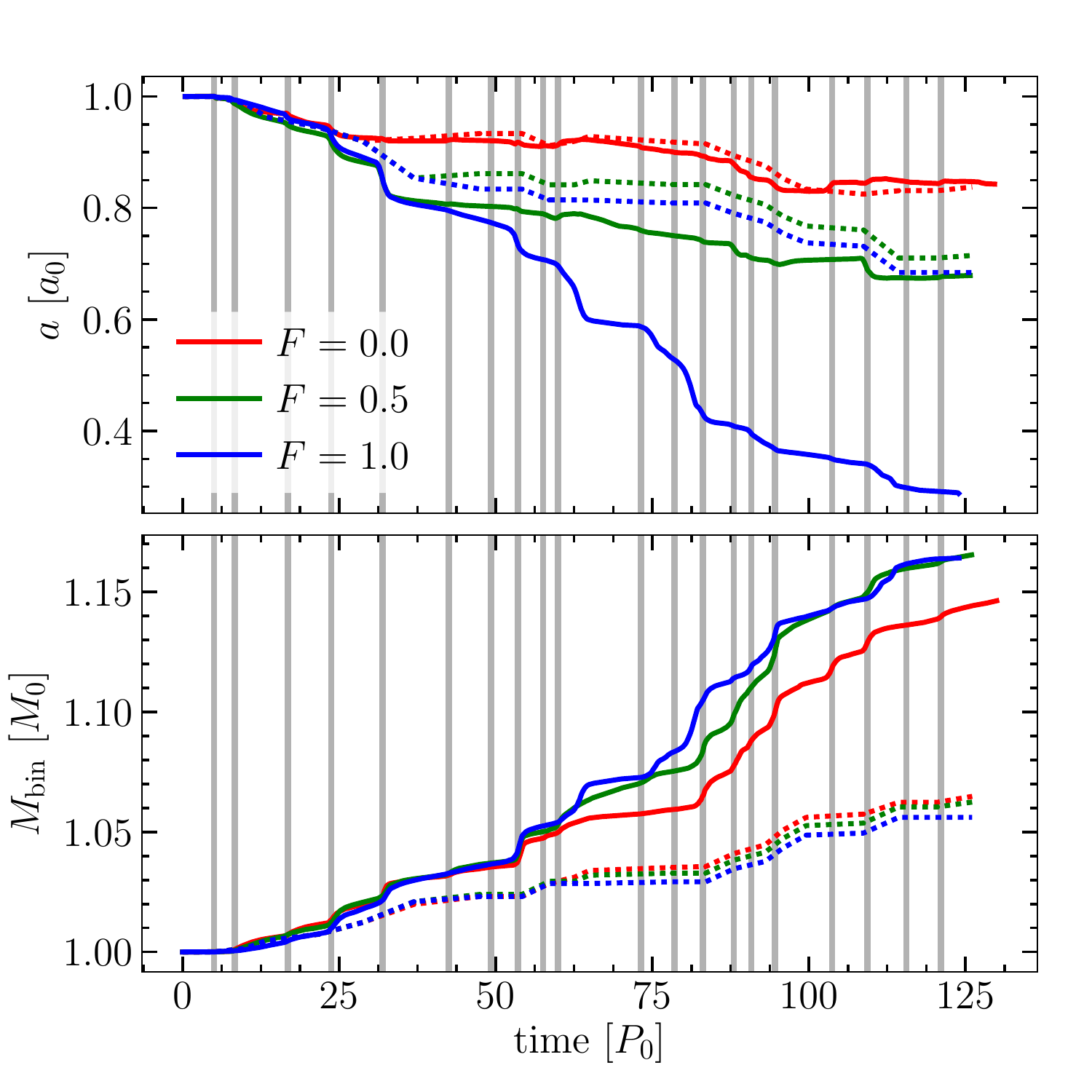}
\caption{Time evolution of the binary semimajor axis (top panel) and mass
  (bottom panel) in the \texttt{50k} runs for the different cloud
  distributions, as indicated in the legend. The solid lines depict the
  evolution measured directly from the simulations, while the dashed lines
  correspond to predictions based on the results from the single cloud model
  presented in \paperII. Similar to previous figures, the shaded vertical
  areas illustrate the start of the interaction of the binary with a new
  cloud.}
\label{fig:mcs_sim_events}
\end{figure}

To highlight the differences between distributions, in the upper panel of
Fig.~\ref{fig:mcs_sim_events} we display the semimajor axis evolution for
the three levels of anisotropy, while the lower panel shows the mass evolution.
The solid lines on this figure represent the evolution measured from the
simulations. As expected, the binary shrinks more efficiently when we
increase the fraction of retrograde events. For instance, after the infall of
10 clouds the orbit shrinks by $\Delta a/a\approx 40\%$ in the \F{1.0}
simulation, by $\Delta a/a\approx 20\%$ in the \F{0.5} model and by a mere
$\Delta a/a\approx 8\%$ for \F{0.0} (see Table~\ref{tab:deltasF}).
This can be understood in terms of the exchange of angular momentum with the
captured material, as studied in detail in \paperII. Because retrograde clouds
bring negative momentum to the binary, any captured and accreted material
would immediately subtract angular momentum,
while
{clouds with angular momentum closely aligned to $L_{\rm bin}$}
tend to actually expand the MBHB orbit {(see Fig.~6 in \paperII)}.
In contrast with the semimajor axis, the total mass evolution remains very
similar for the three levels of anisotropy, as the capture of material does
not depend strongly on the orbit of the incoming gas. However, because the
gas exchanges its angular momentum with the accreting black holes, the net
effect in the binary orbit does depend on the different relative inclinations.

In the Monte Carlo models presented in \paperII, we have extrapolated the
results from the single cloud simulations to evolve a binary embedded in a
clumpy environment where individual gas pockets infall with different levels
of anisotropy. One of the critical assumptions made in this approach was
ignoring the effects of the left-over material after the first stages of the
interaction. To investigate this effect we compute the binary evolution
following the same procedure explained in \paperII, i.e. estimating the total
change in mass and semimajor axis based on the cloud's initial orientation and
pericentre distance, using the values extrapolated from the single cloud
models. In this procedure we are ignoring any secular effects from the
non-accreted gas (as well as cloud-cloud interactions), by considering
only the evolution measured during the prompt accretion phase.
The results obtained are shown with dashed lines in
Fig.~\ref{fig:mcs_sim_events}.

During the interaction with the first few clouds the single cloud predictions
agree well with the measured binary evolution, but after few events this
simple model consistently underestimates both semimajor axis and mass
evolution. This departure allows us to directly measure the effects of the
surrounding material, previously unresolved with the Monte Carlo models in
\paperII. In the following subsections, we analyse two effects: i) enhancement
in accretion due to the interaction of the new incoming material with the gas
already in orbit, and ii) gravitational torques from long-lived circumbinary gas
structures.

{It is important to stress that we always expect a departure between the
combined effect of multiple clouds with respect to the extrapolation
from the single cloud models, independent of the infall rate, since the latter
lacks the aforementioned secular effects.
For the relatively high frequency of clouds presented in this paper, this
deviation is already clear, and is dominated by enhanced accretion
as we show below.
On the other hand, for a very low infall frequency, the surviving material
would be allowed to settle into a circumbinary disc that exerts some small
but persistent
gravitational influence on the binary, continuing to extract
its angular momentum \citep[e.g.][]{C09}.}

\subsubsection{Enhanced Accretion}

\begin{figure*}
\centering
\begin{picture}(1,0.3333)
\put(0,0){\includegraphics[width=0.3333\textwidth]{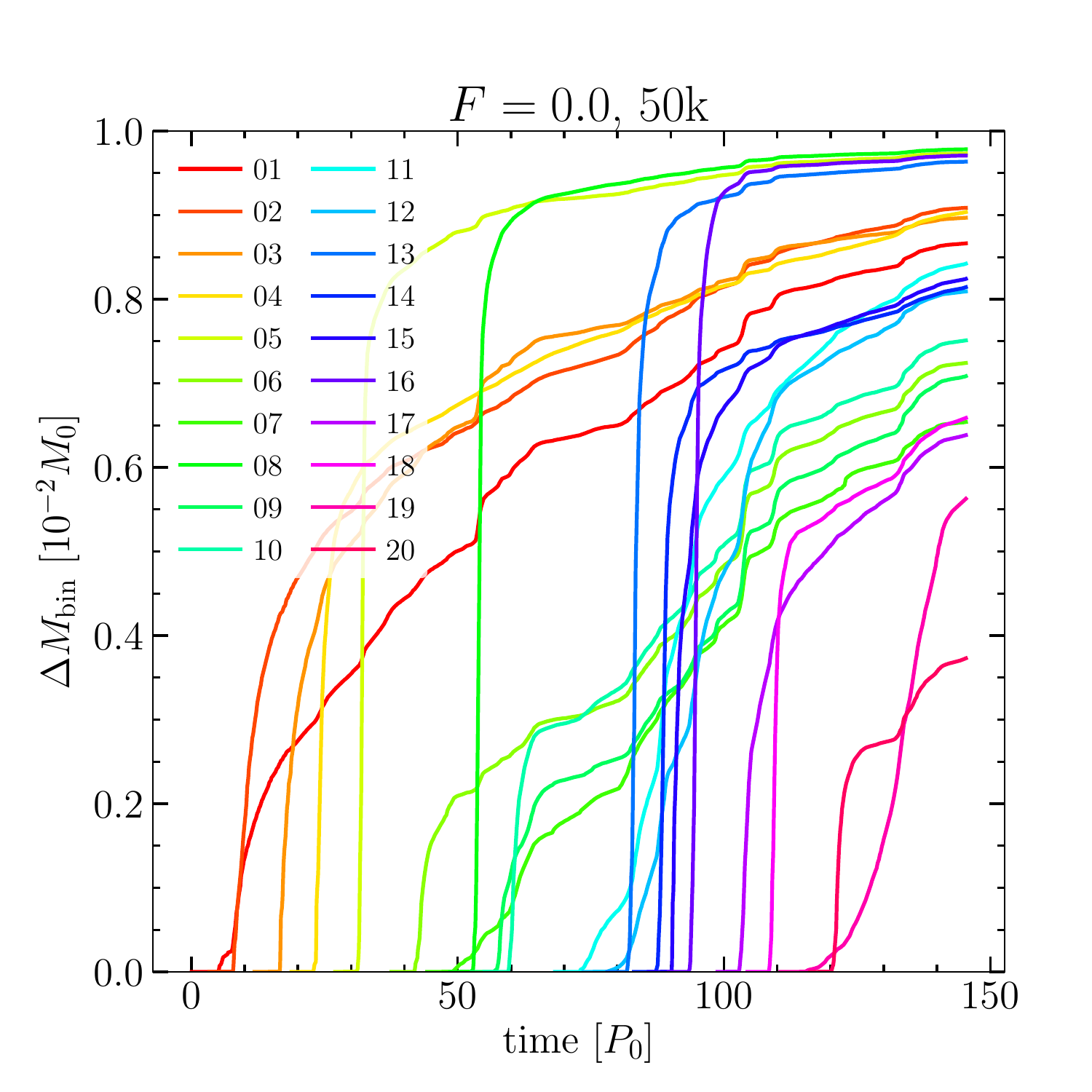}}
\put(0.3333,0){\includegraphics[width=0.3333\textwidth]{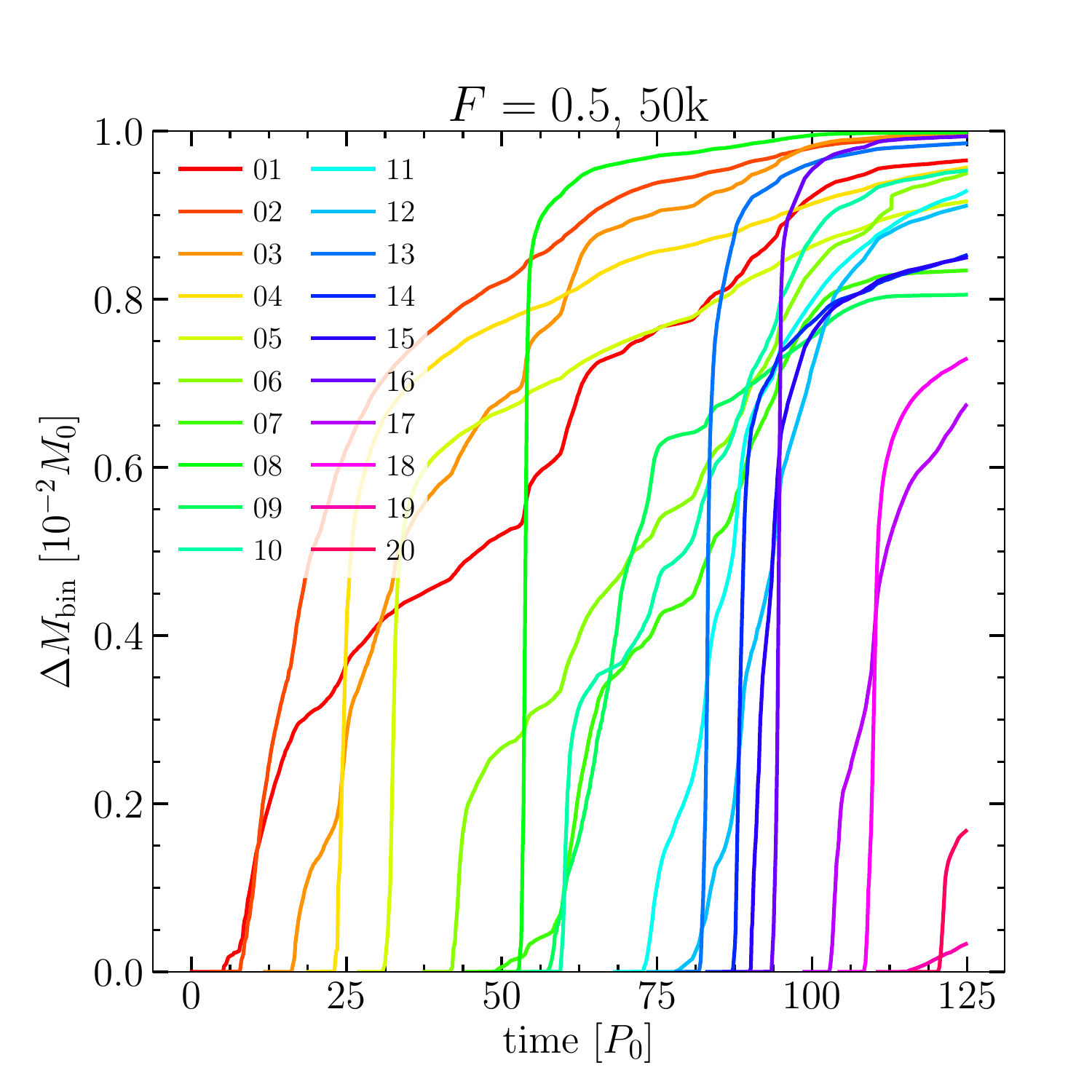}}
\put(0.6666,0){\includegraphics[width=0.3333\textwidth]{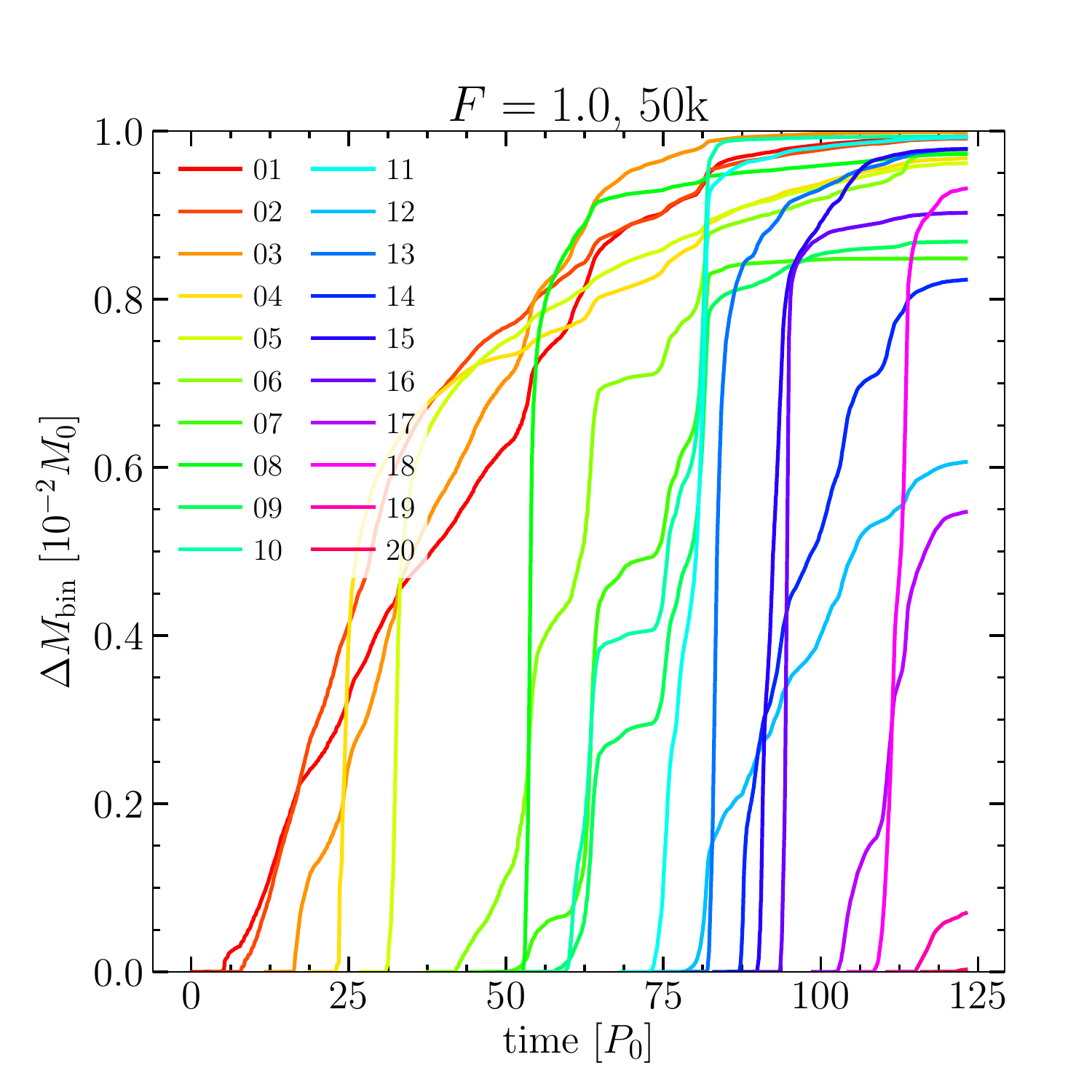}}
\end{picture}
\caption{Time evolution of the accreted mass contributed by each cloud in the
  different runs. Note that the mass is presented in unit of $10^{-2}M_0$, which
  corresponds to the initial mass of each cloud, thus the $y$-axis can be
  interpreted as the fraction of the cloud that has been accreted by the
  binary.
  Here we observe that some of the incoming clouds produce
  spikes also in the previous clouds.}
\label{fig:deltam_clouds}
\end{figure*}

One striking difference with respect to the single cloud extrapolations is
the total accreted mass by the binary. From the lower panel of
Fig.~\ref{fig:mcs_sim_events} we notice that the single cloud model
underestimates the final mass by a factor of $\sim$3. This occurs because
there is an accumulation of material around the binary, and the new incoming
clouds are able to drag some of it, increasing accretion and consequently
the effect on the binary semimajor axis. Note that the deviation of the
measured evolution from the predicted one is episodic rather then continuous.

To corroborate this point, we present the contribution from each individual
cloud to the mass accreted by the binary in Fig.~\ref{fig:deltam_clouds}.
Here it is possible to observe that some of the incoming clouds produce an
accretion spike also in some of the previous clouds. For example, one of the
most noticeable cases is the arrival of cloud 8, which brings to the binary
a significant fraction of clouds 1, 3 and 6. Recall from Fig.~\ref{fig:peris}
that the 8th cloud has the smallest pericentre distance, very close to zero.
This event produces the largest accretion in all distributions.
This extra-accretion triggered by interactions with leftover
gas from previously infalling clouds is obviously not captured by the simple
model derived from the single cloud simulations.
{Additionally, since the capture of retrograde material is more efficient
in shrinking the binary orbit, this enhancement in accretion
translates into larger deviations when increasing the fraction
of counter-rotating clouds, although it is hard to see this trend
due to the scatter introduced by the specific cloud realisations.
For instance, in the \F{0.5} case shown in Fig.~\ref{fig:mcs_sim_events}
the binary evolution is only slightly accelerated with respect to the single cloud
extrapolation, while in the additional set of simulations shown in
Fig.~\ref{fig:RunA_sim_events}, the difference is much larger.
It is important to notice that this is a `one way' effect: interactions with
multiple clouds consistently tend to accelerate the binary evolution.}

\begin{figure}
\centering
\includegraphics[width=0.46\textwidth]{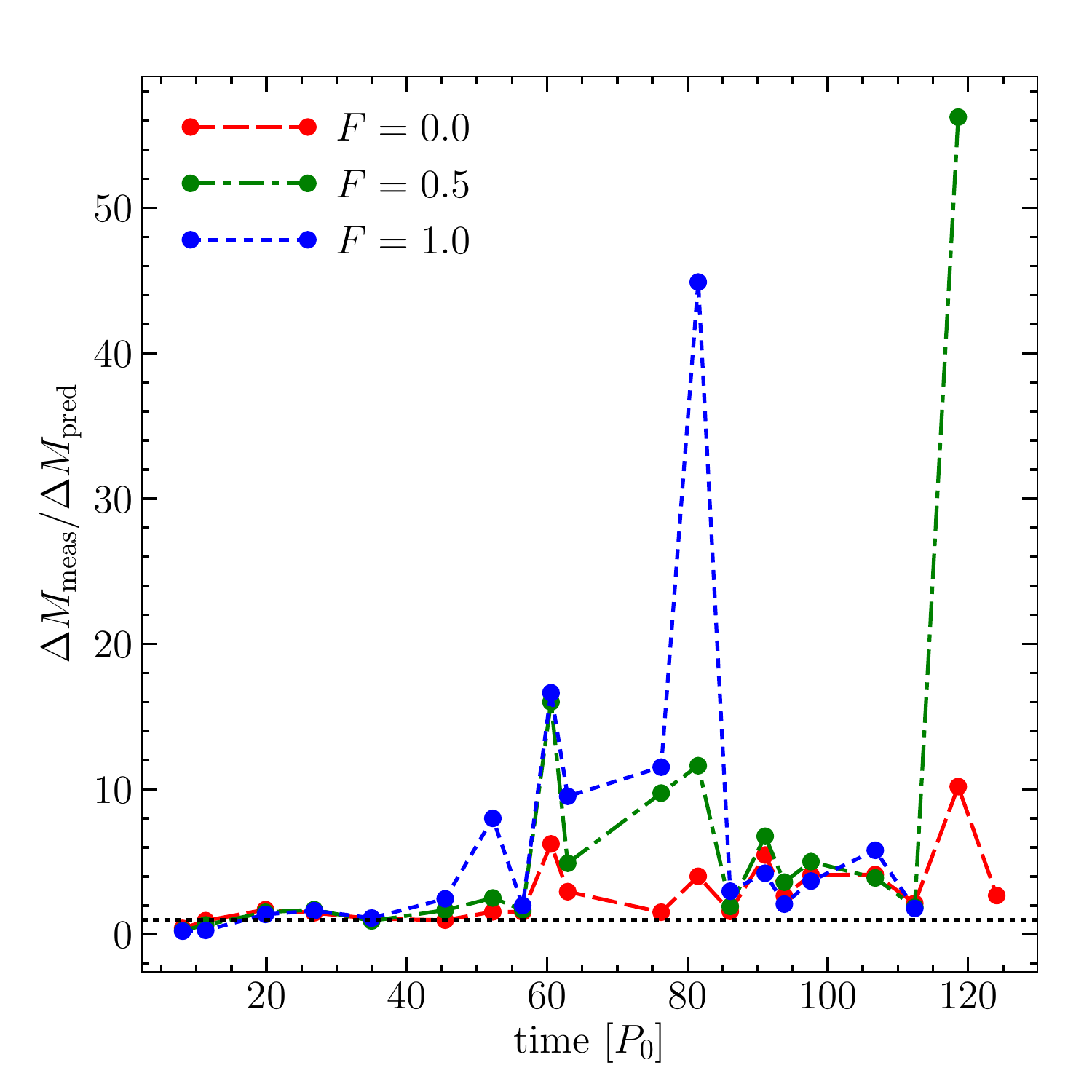}
\caption{Time evolution of the ratio between the total accreted mass measured
  from the simulations and the value predicted from the single cloud models.
  The horizontal dotted line indicates $y=1$.
  {This plot shows that the departure from the predicted accreted mass
  is stronger when there are circumbinary structures present.}}
\label{fig:acc_mass}
\end{figure}

Finally, in order to quantify the total increase of the accretion due to the
accumulation of gas around the binary, we compute the evolution of the ratio
between the total accreted mass measured from the simulation and the value
predicted using the single cloud models, which is shown in
Fig.~\ref{fig:acc_mass}. As previously observed, the prediction and the actual
evolution agree reasonably well during the first few clouds, and this ratio
oscillates around 1. After the build up of material around the binary, however,
there are large bursts of accretion, reaching values up to $\gtrsim50$ times
the predicted level. This picture becomes stronger if we consider that around
$t=60P_0$, following the infall of the 10th cloud, there are circumbinary
discs formed in all the 3 distributions (see next subsection).
These discs are disrupted at least partially by the new incoming clouds,
promoting accretion onto the binary.

In a companion paper \citep{MaureiraFredes2018},
we are using the same suite of simulations
to study in detail the formation and evolution of gaseous structures arising
from these sequences of infalling clouds onto a binary. In particular, we focus
our analysis in the dependency with the level of anisotropy of the incoming
clouds (expressed in terms of the $F$-distributions) and the time
difference between events. There we show that is typically very difficult
to build a substantial circumbinary disc in this scenario
of incoherent accretion events, especially with shorter times in between clouds,
because the incoming gas constantly perturbs the arising structure.
Nevertheless, we do observe the formation of transient circumbinary structures
that exert some gravitational influence on the binary's orbit, as we discuss
below.

\subsubsection{Circumbinary disc formation and secular evolution}

As thoroughly discussed in \paperII, another effect expected from the left-over
gas is the formation of circumbinary structures. This represents one of the
most important limitations of the models based on the single cloud simulations.
For instance, in the extreme case where every cloud comes exactly coplanar and
corotating with the binary (sometimes referred to as the `prolonged accretion'
scenario), the extrapolation from the single cloud models predicts that the
semimajor axis would only increase. In reality, however, following several of
such accretion events a massive circumbinary disc is expected to form. This
structure can continue extracting and transporting the binary angular momentum,
evolving the binary towards coalescence.

\begin{figure*}
\centering
\begin{picture}(1,0.3333)
\put(0,0){\includegraphics[width=0.3333\textwidth]{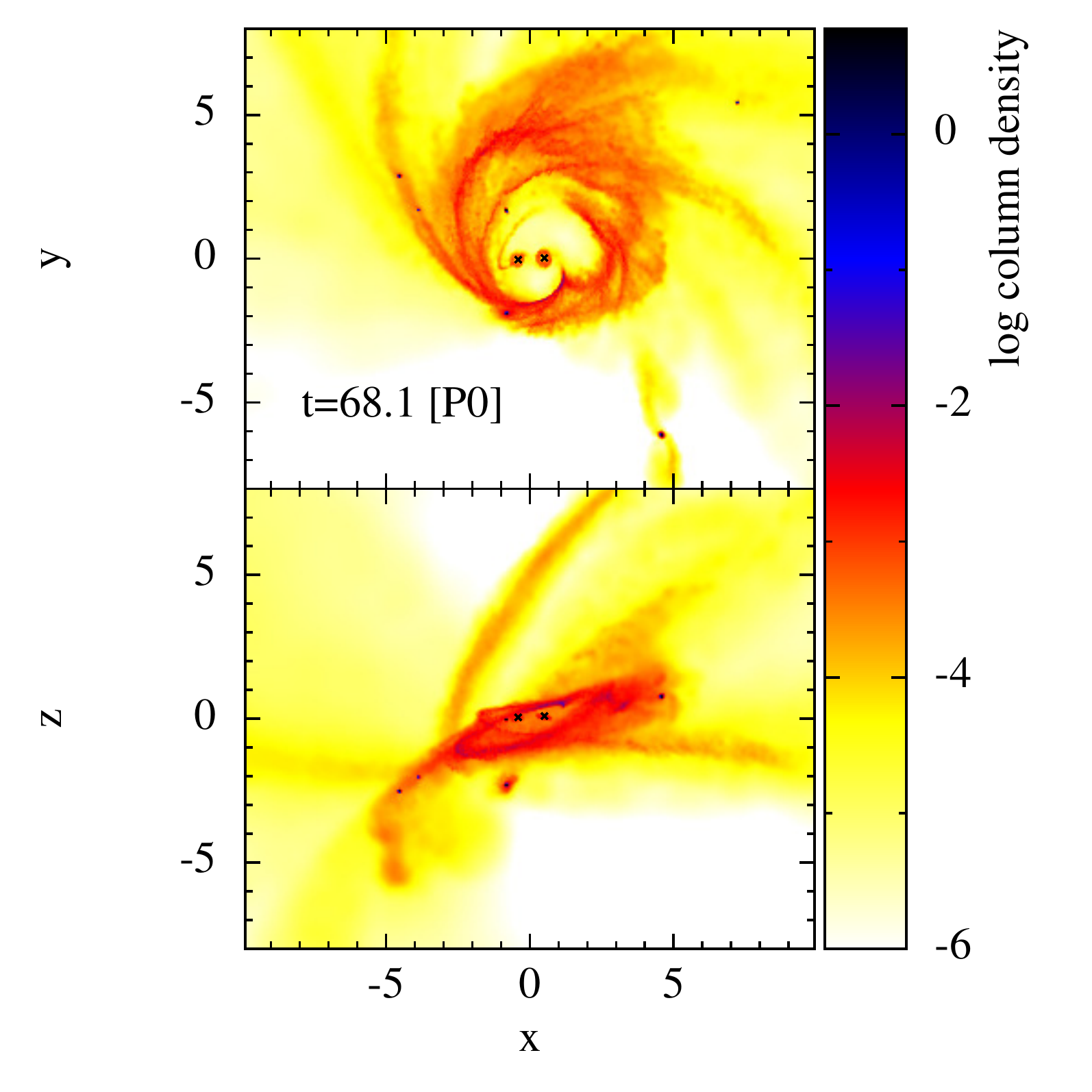}}
\put(0.3333,0){\includegraphics[width=0.3333\textwidth]{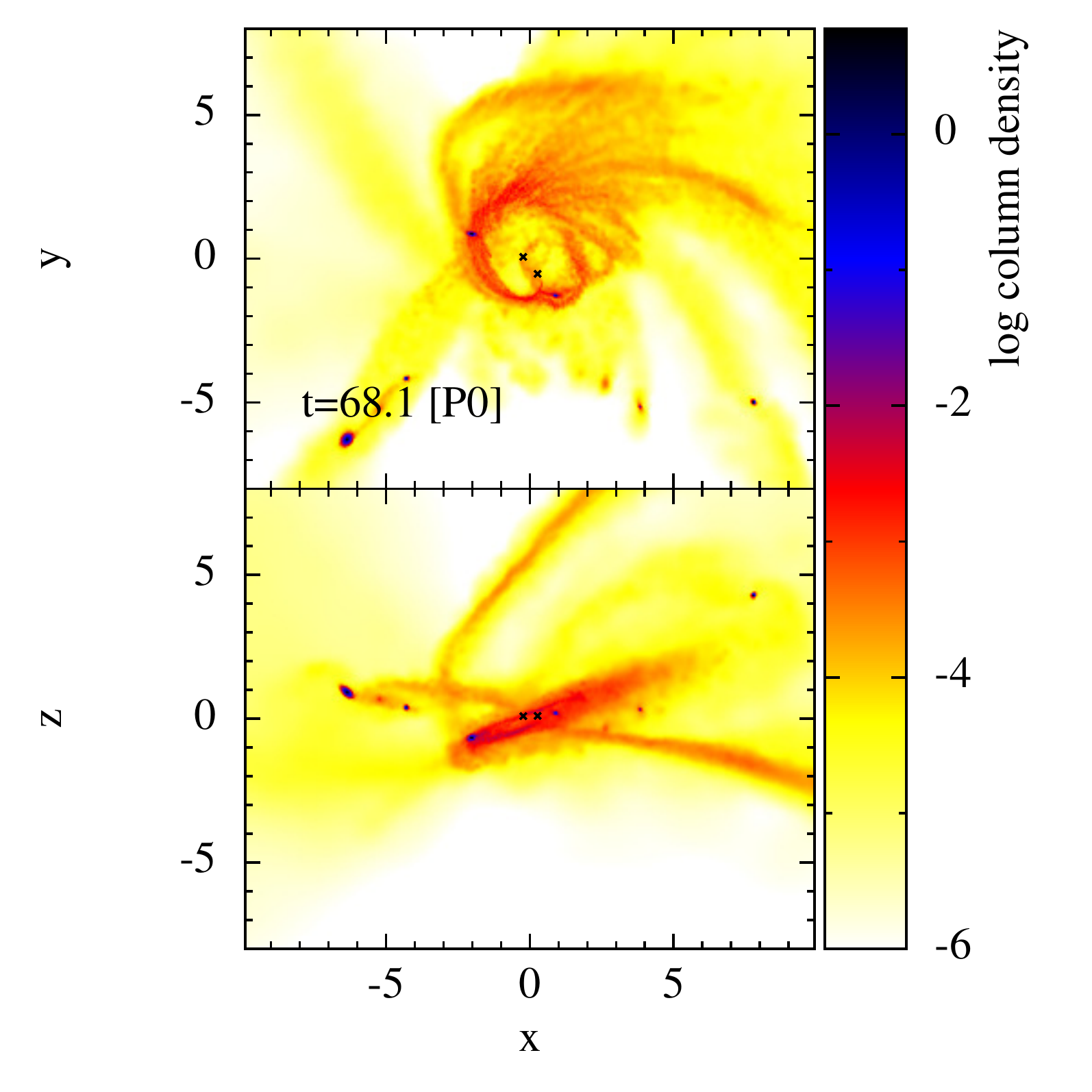}}
\put(0.6666,0){\includegraphics[width=0.3333\textwidth]{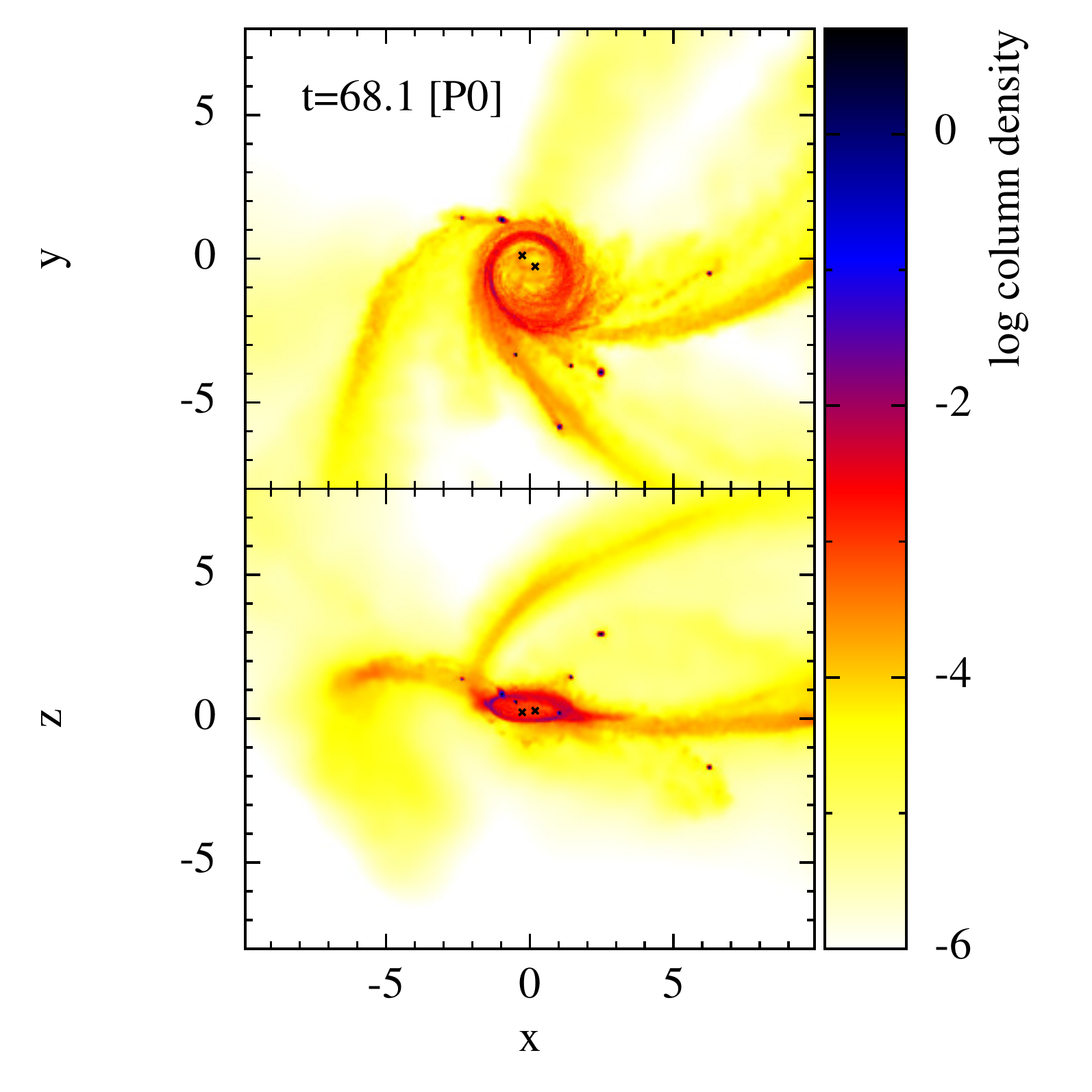}}
\end{picture}
\caption{Column density maps of an intermediate output from models \F{0.0}
  (left), \F{0.5} (centre) and \F{1.0} (right). For reference, the binary
  lies on the $x$-$y$, rotating counter-clockwise. The top and bottom panels
  show a face-on and edge-on view of the circumbinary structure formed after
  the interaction with 10 gaseous clouds. For the \F{0.0} and \F{0.5} models,
  the disc is prograde with respect to the binary orbit, while for the
  \F{1.0} model the structure is retrograde.}
\label{fig:circumb_discs}
\end{figure*}

Following the arrival of the 10th cloud, there is enough time before the next
event for the gas to settle in a well-defined coplanar circumbinary structure
in the 3 distributions. This can be clearly seen on the density maps shown in
Fig.~\ref{fig:circumb_discs}. For the \F{0.0} and \F{0.5} models, we observe
the formation of a corotating circumbinary disc, while for the \F{1.0} model
the structure is a retrograde and much narrower, ring-like circumbinary disc.

\begin{figure*}
\centering
\begin{picture}(1,0.3333)
\put(0,0){\includegraphics[width=0.3333\textwidth]{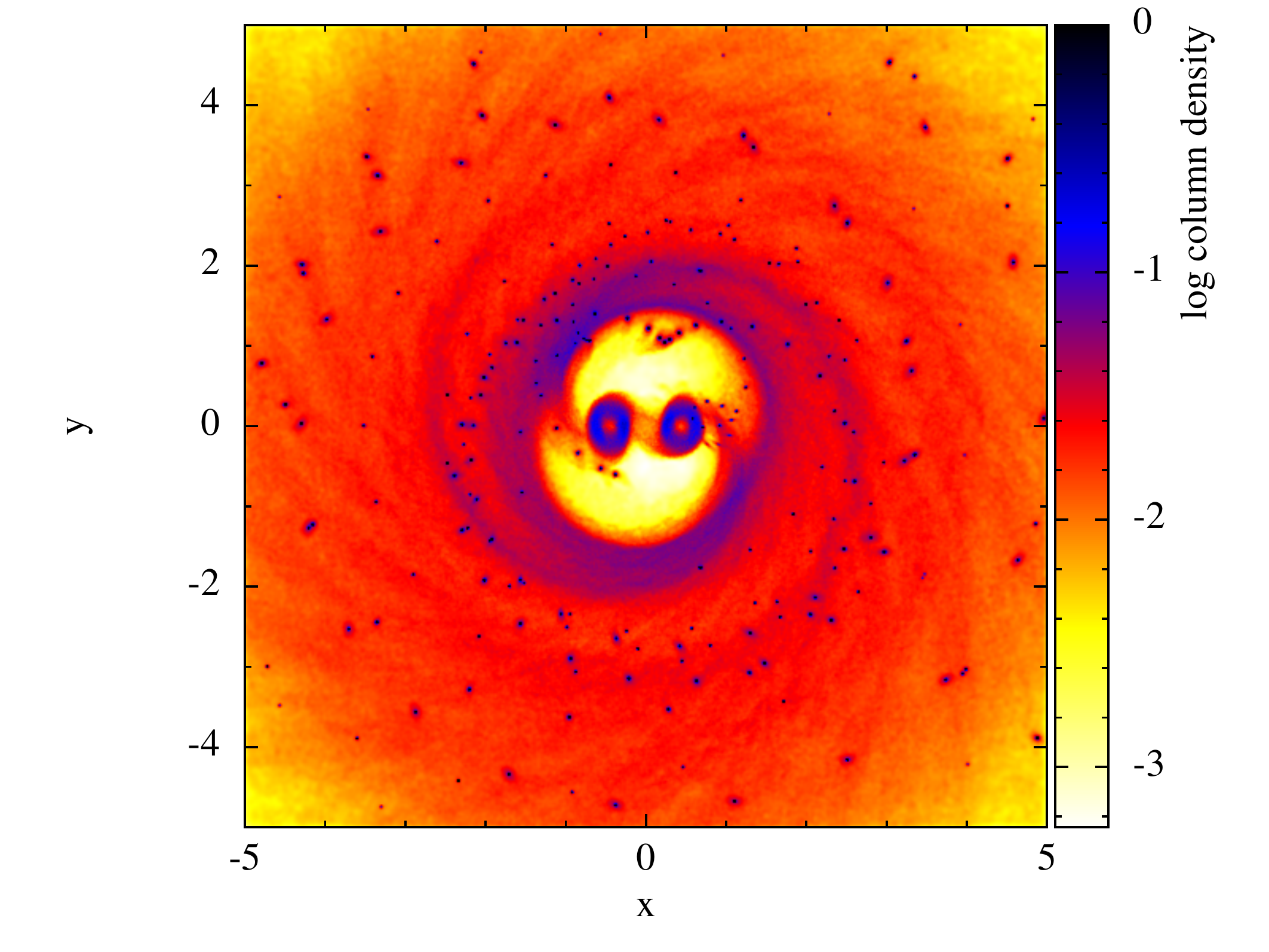}}
\put(0.3333,0){\includegraphics[width=0.3333\textwidth]{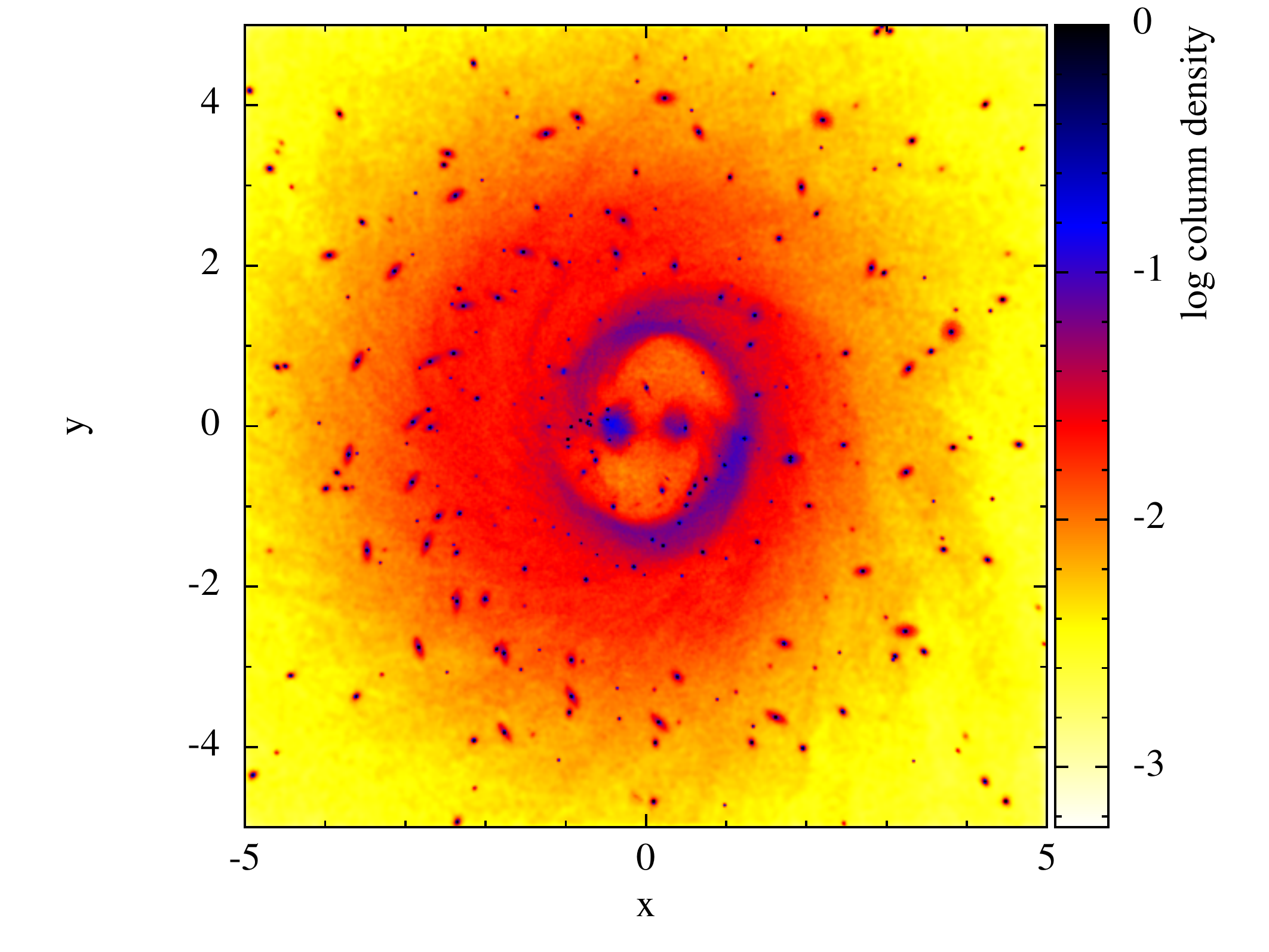}}
\put(0.6666,0){\includegraphics[width=0.3333\textwidth]{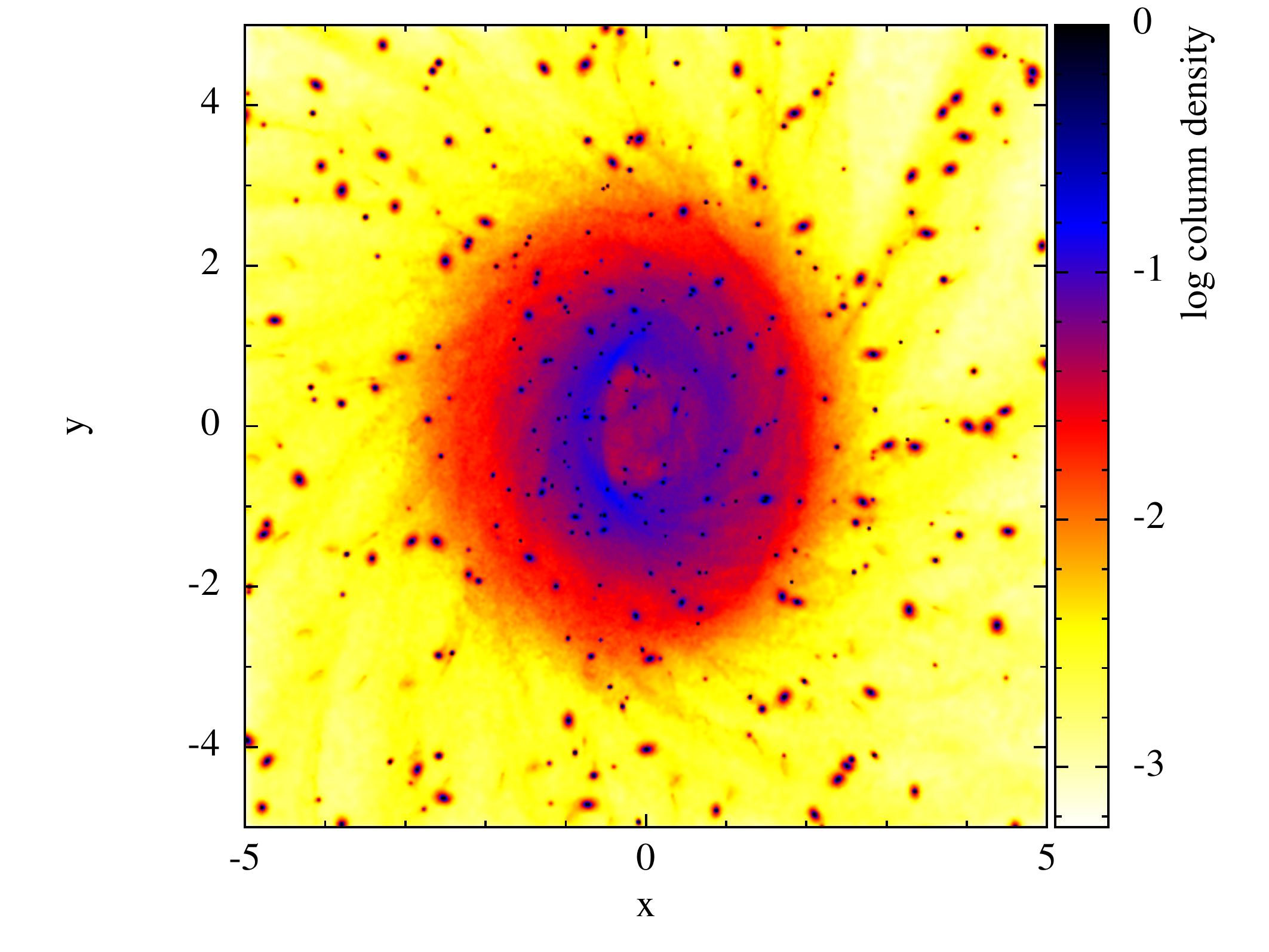}}
\end{picture}
\caption{Density maps averaged over 5 binary orbits for the \F{0.0} (left),
  \F{0.5} (centre) and \F{1.0} (right) models. These maps were computed in a
  reference frame corotating with the binary. Note that the large amount of
  clumps seen in these panels are due to few clumps caught at different
  orbital phases when creating the maps by superimposing the snapshots.
  For the \F{0.0} and \F{0.5} distributions, there are gaseous streams leaking
  through a cavity sustained by the gravitational influence of the binary.
  In contrast, for the \F{1.0} distribution the gas is able to orbit much closer,
  directly impacting the black holes.}
\label{fig:average_dens}
\end{figure*}

To study any persistent structures of these developing discs we present the
column density maps averaged over five orbits (100 snapshots) in
Fig.~\ref{fig:average_dens}. These maps were computed in a reference frame
corotating with the binary, with the black holes aligned on the $x$-axis.
It is clear that the gaseous structures are better defined in the \F{0.0}
simulation, where the binary has opened a clear cavity and each MBH is
surrounded by prominent mini-discs. In contrast, the less coherent infall of
material in the \F{0.5} simulation delays the emergence of these features.
Nevertheless, in both cases there is a dense, roughly circular ring of gas
located at $r\approx 2a$. In this region the material piles up due to
resonances with the gravitational potential of the binary
\citep{Artymowicz1994}. Although the gravitational forces keep most of the gas
confined at this radius, some material leaks thorough the cavity in the form
of narrow streams that reach the black holes, a result that has been
extensively reported in the literature \citep[e.g.][]{Artymowicz1996, MacFadyen2008,
C09, Roedig2012, DOrazio2013, Farris2015b, Dunhill2015, Tang2017}.
Such inflowing gas can exert a net negative torque inside the cavity region,
shrinking the binary more effectively than the resonant torques.

In contrast, we observe the formation of a counter-rotating circumbinary ring
in the \F{1.0} run. Because of the absence of gravitational resonances
for counter-rotating material, the gas is able to orbit much closer to the
binary, directly impacting onto the black holes. As a consequence, the
morphology of the gas is quite different, as there are no mini-discs, nor
gas in the form of streams. These features are extensively discussed in
\papergas.

\begin{figure*}
\centering
\begin{picture}(1,0.3333)
\put(0,0){\includegraphics[width=0.3333\textwidth]{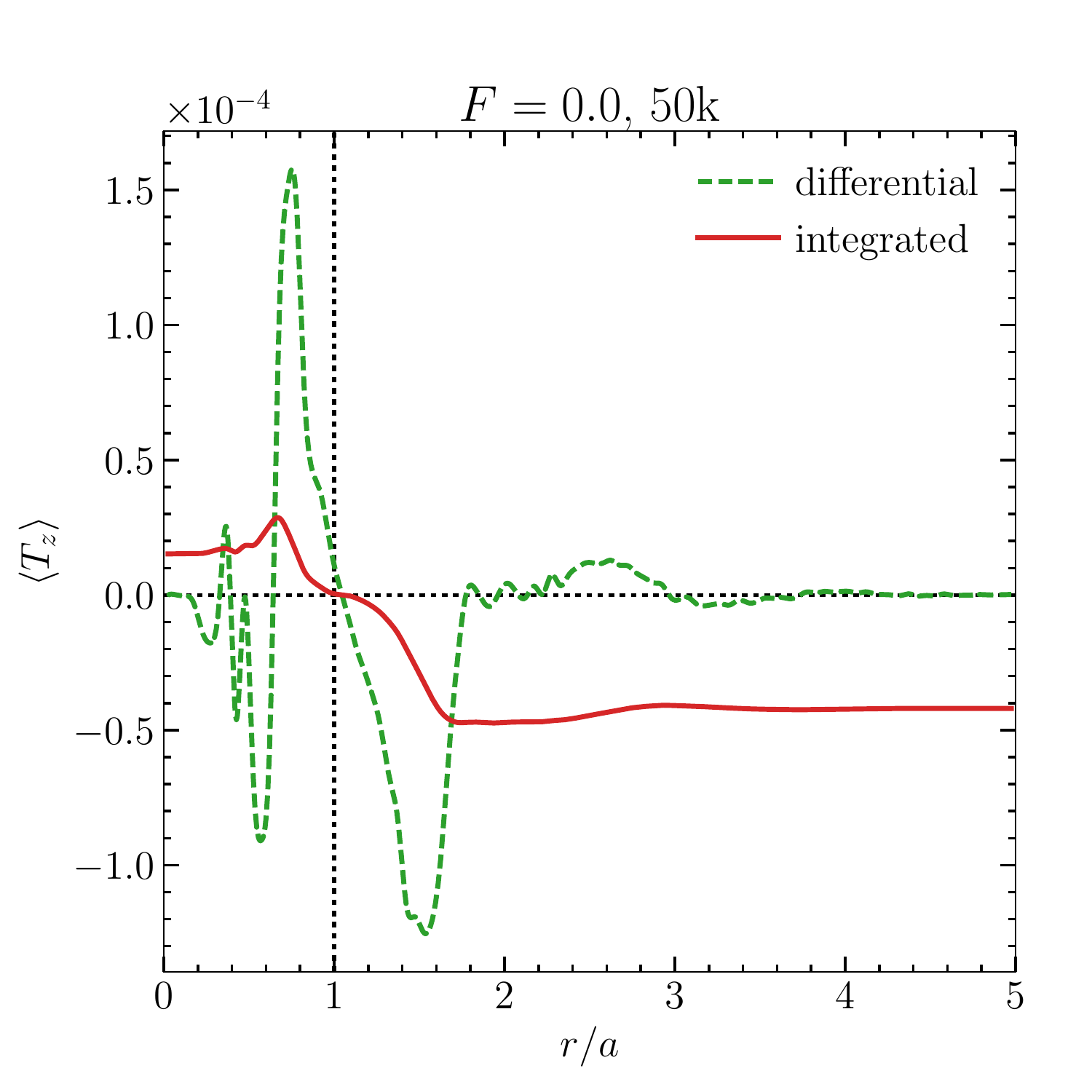}}
\put(0.3333,0){\includegraphics[width=0.3333\textwidth]{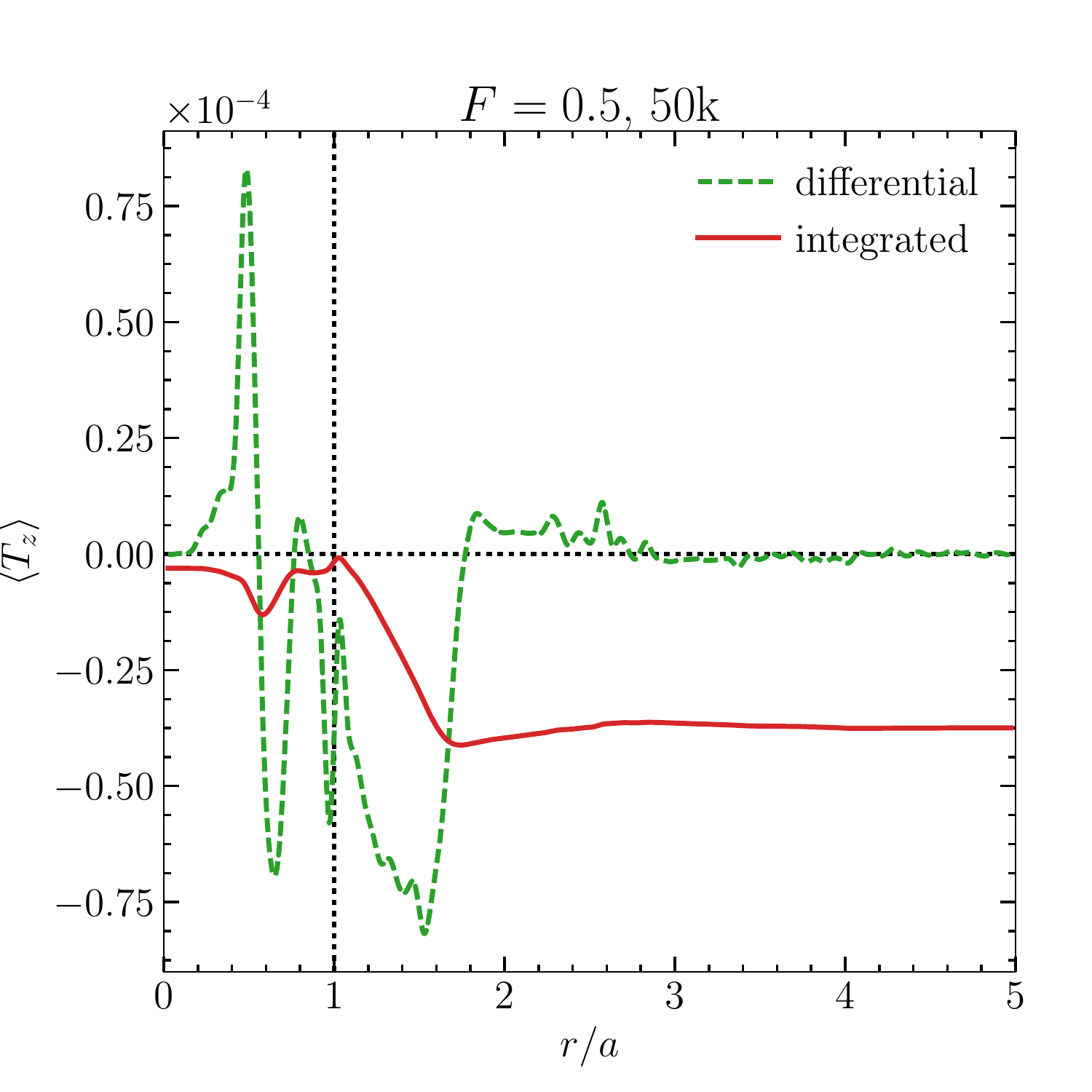}}
\put(0.6666,0){\includegraphics[width=0.3333\textwidth]{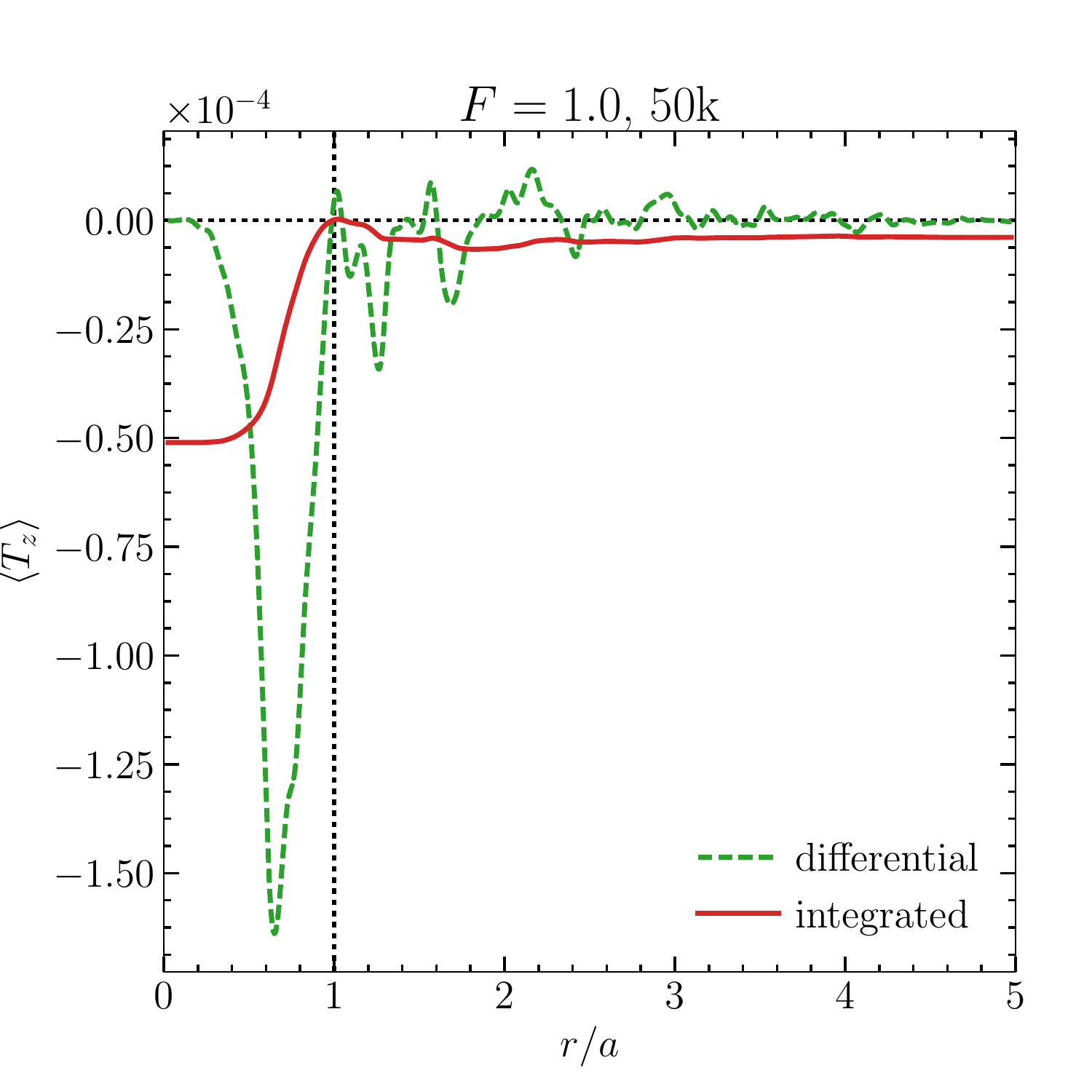}}
\end{picture}
\caption{Torque radial profiles averaged over the last five binary orbits.
  In each panel, the green dashed line represents the differential torque,
  while the red solid line is the total integrated torque. This latter is
  integrated starting from the corotation radius (vertical dotted line)
  inwards and outwards. The quantities are presented in code units, namely,
  $[GM_0^2a_0^{-2}]$ for the differential torque and $[GM_0^2a_0^{-1}]$ for the
  integrated profile.}
\label{fig:torque_profile}
\end{figure*}

To establish the role of the discs in the binary evolution we compute the
gravitational torques exerted by the gas. The total gravitational torque is
given by directly summing the individual particle torques onto each MBH as
follows
\begin{equation}
\vb{T}_g=\sum_{i=1}^N\sum_{j=1}^2 \vb{r}_j\cross \frac{GM_jm_i}{\abs{\vb{r}_i-\vb{r}_j}^3}(\vb{r}_i-\vb{r}_j),
\end{equation}
where the index $i$ runs over all the gas particles in the simulation,
and $j=1,2$ indicates the two black holes. Using this expression, we compute
the torque radial profiles shown in Fig.~\ref{fig:torque_profile},
averaged over 5 orbits of the simulation.
 As the binary's inclination slightly evolves in our simulations, we compute
  these torques in a reference frame in which the binary angular momentum is
  exactly aligned with the $z$ axis. With this choice, we can directly link
  the $z$ component of the torque with the binary orbital evolution
  (eq.~\ref{deltaL}).
Because the torques are null at the corotation radius, the integration of the
differential profile is done from that point outwards and inwards.

From the profiles shown in Fig.~\ref{fig:torque_profile} it is clear that
the net effect of the circumbinary material is to produce a negative torque,
hence to extract some angular momentum from the binary.  For the co-rotating
discs, the largest peak is located between $a$ and $2a$, i.e. inside
the cavity wall. This peak basically determines the total strength of the
torque, which is responsible for the shrinking observed in the binary orbit.
Beyond the cavity region the torques oscillate between positive and negative,
cancelling each other out, and consequently making the contribution of the
disc negligible.
For the counter-rotating disc, conversely, the gravitational torques come
from the material inside the corotation radius, very close to the MBHs, and
the outward torques are essentially absent because of the lack of
Lindblad resonances.

\begin{figure*}
\centering
\begin{picture}(1,0.3333)
\put(0,0){\includegraphics[width=0.3333\textwidth]{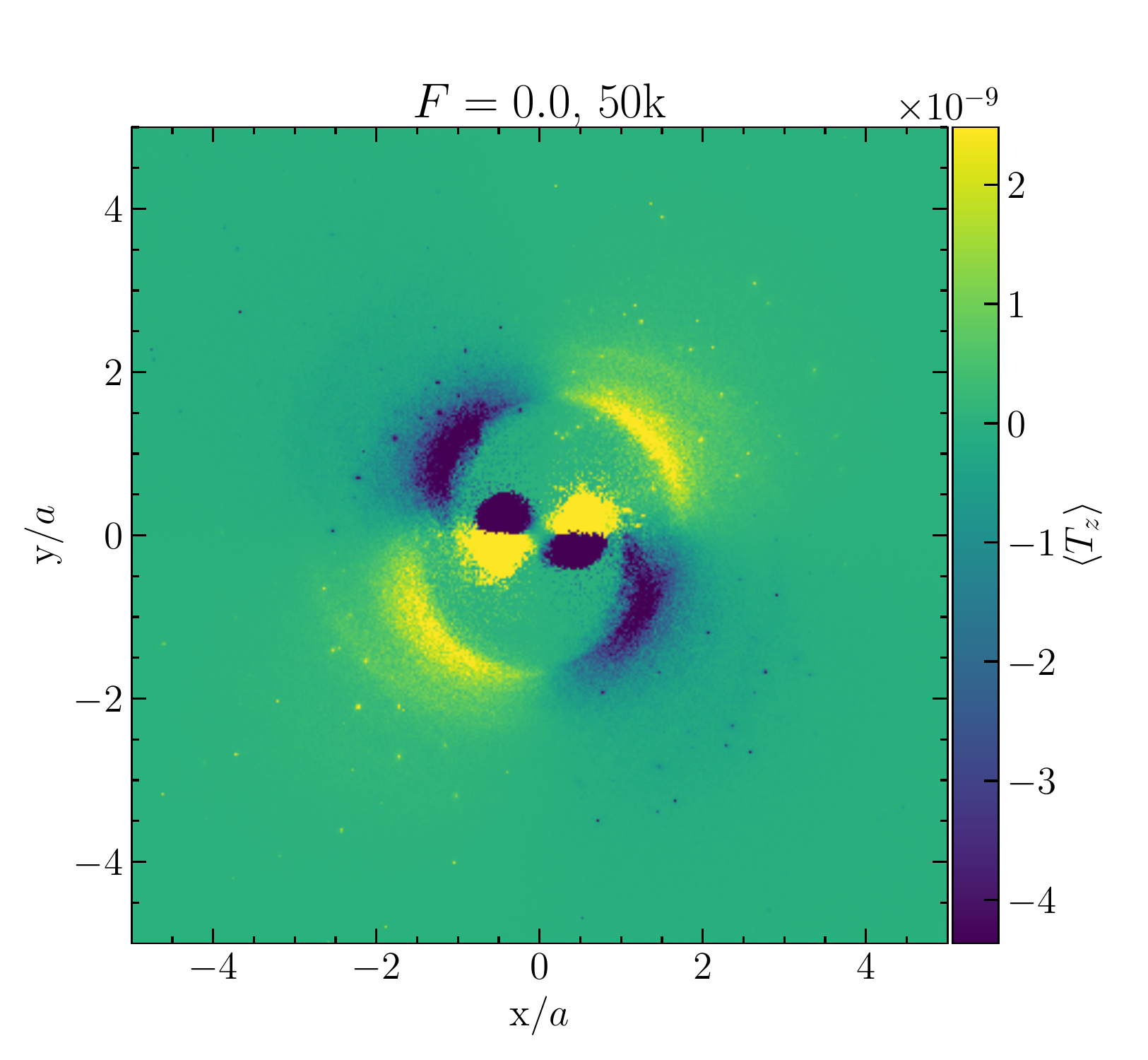}}
\put(0.3333,0){\includegraphics[width=0.3333\textwidth]{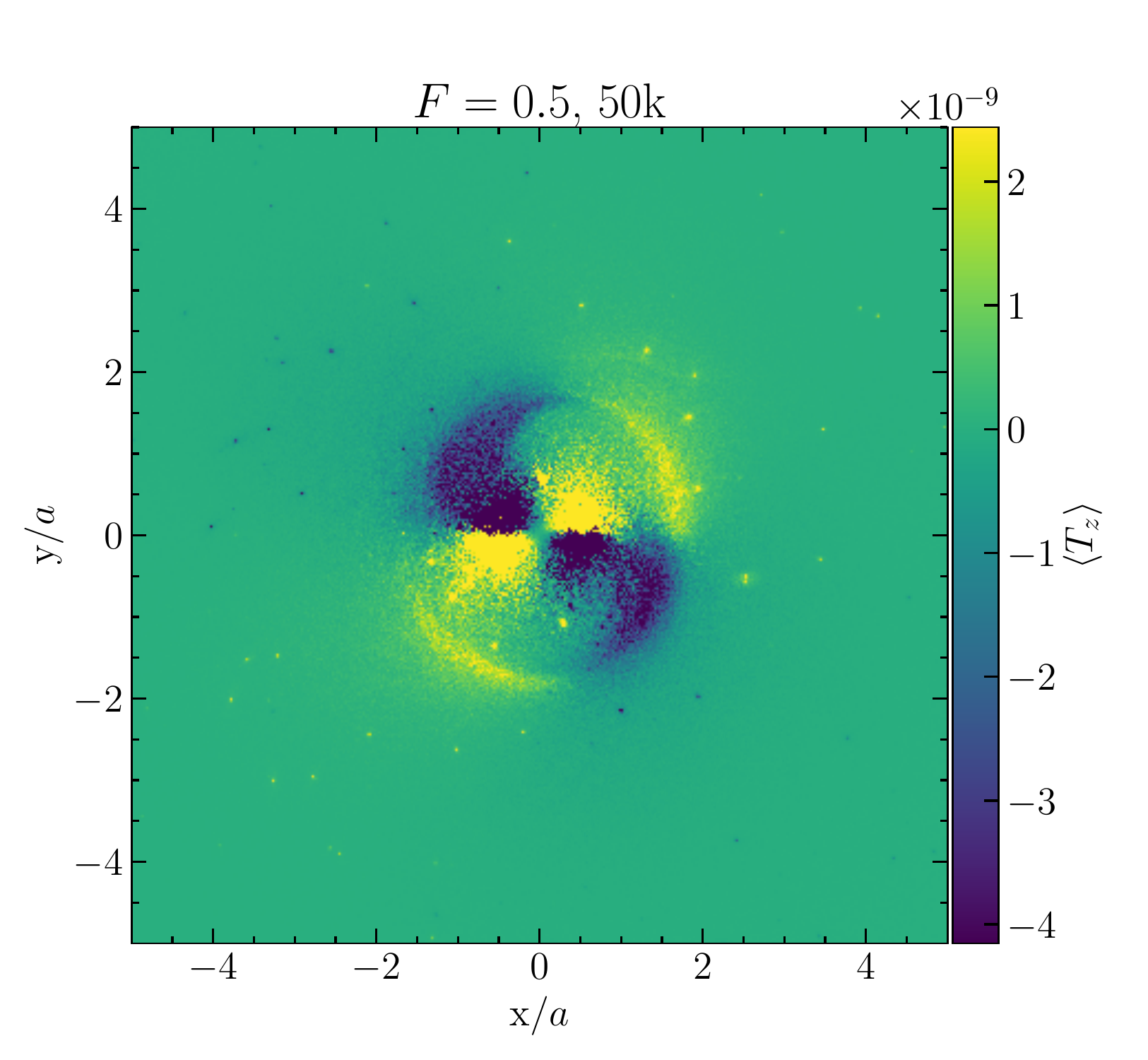}}
\put(0.6666,0){\includegraphics[width=0.3333\textwidth]{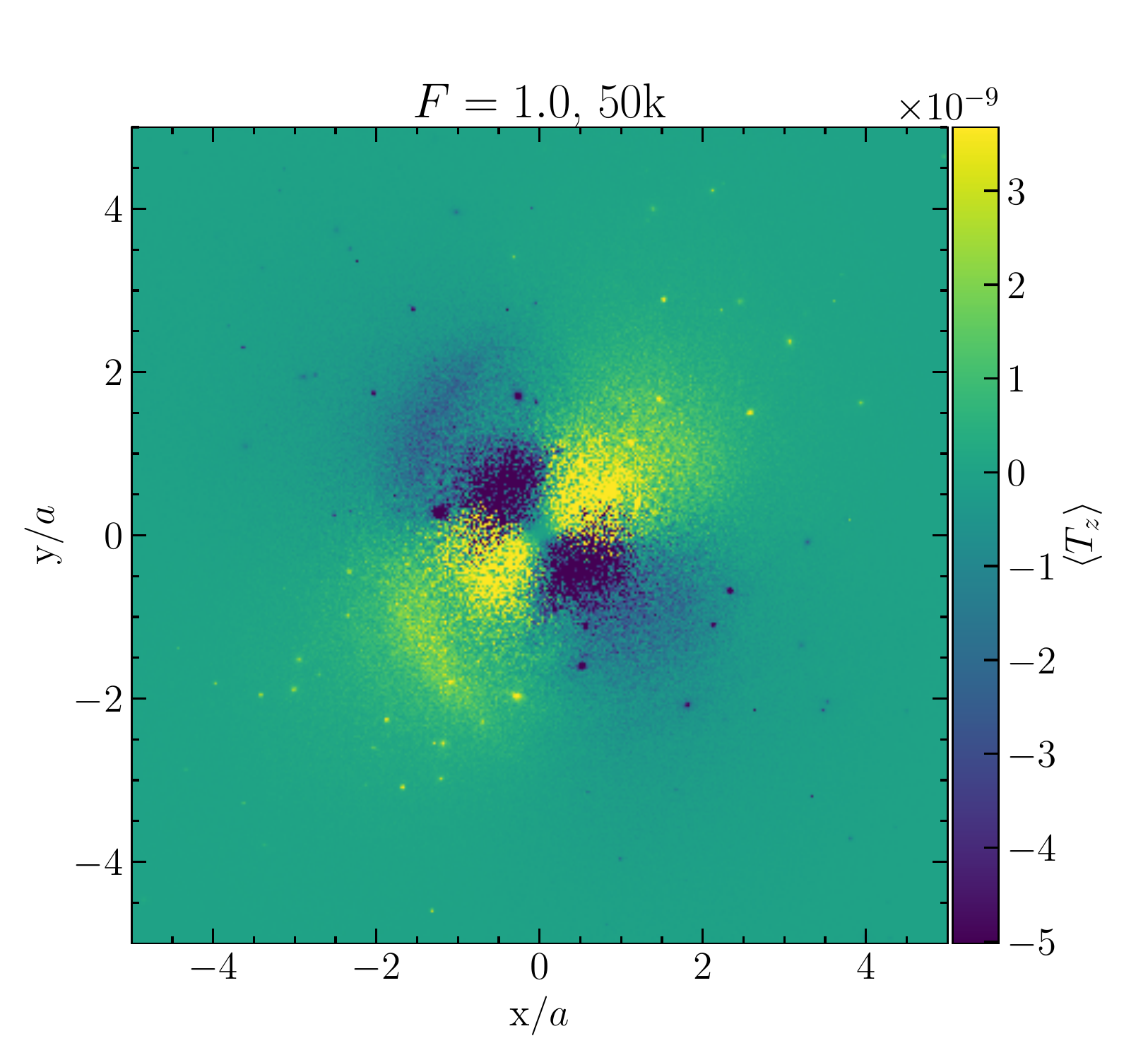}}
\end{picture}
\caption{Averaged surface density torque exerted by the gas onto the binary
  for the \F{0.0} (left), \F{0.5} (centre) and \F{1.0} (right) models. Similar
  to the previous figures, these maps were averaged over five orbits. Note
  that the torque values for the minidiscs are off the scale in both \F{0.0}
  and \F{0.5} models.
  For the prograde discs this shows that the negative torques come from the
  gaseous streams reaching the binary.
  Conversely, in the counter-rotating disc these torques are
  located much closer to the binary.}
\label{fig:average_torque}
\end{figure*}

To identify the location of the negative torques, we show the averaged surface
density torque of the discs in Fig.~\ref{fig:average_torque}. For the
co-rotating circumbinary discs, and similar to the gas distribution shown in
Fig.~\ref{fig:average_dens}, the location of the torques are clearer in
the \F{0.0} model, where the cavity has been almost completely depleted of
gas. Nonetheless, in both \F{0.0} and \F{0.5} cases the gaseous streams
inside the cavity provide a strong source of negative torque. This confirms
that the negative peak observed in the profiles of
Fig.~\ref{fig:torque_profile} comes from the material leaking through the
cavity wall. Rather than being entirely captured by the binary, a fraction
of this gas is flung back to the disc, via gravitational slingshot, carrying
away angular momentum from the binary. This physical effect has been
identified and extensively studied in the literature, mainly by means of
numerical simulations \citep[e.g.][]{Roedig2012, DOrazio2013, Farris2015b, Dunhill2015, Tang2017}.

On the other hand, gravitational torques in the counter-rotating disc are
located much closer to the binary, with no features resembling the gaseous
streams as in the previous cases. This material extracts some of the binary's
angular momentum, but because it is not able to transport it outwards, this
material is eventually accreted by the MBHs. As this gas is counter-rotating
with respect to the binary, its capture and subsequent accretion subtracts
angular momentum much more efficiently than in the co-rotating case
\citep{Nix11a}.

The profiles shown in Fig.~\ref{fig:torque_profile} can be directly
compared with results from previous studies of \enquote{standard} circumbinary
discs. In particular, the simulations of a massive self-gravitating
circumbinary disc by \citet{Roedig2014} find very similar torque profiles
(cf. their Fig.~4), the main distinction being the strength of the peaks --
the torques found in our simulations are noticeably larger.
For instance, for prograde discs, Roedig et al. show a total outwards
torque of $\sim 2\times 10^{-5}$ $[GM_0^2a_0^{-1}]$, while in our simulations
we find it to be $\sim4\times 10^{-5}$ $[GM_0^2a_0^{-1}]$,
hence a factor of $\sim 2$ larger.
In the retrograde case, Roedig et al. find a total integrated torque of
$\sim10^{-5}$ $[GM_0^2a_0^{-1}]$, while we find
$\sim5\times 10^{-5}$ $[GM_0^2a_0^{-1}]$, a factor of $\sim5$ larger.
Even though these values might not seem much higher, note that we are
comparing very massive circumbinary discs with 20\% the binary mass to
the much lighter discs formed in our simulations, that reach at most a few
percent of the MBHB mass \citep{MaureiraFredes2018}.

These large torques are due to the transient nature of our discs; during
the episodes of disc disruption, the binary clear its orbit of gaseous
material, enhancing the torques.
If these circumbinary discs were to settle in a more steady-state, such as the
model of \citet{Roedig2014}, the gaseous streams will be determined mainly by
the internal viscous torques, and we expect much less mass interacting directly
with the binary. This would decrease the efficiency of the gravitational
torques, especially considering their low mass with respect to the MBHs.
To confirm this point, we run a separate set of simulations of these binaries
surrounded by their discs without adding any more clouds -- we referred to
these runs as `forked' simulations (see Section~5.4 in
\citealt{MaureiraFredes2018}).
{Because of the prominence of the circumbinary discs in all distributions,
we analyse simulations `forked' after the 10th cloud.}

\begin{figure*}
\centering
\begin{picture}(1,0.3333)
\put(0,0){\includegraphics[width=0.3333\textwidth]{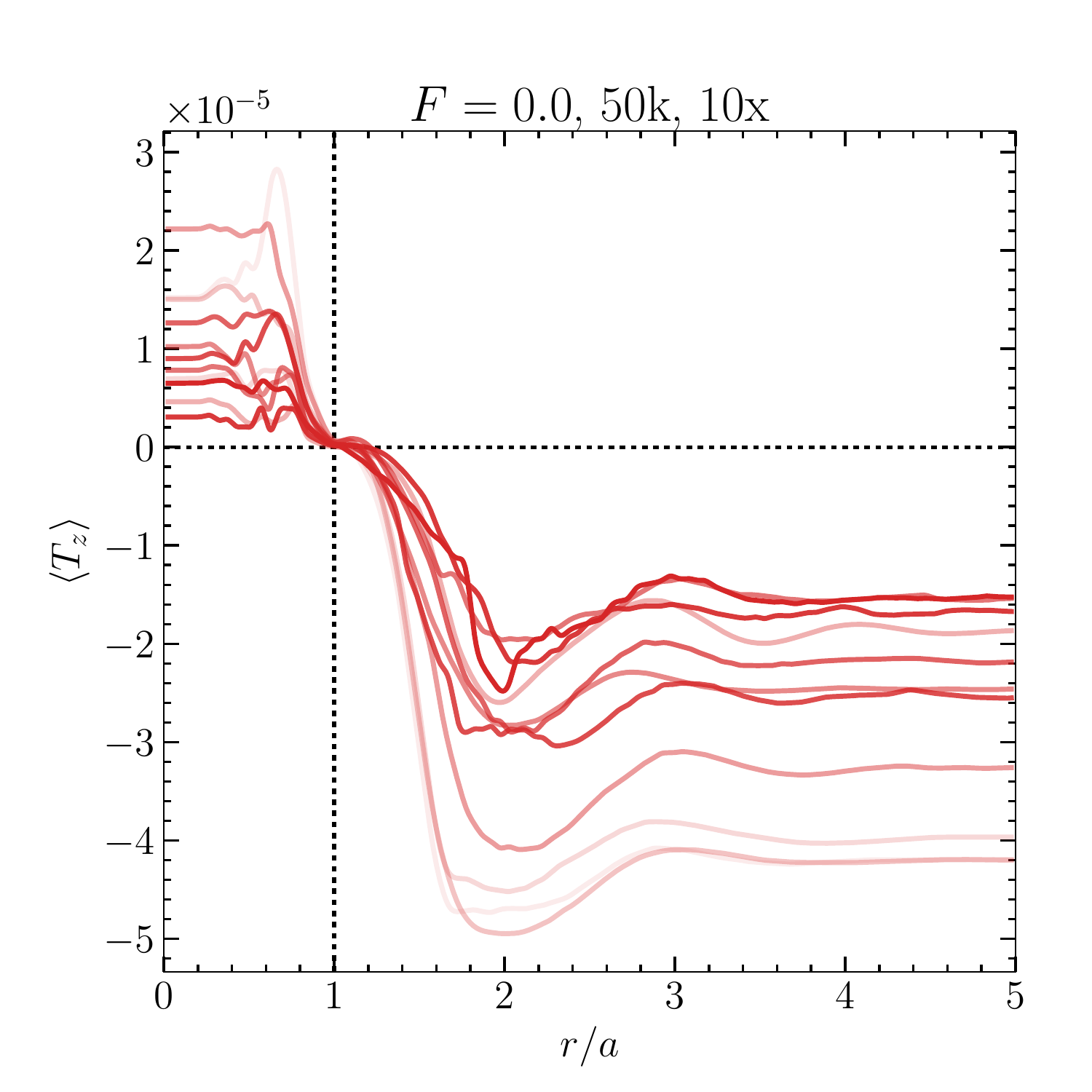}}
\put(0.3333,0){\includegraphics[width=0.3333\textwidth]{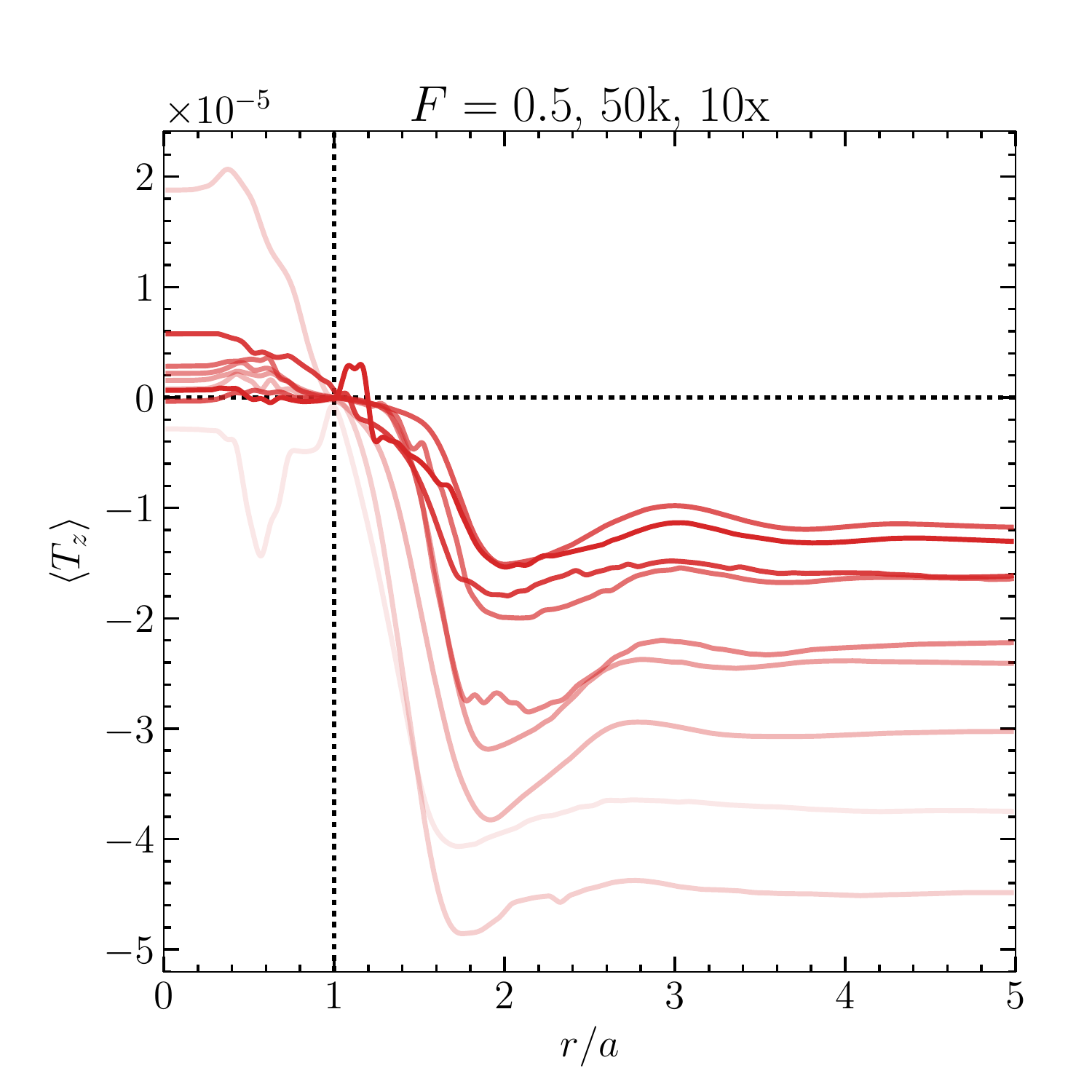}}
\put(0.6667,0){\includegraphics[width=0.3333\textwidth]{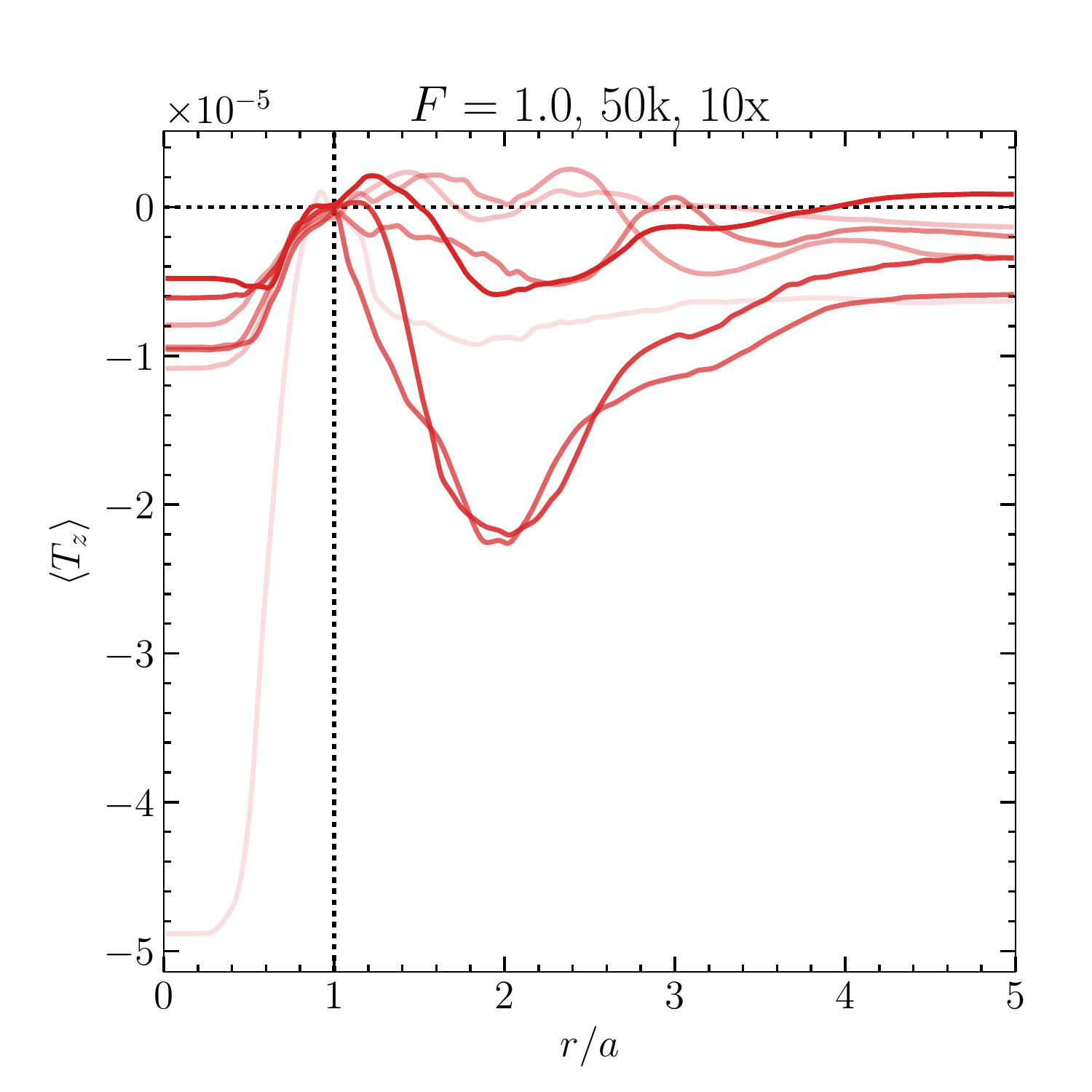}}
\end{picture}
\caption{Time evolution of the integrated torque profile of the circumbinary
  discs, for 'forked 'simulations. Each line corresponds to a
  time average over 5 orbital periods, taken further in the binary
  evolution from the lightest to the darkest line. From left to right we show
  the \F{0.0}, \F{0.5}, and \F{1.0} distributions.
  The trend shows decreasing gravitational torques as a function of
  time.}
\label{fig:prof_evolution}
\end{figure*}

\begin{figure}
\centering
\includegraphics[width=0.5\textwidth]{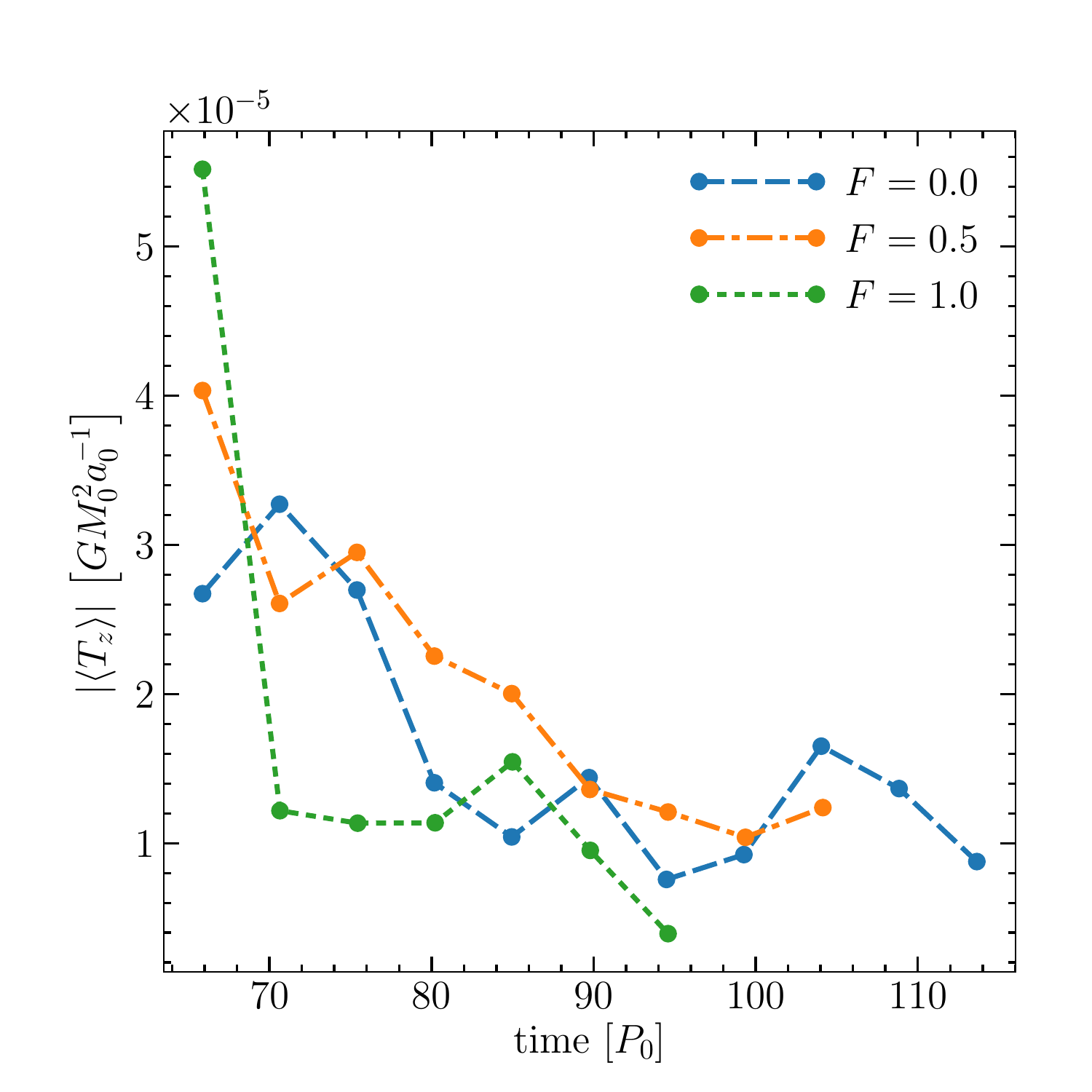}
\caption{Total integrated torque evolution measured during the `forked'
  simulations. Points correspond to the average of the profiles shown in
  Fig.~\ref{fig:prof_evolution}. The light blue line shows the evolution of
  the \F{0.0} distribution, the orange line is the \F{0.5} distribution, and
  the green lines is the \F{1.0} distribution.
  This shows that the slow relaxation of the discs translates into decreasing
  gravitational influence.}
\label{fig:tot_torque}
\end{figure}

Using these simulations we measure how the total integrated torque profile
evolves in time for each of the discs, as we show in
Fig.~\ref{fig:prof_evolution}. Similar to Fig.~\ref{fig:torque_profile}, each
line is the radial profile of the torque's $z$ component averaged over 5
binary orbital periods. From this figure it is clear that the gravitational
influence of each disc tends to decrease as time passes (from lighter to
darker lines), showing that the measured evolution driven by torques is still
transient and not a long term effect. For the co-rotating circumbinary discs
(left and middle plots of Fig.~\ref{fig:prof_evolution}) the shape of the
profile remains rather unperturbed, which means that the discs evolve
smoothly. This is not the case for the counter-rotating disc (right plot of
Fig.~\ref{fig:prof_evolution}), where we observe strong changes in the
profile outside the corotation radius. This occurs because this circumbinary
ring-like structure has little angular momentum, and hence it can be perturbed
much more easily by gaseous clumps and streams reaching the binary from
previous interactions. Nevertheless, the profile \emph{inside} the binary's
orbit remains roughly unperturbed, just decreasing its strength with time.
Consequently, the evolution of the torque profiles displayed in
Fig.~\ref{fig:prof_evolution} clearly shows the transient nature of the large
torques found in our simulations.

The relaxation of the discs becomes even more apparent when we compute the
total $z$ torque (averaged over five orbits) as a function of time, as
shown in Fig.~\ref{fig:tot_torque}.
The trend is clear: the net gravitational effect of the discs decreases in
time. This damping of the torque is occurring somewhat slowly in our
simulations, but this is likely due to the fact that we are not including any
source of viscosity on top of the artificial term to capture shocks. This
artificial viscosity acts as an effective viscosity within the disc, but is
typically very small \citep{Lodato2010}. We would expect these discs to relax
much faster with a more realistic prescription for the viscosity.

Nonetheless, we can use the formalism presented in \citet{Roedig2014} to make
an order of magnitude estimation of total effect of such discs in the binary
orbit. If we take the simple assumption that the gravitational torques
evolve only the semimajor axis, we can write
\begin{equation}
\frac{\dot a}{a_0}\sim \frac{2T_{z}}{L_{z,0}},
\end{equation}
where $L_{z,0}=0.25M_0\sqrt{GM_0a_0}$ for our equal-mass and circular binary.
By taking an average torque of $T_z\sim 10^{-5}$, this expression yields
\begin{equation}
\dot a\sim - 8\times 10^{-5}a_0\Omega_0,
\end{equation}
which over the $\approx 900$ dynamical times that span our \textsc{50k}
simulations translates onto a total evolution of
\begin{equation}
\Delta a_{\rm disc} \sim 7\times 10^{-2}a_0.
\end{equation}
Comparing this value with the final $\Delta a$ of each distribution
(Table~\ref{tab:deltasF}), we find that corresponds to $\sim$40\% for the
\F{0.0}, $\sim$20\% for the \F{0.5}, and $\sim$10\% for the \F{1.0}
distribution. Even though this seems to indicate that the circumbinary
structures contribute with an important fraction of the evolution, we arrived
to this number by considering the presence of the circumbinary disc during the
entirety of the run, which is certainly not true. As a matter of fact, these
discs arose after the arrival of the 10th cloud, and only the one from the
\F{0.0} distribution survives the interaction during the rest of the
simulation, the other two are destroyed by the interaction following a few
clouds \citep{MaureiraFredes2018}. Consequently, this estimate demonstrates that the
extraction of angular momentum via gravitational torques from a circumbinary
disc is subdominant with respect to capture and accretion of material,
particularly for the \F{0.5} and \F{1.0} distributions, where these structures
are unlikely to survive.

\section{Additional run: the effects of shorter arrival times}
\label{sec:comparison}

\begin{figure}
\centering
\begin{tabular}{c}
    \includegraphics[width=0.45\textwidth]{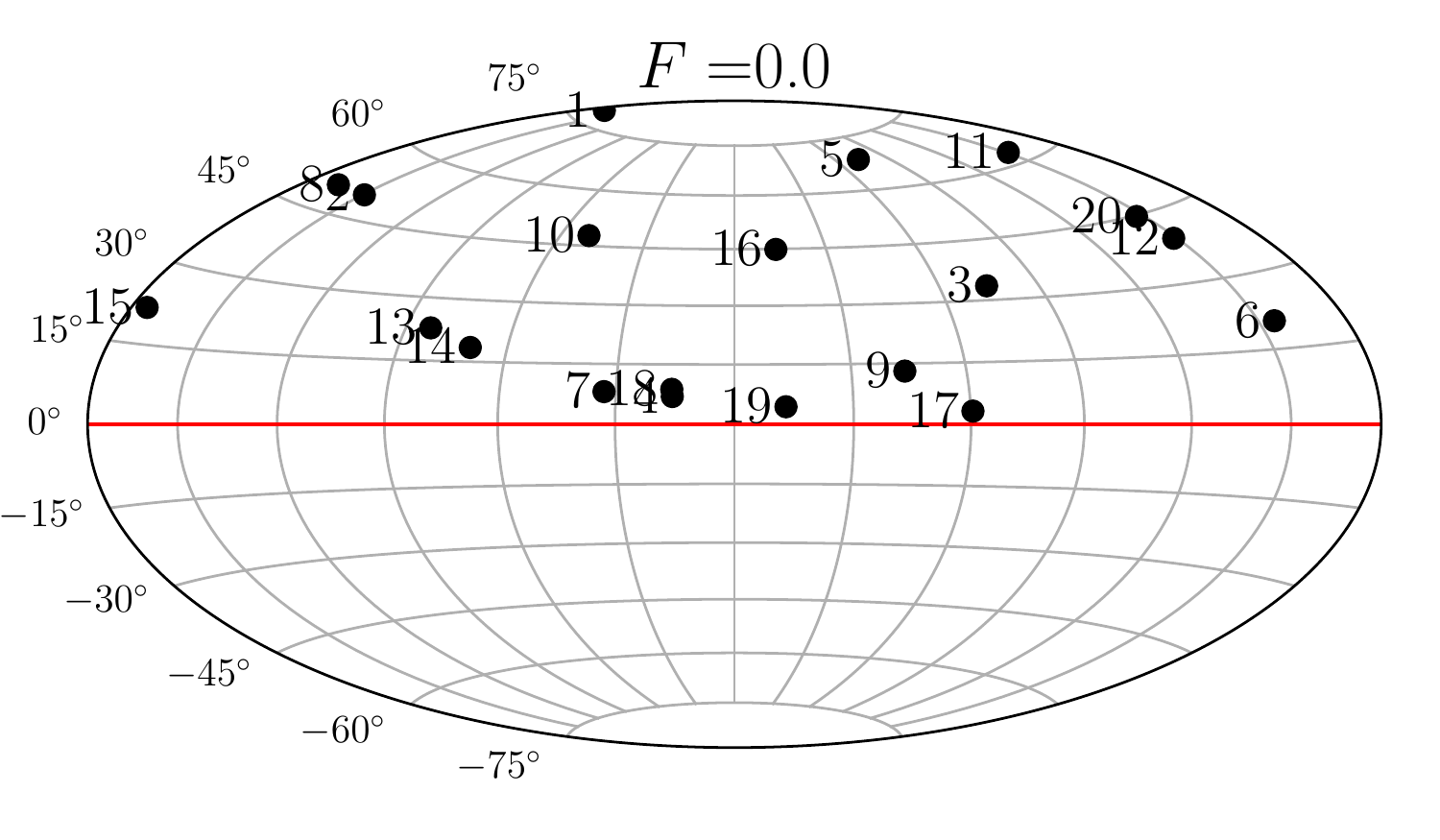} \\
    \includegraphics[width=0.45\textwidth]{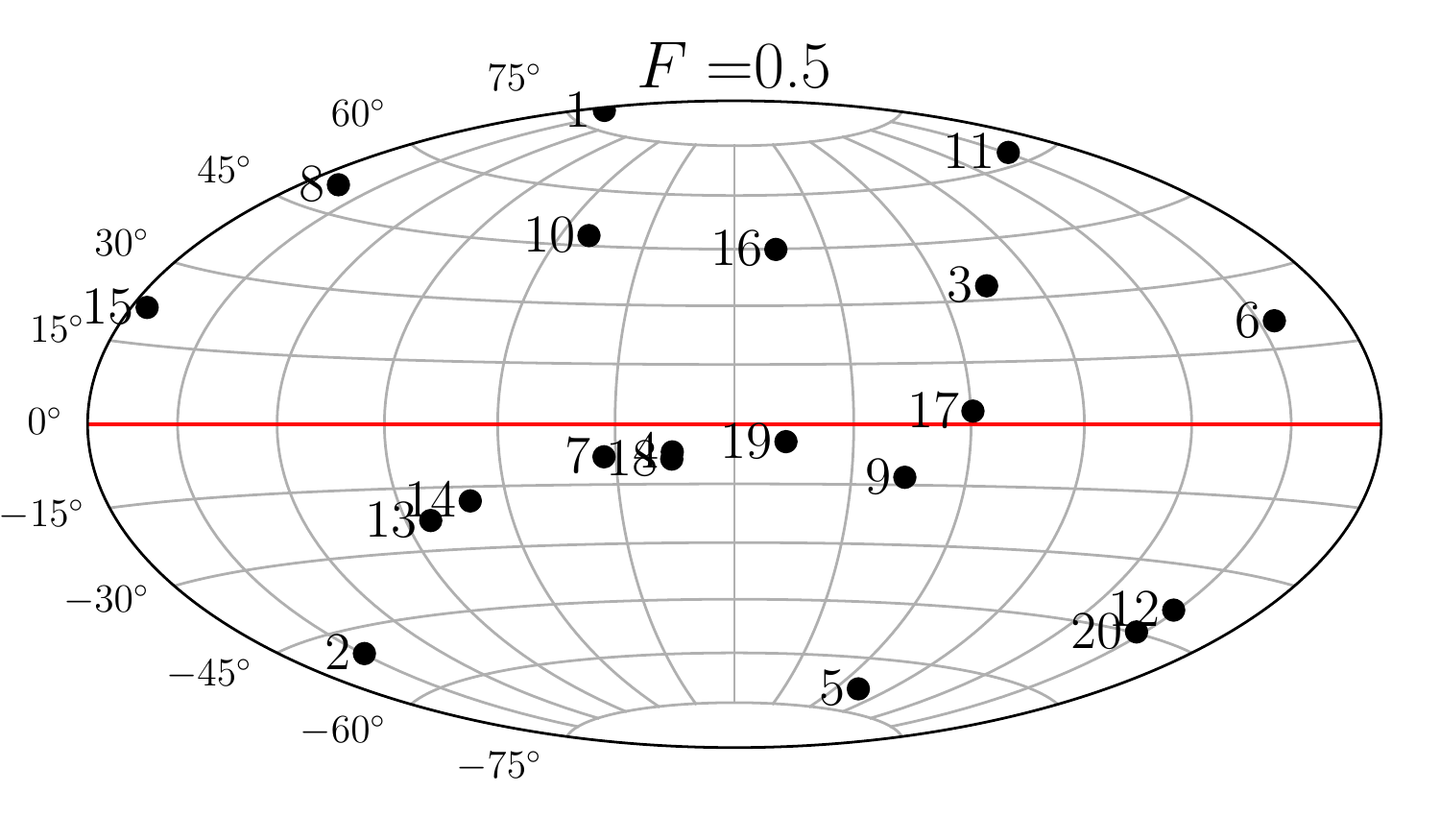} \\
    \includegraphics[width=0.45\textwidth]{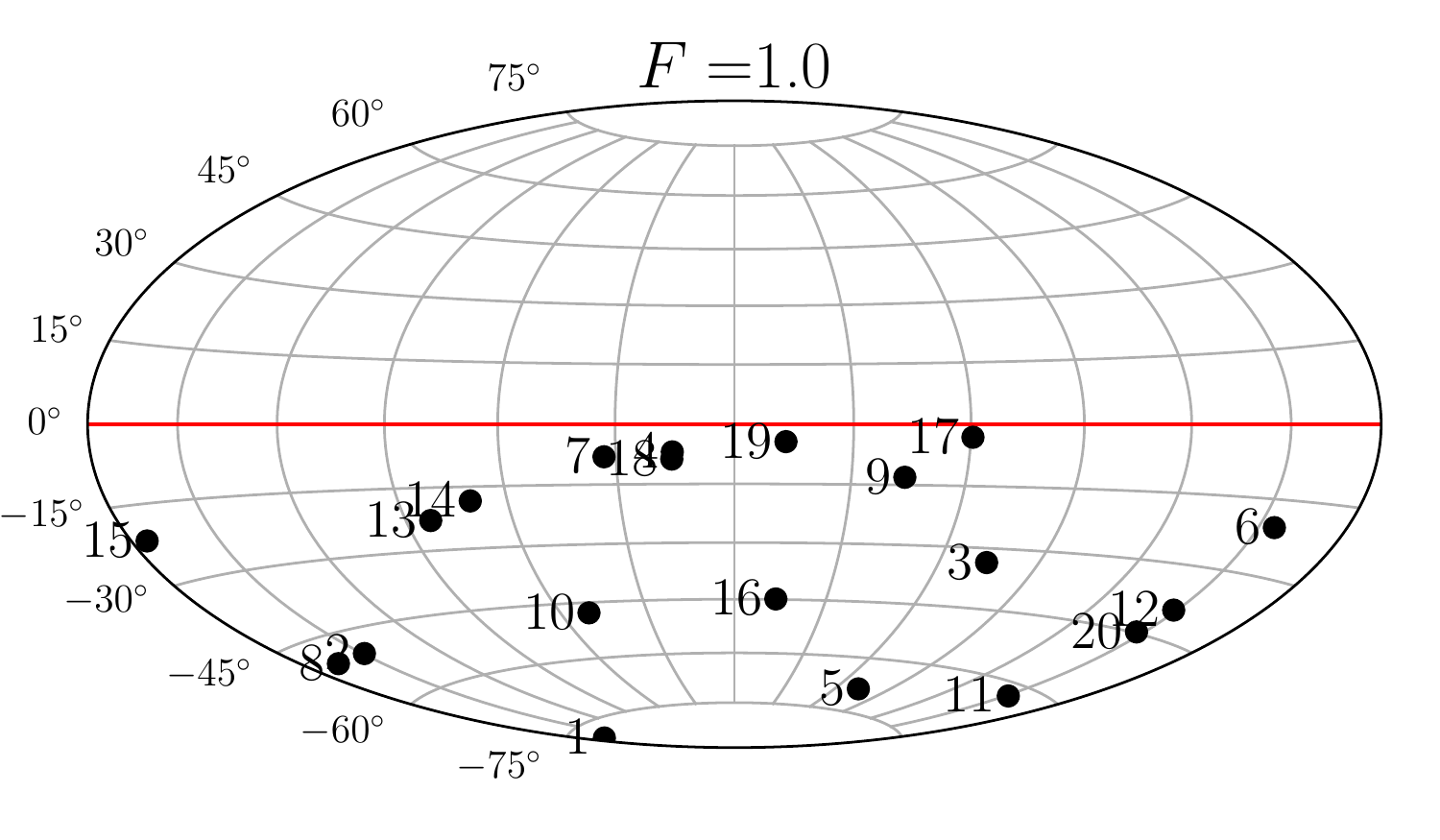} \\
    \includegraphics[width=0.45\textwidth]{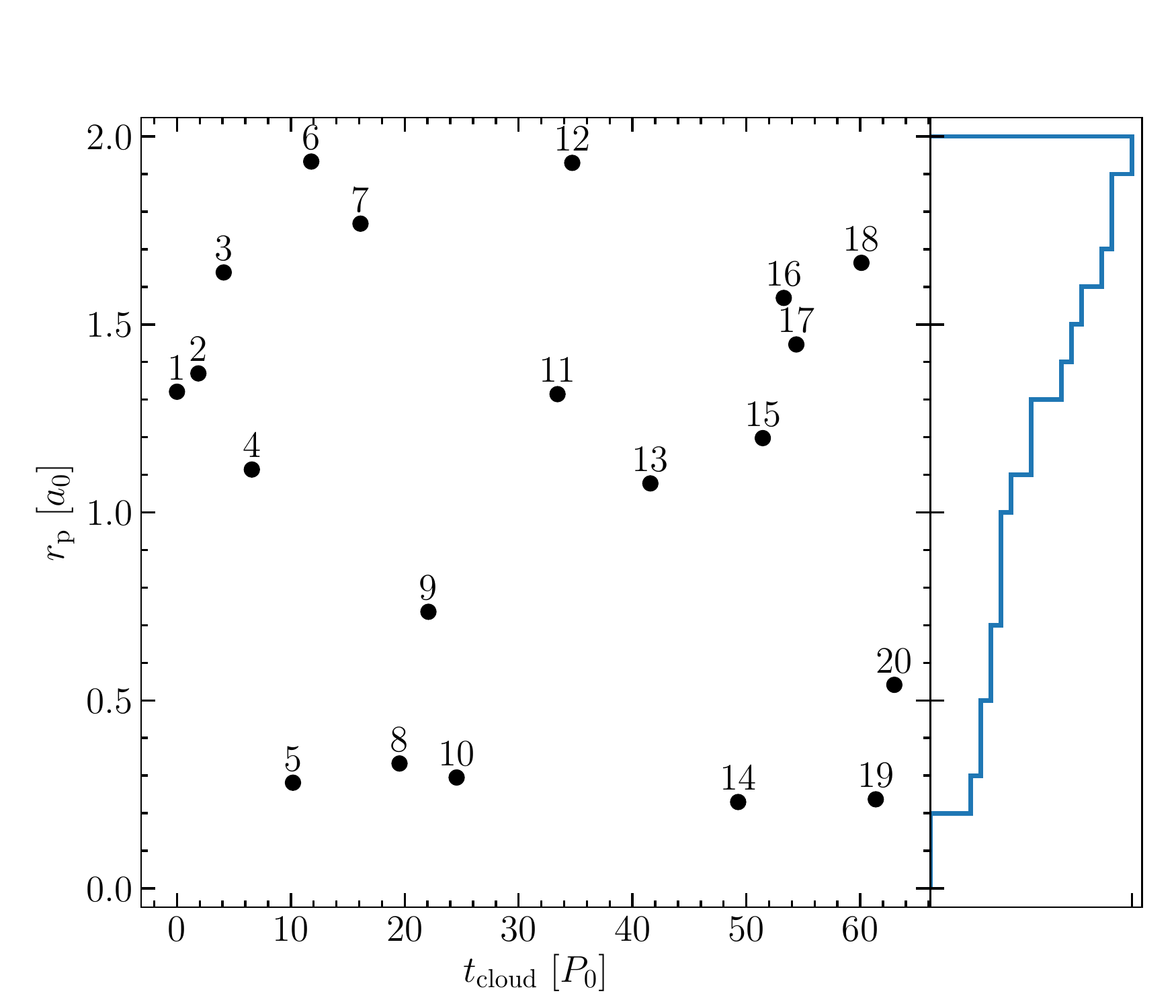} \\
\end{tabular}
\caption{Compilation of the initial conditions for \runa.
This is similar to the distributions shown in Figs.~\ref{fig:ics}
and~\ref{fig:peris}, where on the three top plots we display the cloud's
angular momentum orientations for the different levels of anisotropy,
while the bottom plot shows the distribution of pericentre distances
and the as a function of the time of inclusion.}
\label{fig:RunA_ics}
\end{figure}

Due to the stochasticity intrinsic to the low number of injected
clouds, significant scatter in the MBHB evolution can be expected in
different realisations of the interacting cloud population. We thus
analysed a second suite of simulations, corresponding
to \texttt{RunA} in \papergas. Initial conditions for this set of runs
are shown in Fig.~\ref{fig:RunA_ics}. As described in \papergas, due
to computational limitations we were unable to achieve the 20th cloud
for all the $F$ distributions, and we integrated {\runa} \F{0.0} up
to cloud 15, {\runa} \F{0.5} up to cloud 22 and  {\runa} \F{0.0} up
to cloud 12. To facilitate the comparison with the previous results,
here we present only up to cloud 20.

Besides featuring a different random
seed for the generation of initial pericentre and angular
momentum distributions, this set of simulations is characterised
by a much shorter average elapsed time in between clouds.
This can be clearly seen by comparing the time of the final cloud
($t_{\rm 20th}$) in
both runs -- while in \runb~the the 20th cloud is introduced
at around $t_{\rm 20th}\approx$120~$P_0$ (Fig.~\ref{fig:peris}), for
\runa~this time is $t_{\rm 20th}\approx$60~$P_0$
(Fig.~\ref{fig:RunA_ics}, bottom plot). This  translates onto
an average time difference a factor of 2 smaller for the latter.
\begin{figure}
\centering
\includegraphics[width=0.5\textwidth]{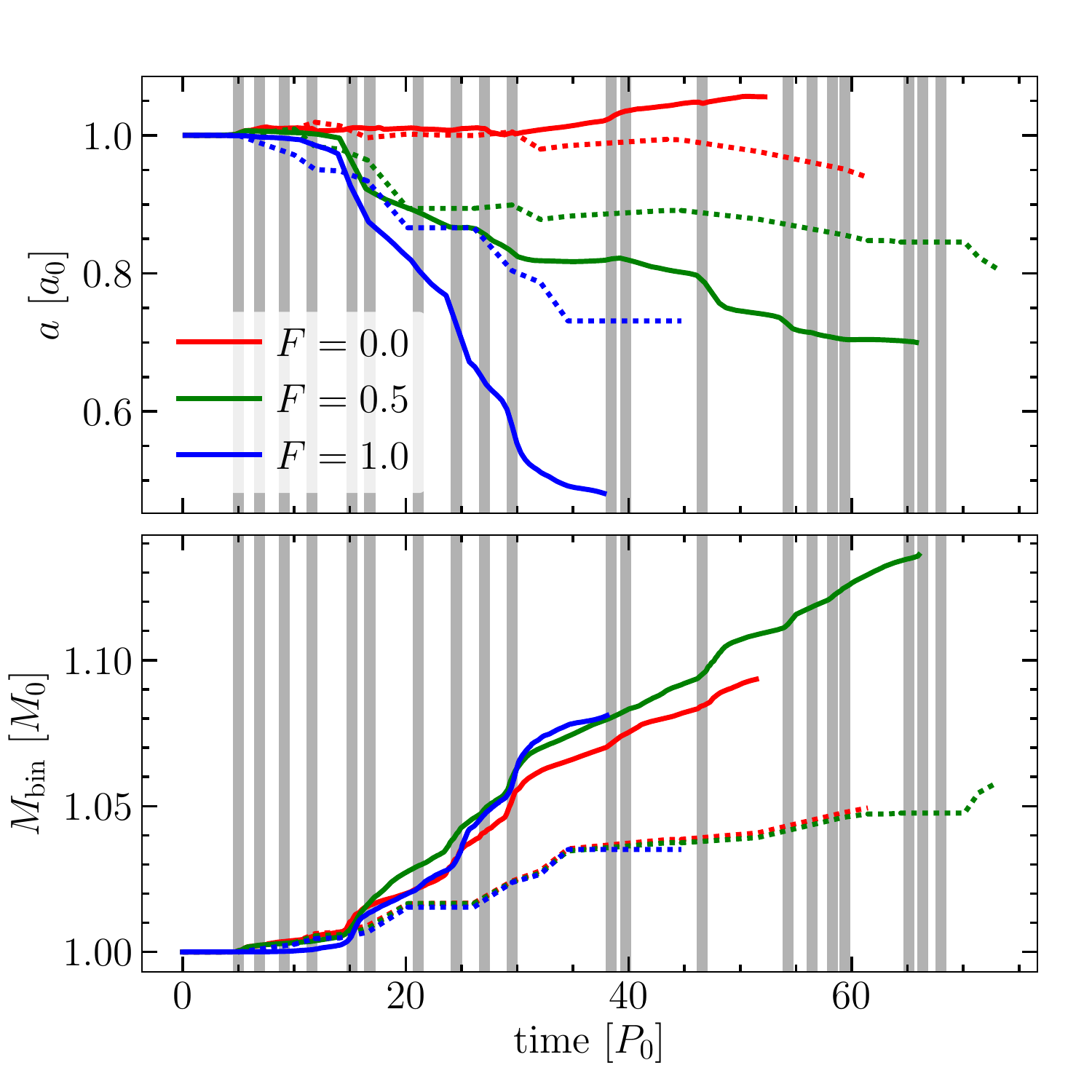}
\caption{Same as Fig.~\ref{fig:mcs_sim_events}, but for \runa.
}
\label{fig:RunA_sim_events}
\end{figure}

One of the most significant results presented in the previous
section is the deviation of the binary evolution from what predicted by
the single cloud model. A difference that clearly shows the effects of the
non-accreted material, which was not taken into account in that model.
We repeated the  analysis for \runa, results are shown in
Fig.~\ref{fig:RunA_sim_events}. Most of the conclusions drawn for
\runb~also apply to this case. For instance, although the
the single cloud model is able to roughly reproduce the evolution
driven by interaction with the first few clouds, it consistently
underestimates the cumulative binary shrinking as further clouds are
added to the system. A striking difference between the actual
evolution and the prediction of the single cloud model
is observed in the \F{0.0} case, in which the binary slightly
expands instead of shrinking its orbit.

\begin{figure}
\centering
\includegraphics[width=0.5\textwidth]{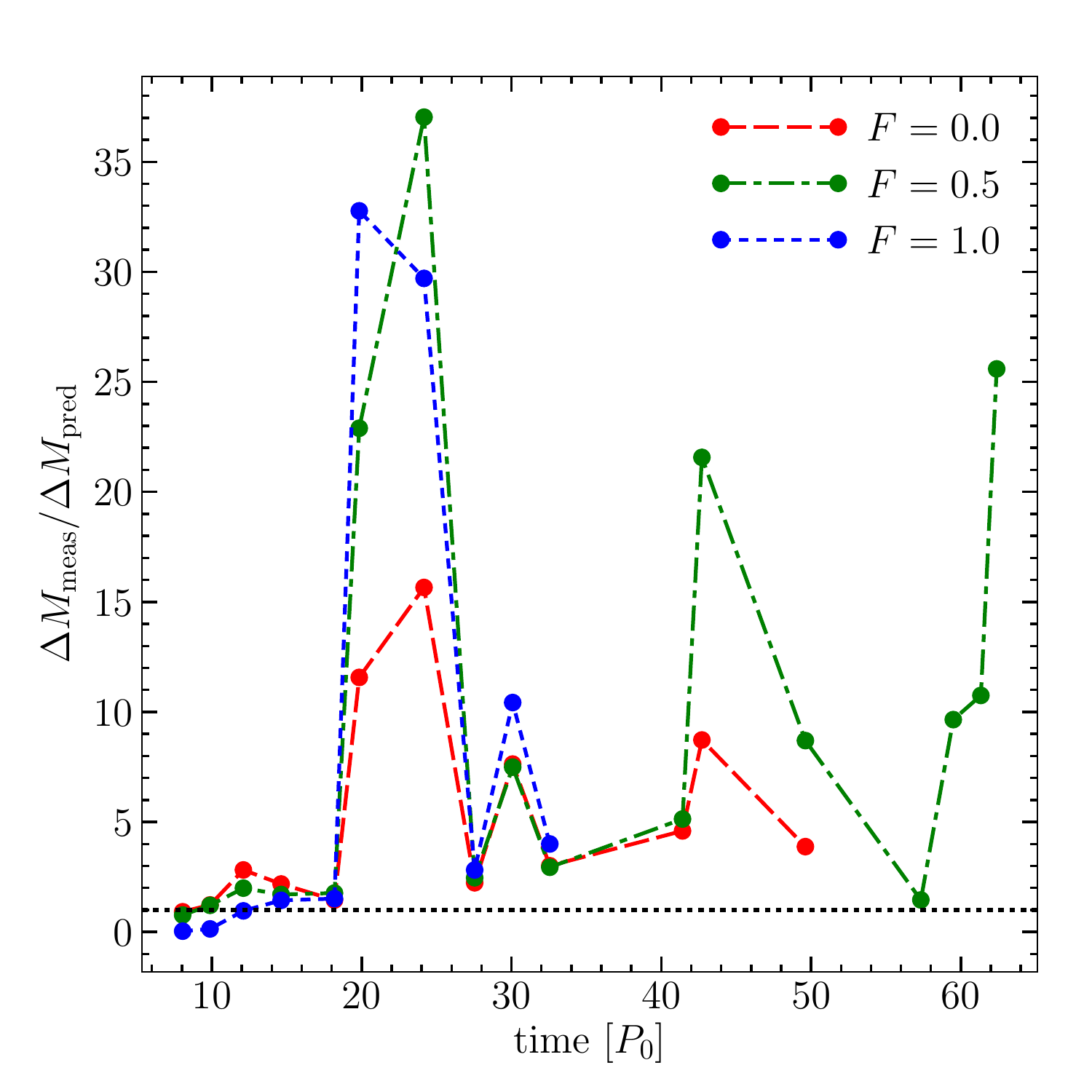}
\caption{Same as Fig.~\ref{fig:acc_mass}, but for \runa.
}
\label{fig:RunA_acc_mass}
\end{figure}

Similar to \runb, we observe that the total accreted mass measured
in the multicloud simulation is about a factor 3 larger than the
value predicted by
the single cloud models. Consequently, the dominant effect of the surrounding
gas on \runa~also appears to be the enhanced accretion caused
by the interaction with the new incoming clouds. To quantify this increase
in accretion, we directly compare figures predicted by the single cloud
model with those measured from the simulations in Fig.~\ref{fig:RunA_acc_mass}.
This comparison clearly shows that the single cloud model underestimates
the accreted mass onto the binary after the first few clouds even more
consistently than in \runb. This occurs because the typically shorter
times between events does not allow for material to settle into a disc.
This is one of the main results shown in \papergas~-- in order for the gas
to settle into a more stable circumbinary configuration it needs to gain
some angular momentum, which takes a few binary orbits. When clouds interact
frequently with the binary, new incoming clouds strongly perturb the
surrounding gas structures, prompting a lot of accretion onto the binary.
In agreement with \runb, this is the main source of deviation with respect
to the single cloud predictions shown in Fig.~\ref{fig:RunA_sim_events}.

\begin{figure*}
\centering
\begin{picture}(1,0.3333)
\put(0,0){\includegraphics[width=0.3333\textwidth]{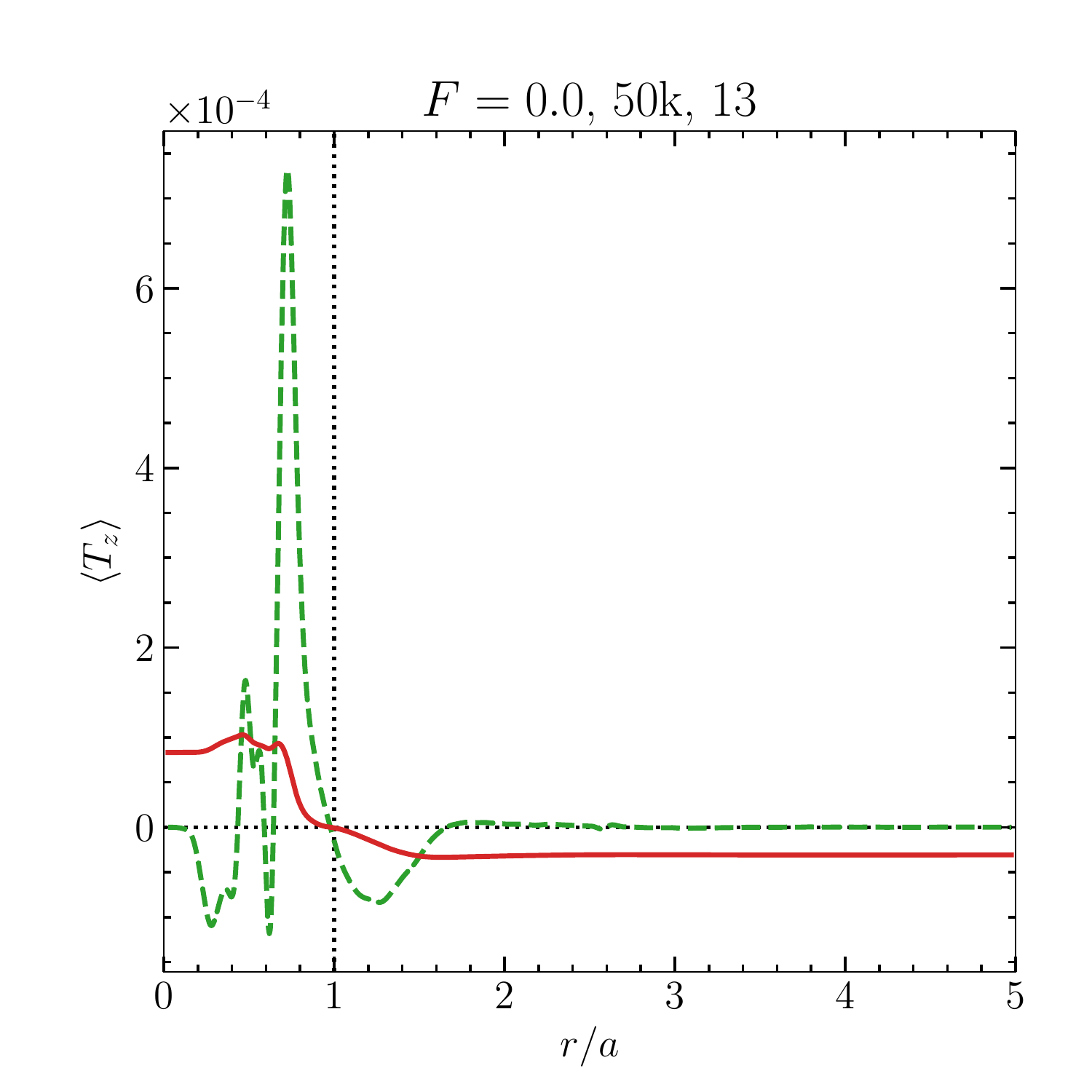}}
\put(0.333,0){\includegraphics[width=0.3333\textwidth]{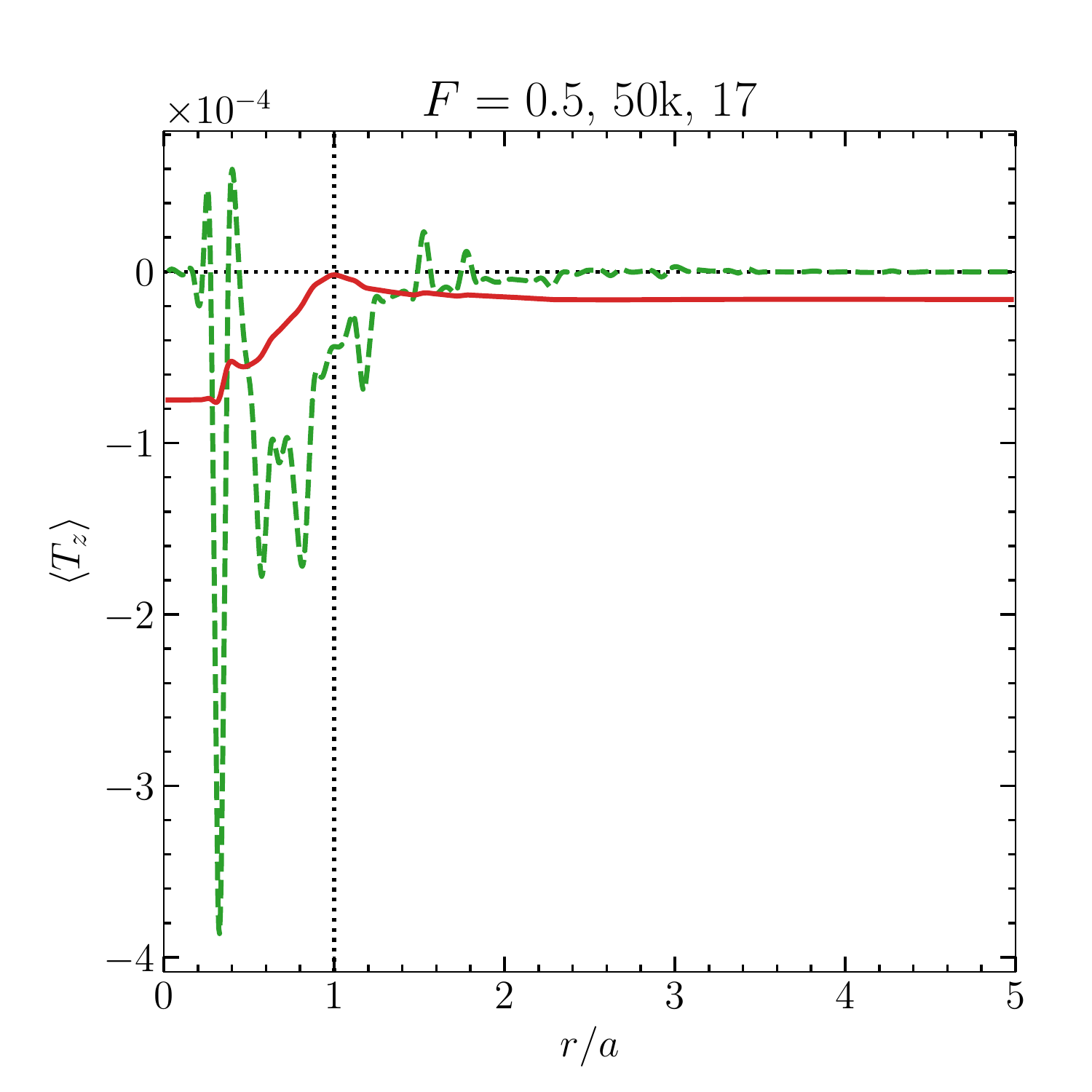}}
\put(0.6667,0){\includegraphics[width=0.3333\textwidth]{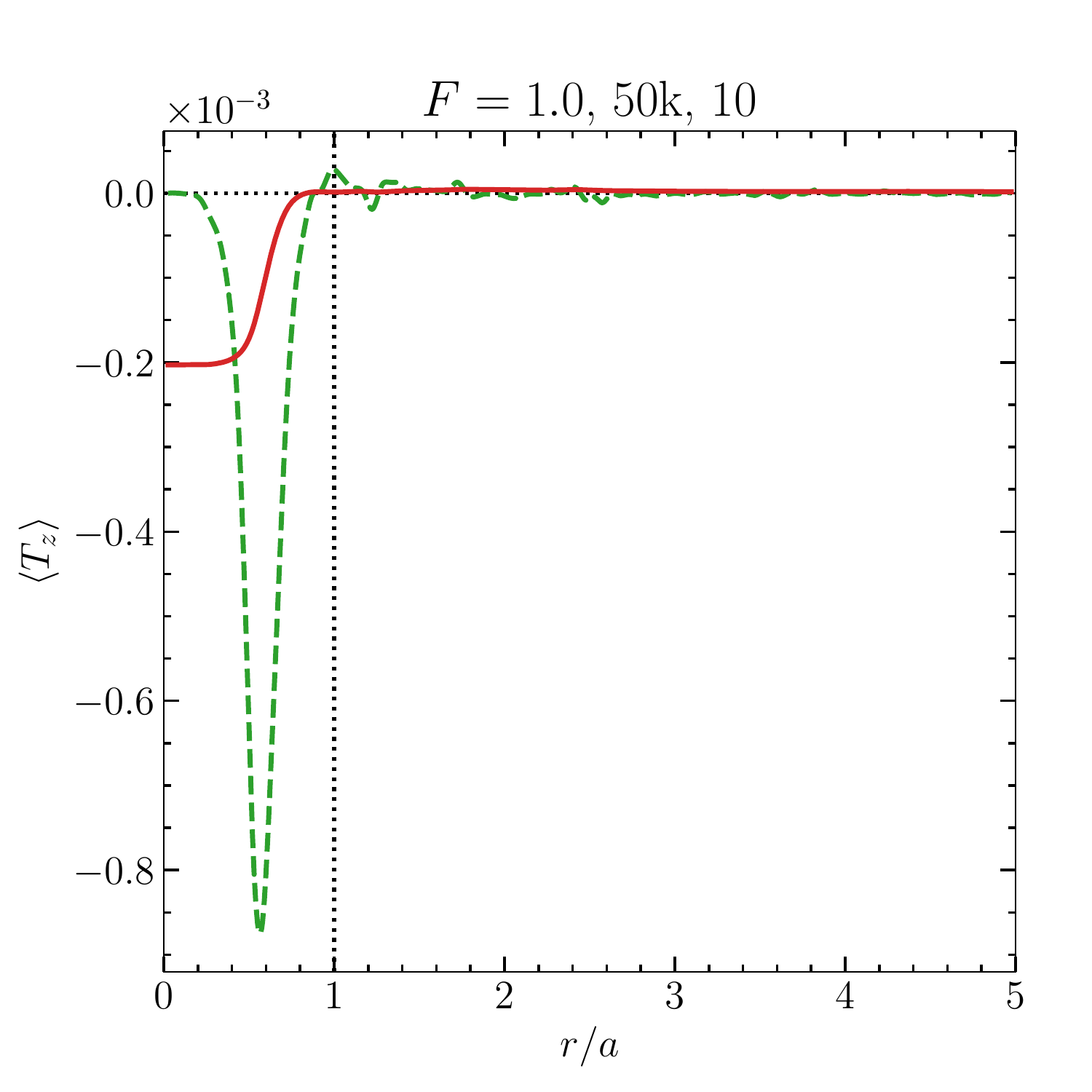}}
\end{picture}
\caption{Same as Fig.~\ref{fig:torque_profile}, but for the circumbinary
  structures observed during \runa. Please note that the
  circumbinary dics responsible for the represented torques
were identified on different times depending on the distributions:
after the 12th cloud for \F{0.0}, 16th for \F{0.5} and 10th for \F{1.0}.
}
\label{fig:RunA_profiles}
\end{figure*}

Despite the higher infall rate of clouds in \runa, we do observe also
the formation of somewhat coplanar circumbinary discs, albeit they get
continuously destroyed by the interaction with new incoming gas
\citep{MaureiraFredes2018}.
Opposite to \runb, those discs are less prominent and on average much lighter.
Nonetheless, we can compute the gravitational influence of these
circumbinary structures through the torque that the gas exerts on the binary.
The differential and integrated radial torque profiles are shown in
Fig.~\ref{fig:RunA_profiles}.

Compared to the profiles shown in Fig.~\ref{fig:torque_profile} we observe
similar features, confirming the robustness of the dynamics against
different realisations of the interacting cloud population.
In particular, in the \F{0.0} and \F{0.5} runs, material outside the
corotation radius is transporting some of the binary's angular momentum outwards via
gravitational slingshot, while for the \F{1.0} distribution all the negative torques come
from the material inside the binary orbit.

  Interestingly, in the \F{0.0} simulation the positive torque inside the
  corotation radius is larger than the outwards torque, which implies that the
  net effect of the gravitational forces is to add momentum to the binary,
  expanding its orbit. This is likely why the final semimajor axis during this
  simulation is larger than the initial value, in contrast to the predicted
  evolution which is a slight shrinking of the orbit.
  The continuous disruption of arising structures
due to frequent infall of clouds, supplies a larger amount of gas
inside the binary's orbit, enhancing these positive torques.
If the accretion events were to stop, or decrease in frequency,
the surviving material would settle into a circumbinary disc,
and as the binary opens up a cavity, gravitational slingshot of gaseous
streams would becomes the dominant effect, shrinking its semimajor axis.

{Overall, the main conclusions found in the previous section  apply to
this run. The dominant effect
of the non-accreted material on the binary evolution comes from the
enhanced accretion produced by stream interactions. Due to the higher
infalling rate we find that is even more difficult to build massive and
stable circumbinary structures \citep{MaureiraFredes2018},
which further decreases the
influence of the angular momentum transport by gravitational torques.}
{The net result is that the dynamics of the system is mostly driven by
capture and accretion of mass, which has the interesting effect of making
the evolution of the binary even more sensitive to the details of the
cloud distributions. In general, for isotropic or counter-rotating
distributions, the mutual partial cancellation of the MBHB and the accreted gas
angular momenta accelerates the binary shrinking, driving the binary
to a faster coalescence. In the co-rotating case, the resulting binary
evolution depends on a subtle balance between accreted mass and angular
momentum, with the net effect that the MBHB can also slightly expand.
A longer simulation is required to confirm whether the
expansion seen in {\runa} \F{0.0} is only transient or if it holds
in the long term.}

\section{Summary and discussion}
\label{sec:summary}

On this paper we have introduced a new suite of simulations to study the
evolution of a binary embedded in a turbulent environment, where we expect
several gaseous clumps to infall and dynamically interact with the binary.
The main goal was to overcome the limitations of the
Monte Carlo evolution presented in \paperII~by directly modelling the
interaction of a MBHB with a sequence of incoherent accretion events.
These models combine the `discrete' evolution due to the prompt interaction
of each individual cloud, with the secular torques exerted by the left-over
gas. The latter could not be numerically resolved in the single cloud models,
but could have a role on the dynamics of both gas and binary.

In order to directly compare with the results of the Monte Carlo evolution
of \paperII, the cloud orbits where sampled using the same probability density
functions. In particular, we have used the $F$-distributions for the angular
momentum orientation which are an indication of the anisotropy levels in the
surrounding medium. These distributions have been linked to the morphological
and kinematical properties of the host galaxy \citep{Sesana2014}.

Our main findings can be summarised as follows:
\begin{enumerate}
\item The MBHB angular momentum evolution follows the injection of momentum
  in the form of clouds. That is, the binary monotonically increases its
  angular momentum {(even though it still shrinks)} in the \F{0.0} model
  where each new cloud adds to the total momentum budget, while continuously
  decreasing for \F{1.0} where the opposite occurs. This is consistent with
  the scenario where the orbital evolution is dominated by direct exchange of
  angular momentum through capture and accretion, as demonstrated by the
  single cloud simulations in  \paperII.
\item The binary exhibits an almost identical orbital evolution at every
  investigated resolution. This is because the dynamical evolution of the
  system is dominated by capture and gravitational slingshot, which are
  properly resolved even with a low number of particles.
\item The largest differences are present in the eccentricity evolution due
  to typically low values ($e\lesssim 0.02$), making it much more sensitive
  to noise. Because of this, we focused the analysis on the semimajor
  axis and mass evolution, considering that the binary remains circular
  throughout the whole simulation.
\item Also concordant with the single cloud simulations, the evolution of the
  semimajor axis is strongly dependent on the distribution of the infalling
  gas, being fastest when it features more counter-rotating clouds. Conversely,
  the binary mass evolution is roughly independent of the cloud distribution
  during most of the interaction.
\item By directly comparing the observed binary evolution with a
  semi-analytical model constructed on the basis of the single cloud
  simulations, we found that accretion onto the MBHB increases considerably
  (by about a factor of three). This occurs because new incoming material
  is able to drag inward a fraction of the surrounding gas leftover by
  previous accretion events.
\item Enhanced accretion has a strong impact on the evolution of the
  binary. Because of angular momentum cancellation with a larger
  mass of infalling gas, the binary evolution is accelerated in the
  isotropic and counter-rotating cases (by 20-to-50\% and a factor $\approx$2,
  respectively, see Figs.~\ref{fig:mcs_sim_events} and~\ref{fig:RunA_sim_events}).
  In the co-rotating case, the evolution depends
  on a fine balance of mass and angular momentum accretion, and can lead,
  at least temporarily, to the expansion of the binary semimajor axis.
\item In the \F{0.0} and \F{0.5} models we observe the intermittent formation
  of well-defined, coplanar and co-rotating circumbinary discs.
  The discs present well-known features, in particular, a cavity located at
  $r\approx 2a$ which is maintained by the resonances of the orbiting gas
  with the binary's gravitational potential.
\item Similarly, we observed the formation of intermittent coplanar and
  \emph{counter-rotating} circumbinary discs in the \F{1.0} model. Due to the
  absence of resonances, in this case the gas is able to orbit much closer to
  the binary.
\item By computing the gravitational torques exerted by the surrounding gas,
  we have shown that these discs are able to transport some of the binary's
  angular momentum via gaseous streams that are slingshotted away once they
  get close to the MBHs. This process can continue shrinking the binary orbit
  in the absence of new impacting clouds. However, due to the low mass of
  these discs, their effect on the binary orbital decay is subdominant with
  respect to the capture and subsequent accretion of the incoherent material
  deposited by clouds in the vicinity of the MBHB.
\end{enumerate}

An interesting case is the isotropic scenario \F{0.5} where the total angular
momentum brought by the gas is close to zero, at least after a significant
number of clouds. In this scenario, the intersection of highly misaligned
streams tend to effectively cancel angular momentum, forcing material to
plunge onto the binary. Therefore, we do not expect a steady circumbinary
disc to survive under these conditions. Following a sequence of 10 clouds,
however, there is a clear circumbinary structure rising from the non-accreted
material, which is already producing a noticeable effect on the binary orbit.
This implies that the formation of circumbinary discs is a relevant process
even for the \F{0.5} model, even if these discs are transient and with
negligible self-gravity. One of the assumptions of the Monte Carlo models
presented in \paperII~is that there is no circumbinary disc during the
isotropic scenario due to the low total angular momentum of the gas.
Our results clearly show the inadequacy of this assumption.
This occurs because corotating material is able to gain more angular
momentum via resonances and gravitational slingshot compared to
counter-rotating gas, thus forming meta-stable disc-like configurations.

At this point, it is important to stress that our simulations are
subject to a number of limitations, that we now briefly consider.
The mass of the circumbinary discs intermittently forming throughout the
simulations is small compared to the binary and hence their self-gravity
is negligible \citep{MaureiraFredes2018}. In this situation the outwards transport of
angular momentum throughout the disc will occur through an unresolved viscosity
\citep[see e.g.][]{Syer1995,Ivanov1999,Ragusa2016}.
For geometrically thin and optically
thick discs, such viscosity is commonly parametrised using the $\alpha$-disc
model \citep{SS1973}, where $\alpha$ represents the strength of the internal
viscous stresses. In our simulations, however, the only source of viscosity
is the artificial term introduced to capture shocks, hence the transport of
angular momentum throughout the disc is likely not adequately modelled.
Additionally, the evolution of the binary-disc system might be affected by
the simple thermodynamics adopted in these models. As discussed in
\citet{Roedig2012}, the morphology of the streaming gas has a major impact in
the binary evolution and the structure of the surrounding disc, thus it is
directly influenced by the gas thermodynamics. Moreover,
\citet{Tang2017} using 2D hydrodynamical simulations of circumbinary accretion
discs found that the torques are sensitive to the sink prescription: slower
sinks result in more gas accumulating near the MBHs, driving the binary to
merge more rapidly. Nevertheless, all these effects become relevant for the
\textit{long-term} evolution of a MBHB embedded in a circumbinary disc,
where torques mediated by resonances play an important role in the evolution
of the system. Conversely, during the infall of a sequence of clouds,
the relevant mechanisms driving the binary's orbital response are almost
purely dynamical, namely, gas capture and gravitational slingshot, which are
well resolved with these models.
Therefore, a more physical treatment for the viscosity, thermodynamics and
accretion in required only if one wants to resolve the long-term evolution
of the MBHB surrounded by a circumbinary disc arising in the aftermath of
the accretion events.

{Even though the self-gravity of the discs presented in
this paper is negligible, it could certainly play a role in the evolution of the
system depending on the cloud distribution.
If the leftover material were to settle onto a well defined circumbinary
structure (which depends on the frequency of events, see
\citealt{MaureiraFredes2018}), it could eventually fragment to form
stars, as demonstrated by \citet{Dunhill2014}, especially for higher
cloud masses.
Based on our results, we expect fragmentation on the left-over gas
to marginally decrease the effects of the infalling clouds,
mainly because it would reduce the amount of available material
to interact with the new incoming clouds.
Nonetheless, for these near-radial clouds,
the transient effect during the first impact is still efficient to
evolve the binary (\paperII).
Furthermore, a fragmenting circumbinary disc can continue
shrinking the binary orbit by sending stars into the loss-cone,
as shown by \citet{Pau2013}, although the efficiency
drops considerably with respect to a similar, non-fragmented disc.}

Another delicate point is the simplified treatment of accretion, whereby
every bound particle entering the sink radius is added to the MBHs. When
scaled to our fiducial physical model ($M_{\rm bin}=10^{6}\msun$, $a_0=0.2$ pc),
this results in sparse episodes of highly super-Eddington accretion. The
exact fate of the gas during this `accretion spikes' is uncertain. The
extremely high density of the accretion flow might cause photon diffusion
timescale to be longer than the accretion timescale, thus advecting photons
into the MBHs, in turn allowing super-Eddington accretion rates. Conversely,
radiation pressure in the proximity of the MBHs might cause copious mass
loss through powerful winds, significantly decreasing the amount of accretion.
As noticed in \paperII, the exchange of angular momentum is mostly
mediated by {\it capture} of gas in bound orbits around the individual MBHs
rather than subsequent accretion. Even assuming that the majority of
the gas is accelerated away in form of isotropic winds, the evolution of
the binary is affected at a level of 20-30\%. This conclusion, however,
does not take into account the interaction of the outgoing wind with the
infalling clouds. If winds are isotropic and strong enough to clear the
infalling gas on large scales, this would quench the supply of material
onto the MBHB, eventually stalling the binary. A proper investigation of
this potential issue requires the implementation of a suitable feedback
prescription, and we plan to explore this in future work.

A somewhat different aspect that might be impacting our results is the
resolution elements of the individual clouds. For instance, in their simulations
\citet{Dunhill2015} found that the dynamics of gaseous streams going through
the cavity is resolution dependent. As discussed above, the streams play an
important role transporting angular momentum once the circumbinary disc has
formed, thus the late evolution measured using the lower resolution runs is
likely affected by this issue.
{In fact, we find some differences with resolution in the gravitational
torques studied in \S\ref{sec:evolution}.
For the \F{0.0} and \F{0.5} distributions the differences
are relatively small, $\sim$16\% and $\sim$6\%, respectively,
between the \texttt{50k} and \texttt{1m} models.
On the other hand, for the \F{1.0} distribution,
we found up to a factor of $\sim$2
difference between the \texttt{50k} and \texttt{500k} resolutions.
These larger differences are likely because the material
close to the binary self-interacts more strongly in the retrograde case
compared with the other two distributions, since the absence of resonances
does not allow the gas to gain  angular momentum.
In this scenario, the interaction of the different gas streams determines
the amount of material in the circumbinary disc, and this is more sensitive
to the  resolution of the clouds.
Since we are not observing strong resolution dependence in the binary
semimajor axis, these effects are still sub-dominant with respect to the
accretion of gas. Consequently, resolution would become much more important
for low infall frequency  of clouds, where the material is expected to settle
into a more `stable' circumbinary disc.}

{Another relevant aspect that might have implications for our conclusions is
the chosen mass ratio between the binary and the incoming clouds.
As explained in \paperI, we based our study in the simulations presented by
\citet{BR08}, where they numerically modelled the infall of a molecular cloud
onto single MBHs with the goal of explaining the stellar distribution observed
in our Galactic Centre.
Their fiducial model is a $10^4M_\odot$ cloud infalling onto a $10^6M_\odot$
black hole, hence we adopted the same mass ratio in our simulations.
Based on the numerical studies of multiphase gas in the interstellar medium
\citep[e.g.][]{Hobbs11,Gaspari2013,Gaspari2015,Fiacconi2018},
we expect the orbital parameters and internal properties of the cold
clumps to be independent on the
mass of the central object, as they are determined by large-scale
physical processes in the galaxy such as turbulence, rotation, cooling,
heating, among others. So, in principle, the relative effect that each
cloud has on the binary evolution should scale linearly with the
MBHB-to-cloud mass ratio.
We also deliberately chose identical clouds infalling
onto the MBHBs, in order to single out the impact of their different
orbits. This is clearly not a realistic setup, as a distribution of
clump masses is expected. More sophisticated initial conditions can
be derived from numerical models of realistic
galaxies that include all the relevant physical mechanisms. We plan
to extend our simulation setup in this direction in future work.}

{Despite the aforementioned limitations,}
the results exhibited in this paper suggest that the secular
effects of the left-over material further increase the efficiency of the
incoherent infalling clouds to bring the MBHs towards coalescence respect
to a scenario that only considers the evolution during the prompt accretion
phase, as the one presented in \paperII.
Consequently, provided there is a reasonable rate of events,
these models confirm that infalling clouds present a viable mechanism to
efficiently merge these binaries within less then a Gyr.

\section*{Acknowledgements}

{We are grateful to the anonymous referee for a very insightful report
that helped improving the clarity of the paper.}
We also thank Volker Springel for his suggestions on how to improve the conservation
of angular momentum of these models, and
Johanna Coronado for comments on the manuscript.
The simulations were performed partially
between the \textit{sandy-bridge} nodes at HITS, and \textit{datura} and
\textit{minerva} clusters at the AEI.  FGG acknowledges support from the
CONICYT-PCHA Doctorado Nacional scholarship,
DAAD in the context of the
PUC-HD Graduate Exchange Fellowship,
the European Research Council
under ERC-StG grant EXAGAL-308037, and the Klaus Tschira Foundation.
CMF acknowledges support from the Transregio 7 ``Gravitational Wave Astronomy''
financed by the Deutsche Forschungsgemeinschaft DFG (German Research Foundation).
CMF acknowledges support from the DFG Project  ``Supermassive black holes,
accretion discs, stellar dynamics and tidal disruptions'', awarded to PAS, and
the International Max-Planck Research School.
PAS acknowledges support from the Ram{\'o}n y Cajal Programme of the Ministry of
Economy, Industry and Competitiveness of Spain, as well as the COST Action
GWverse CA16104. This work has been partially supported by the CAS President's
International Fellowship Initiative.
JC acknowledges support from CONICYT-Chile through FONDECYT (1141175) and Basal (PFB0609) grants.
This work was partially developed while
JC was on sabbatical leave at MPE.  JC and FGG acknowledge the kind hospitality
of MPE, and funding from the Max Planck Society through a ``Partner Group''
grant.


\bibliographystyle{style/mnras}
\bibliography{bib/refs} 

\label{lastpage}

\bsp    

\end{document}